\documentclass[structabstract]{aa}  
\usepackage{graphicx}
\usepackage{natbib}
\usepackage{mathrsfs}
\newcommand{\hcn}{H$^{13}$CN}
\newcommand{\hco}{H$^{13}$CO$^+$}
\newcommand{\hcop}{HCO$^+$}
\newcommand{\kms}{km~s$^{-1}$}

\begin{document}
\bibliographystyle{aa}

   \title{Gas dynamics in Massive Dense Cores in Cygnus-X}
  \authorrunning{T. Csengeri et al.}
   \subtitle{}

   \author{T. Csengeri \inst{1},
          S. Bontemps \inst{2},
          N. Schneider\inst{1}, 
          F. Motte\inst{1},
          S. Dib\inst{1,3}
          }

     \institute{Laboratoire AIM, CEA - INSU/CNRS - Universit\'e Paris Diderot, IRFU/SAp CEA-Saclay, 91191 Gif-sur-Yvette, France    
                   \email{timea.csengeri@cea.fr}
            \and
            OASU/LAB-UMR5804, CNRS, Universit\'e Bordeaux 1, 33270 Floirac, France    
            \and
            Universit\"{a}ts-Sternwarte M\"{u}nchen, Scheinerstra$\ss$e 1, D-81679 M\"{u}nchen, Germany
          }

   \date{Received ; accepted }

 
  \abstract
   {The physical conditions in massive dense cores (MDCs) forming
    massive stars and clusters, are not well constrained. Observations are lacking to confront the theories. Recently an extensive study has been started towards the most massive and youngest cores in the Cygnus-X molecular complex and as a first result showed 
 exceptional fragmentation properties in a sample of five cores where individual massive protostars have been recognized.}
   {We study here the kinematic properties of dense gas surrounding massive protostars in these five cores in order to investigate whether turbulent support plays a major role in stabilizing the whole core against a rapid fragmentation into Jeans-mass objects. Alternatively, the observed kinematics could indicate a high level of dynamics suggesting that the cores are actually not in equilibrium and dynamical processes could be the main driver to build up the final stellar masses. }
   {We present IRAM 30m single-dish (\hco and \hcop) data and IRAM Plateau de Bure Interferometer high angular-resolution observations of dense gas tracers ({\hco} and {\hcn}) to reveal the kinematics of molecular gas at scales from 0.03 to 0.1~pc. 
   }
   {Using radiative transfer modeling we show that {\hco} is depleted within the envelopes of massive protostars and traces the bulk of material surrounding the protostars rather than their inner envelopes. {\hcn} shows a better correspondence with the peak of the continuum emission, possibly due to abundance anomalies and specific chemistry in the close vicinity of massive protostars. Analyzing the line-widths we show that the observed line-dispersion of \hco~at the scale of MDCs is smaller than expected from the quasi-static, turbulent-core model. At large-scales, global organized bulk motions are identified for 3 of the MDCs. At small-scales, several spectral components are identified in all MDCs showing filamentary structures and intrinsic velocity gradients towards the continuum peaks. The dynamics of these flows show diversity among the sample and we link this to the specific fragmentation properties of the MDCs. Altogether this points to different initial conditions towards CygX-N3 and -N63 as compared to CygX-N12, -N48 and -N53, which may represent different evolutionary stages.}
   {No clear evidence is found for a turbulence regulated, equilibrium scenario within the sample of MDCs. We propose a picture in which MDCs are not in equilibrium and their dynamics is governed by small-scale converging flows, which may initiate star-formation via their shears. We suggest that dynamical processes are linked to the formation of proto-clusters and high-mass protostars.}


   \maketitle
%

\section{Introduction}
Several models have been proposed to explain what processes dominate and form structures within giant molecular complexes, thus determining the origin of star-forming cores and ultimately of stars. After early scenarios based on quasi-static evolution of molecular clouds, primary regulated by the interplay between gravitation and magnetic fields (e.g. \citealp{MS76}; \citealp{Shu87}; \citealp{Mouschovias99}), the importance of turbulence has been recently emphasized and taken into account (see \citealp{MLK04}; \citealp{MO} and references therein). For high-mass star formation, the consideration of the relative importance of these different supports against gravity has led to two main alternate views. In the first scenario molecular clouds are close to be in equilibrium and are evolving on long time-scales (of the order of a few dynamical crossing times, e. g. \citealp{Larson81}; \citealp{McKee99}). Since turbulence decays within a dynamical time (e.g. \citealp{ML98}; \citealp{PN99}), it needs to be replenished over such long time-scales (e.g. by stellar winds, supernovae explosions, outflows or other feedbacks, \citealp{NS80}; \citealp{Mckee89}) in order to maintain a quasi-static state. As a result, turbulence can regulate the formation of structures within molecular clouds on large scales, while at small scales isotropic, supersonic turbulence could be the main support of cores against collapse, thus complementing their thermal pressure \citep{MT02, MT03}. It naturally offers a description for the formation of high-mass stars as a scaled-up scenario of the low-mass, quasi-static star-formation process. Accretion rates of up to $\sim$ $10^{-3}$ M$_{\odot}$yr$^{-1}$ are reached for these turbulent, massive cores, which is high enough to overcome radiation pressure and enables the protostar to collect a mass larger than 8-10 M$_{\odot}$. The \textit{turbulence regulated, quasi-static scenario} has fundamental predictions: 1.) the existence of massive cores, which are not yet forming stars (an equivalent of low-mass pre-stellar cores); 2.) a high-level of isotropic, supersonic turbulence on the size-scale of massive cores; 3.) the collapse of massive cores directly into a single (or close binary) massive object. To be more precise, gravitationally unstable density fluctuations are predicted, but with a low number of gravitationally bound fragments within a single core.

Alternatively, 
clouds and clumps may be unstable and fast evolving structures with high-level of dynamics (e.g \citealp{VS02}; \citealp{KH09}). Briefly, large-scale turbulent flows create structures by shock-dissipation and, as shown by numerical simulations, supersonic turbulence fragments the gas efficiently in very short time-scales (\citealp{Padoan01}; \citealp{VS07}). This picture of \textit{gravo-turbulent fragmentation} naturally provides the seeds for star-formation by creating gravitationally bound density fluctuations (\citealp{Klessen}; \citealp{Dib07}). In hydrodynamic simulations this results in fragments with mass around the local Jeans mass ($\sim$1 M$_{\odot}$), and thus, it is efficient to produce low mass-fragments, while there is a significant lack of massive ones (\citealp{Bonnel07}; \citealp{Dib10}). Bonnell et al. proposed as a solution that seeds may continue to accrete material from regions which were originally not bound to the protostellar envelope. This scenario is therefore introduced as \textit{competitive accretion} (\citealp{Bonnel01}; \citealp{BB06}). Massive seeds continue to grow in mass via two mechanism: benefitting from the channelling of material on the common cluster potential well or if the relative velocities are too low, the accretion processes are dominated by the tidal field of the protostar itself. 

According to \citet{HT08} {\sl magnetic fields} can lower the level of fragmentation resulting in only a few more massive objects, which continue to gain mass via channelling of material along the field lines. Note, however, that strong perturbations (that must be seeded in the clumps/cores in order to allow them to fragment in the presence of magnetic fields) may also naturally be reproduced in the gravo-turbulent fragmentation of magnetically supercritical and nearly critical clouds as well \citep{Dib10b}. In the analytic description of \citet{HC08} (magnetic field included), the interplay between gravothermal and gravoturbulent fragmentation can naturally reproduce the observed frequency of massive stars - depending on the Mach-number within molecular clouds. The fundamental predictions of the dynamical view are, that 1.) non-isotropic turbulence/velocity fields should be present with dynamical patterns;  2.) 
a high-level of fragmentation is expected with the most massive stars forming in the center of clusters (since they are the most efficient in competing for mass). 

We intend to put observational constrains on the key parameters of current star-formation theories by performing a systematic, high angular-resolution study of massive dense cores (MDCs) in Cygnus-X. \cite{B09} (hereafter Paper I) presented 1mm and 3mm continuum maps of 6 IR-quiet MDCs, which were obtained by the Plateau de Bure Interferometer. Twenty three fragments were detected in a sample of six cores, which challenge the turbulence regulated, monolithic collapse view - according to which these cores should be much less fragmented. However, the most compact MDC, CygX-N63, stands out from the picture since it contains only a single, compact object, and could therefore be a representative of a single massive protostar with an envelope mass of $\sim$60 M$_{\odot}$. In addition, a total of 8 fragments are found to be precursors of OB stars with envelope masses ranging from 6-23 M$_{\odot}$. We found that the level of fragmentation within these MDCs is higher than expected in the quasi-static formation scenario, but lower than predicted by a pure gravo-turbulent scenario. Also, we compared the mass of the cores with the total mass contained in the fragments, which is 28, 44 and 100\% for CygX-N12, CygX-N53 and CygX-N63, respectively. 
This efficiency of turning the cloud mass into protostars is significantly higher than what has been observed for clusters \citep{LL03} and predicted for low-mass stars and clusters \citep{MM00}. 

The origin/physical reason of such diverse fragmentation properties is unknown, and can only be elucidated with kinematic studies using molecular lines.
Here we present a study of high density molecular tracers using single-dish observations of {\hcop}(J=1--0) and {\hco}(J=1--0) line emission from a sample of MDCs together width high angular-resolution, interferometric observations of {\hco}(J=1--0) and {\hcn}~(J=1--0) lines. These molecules have a high critical density ($>$$10^4 $ cm$^{-3}$, e.g. \citealp{Bergin07}) and are known to be tracers of dense gas, therefore allowing to probe the dynamics of gas within MDCs.

 

\section{Source selection and observations}
\label{sec:obs}

\subsection{IR-quiet MDCs in Cygnus-X and sample selection}
  \begin{figure}[]
   \centering
    \includegraphics[width=9cm]{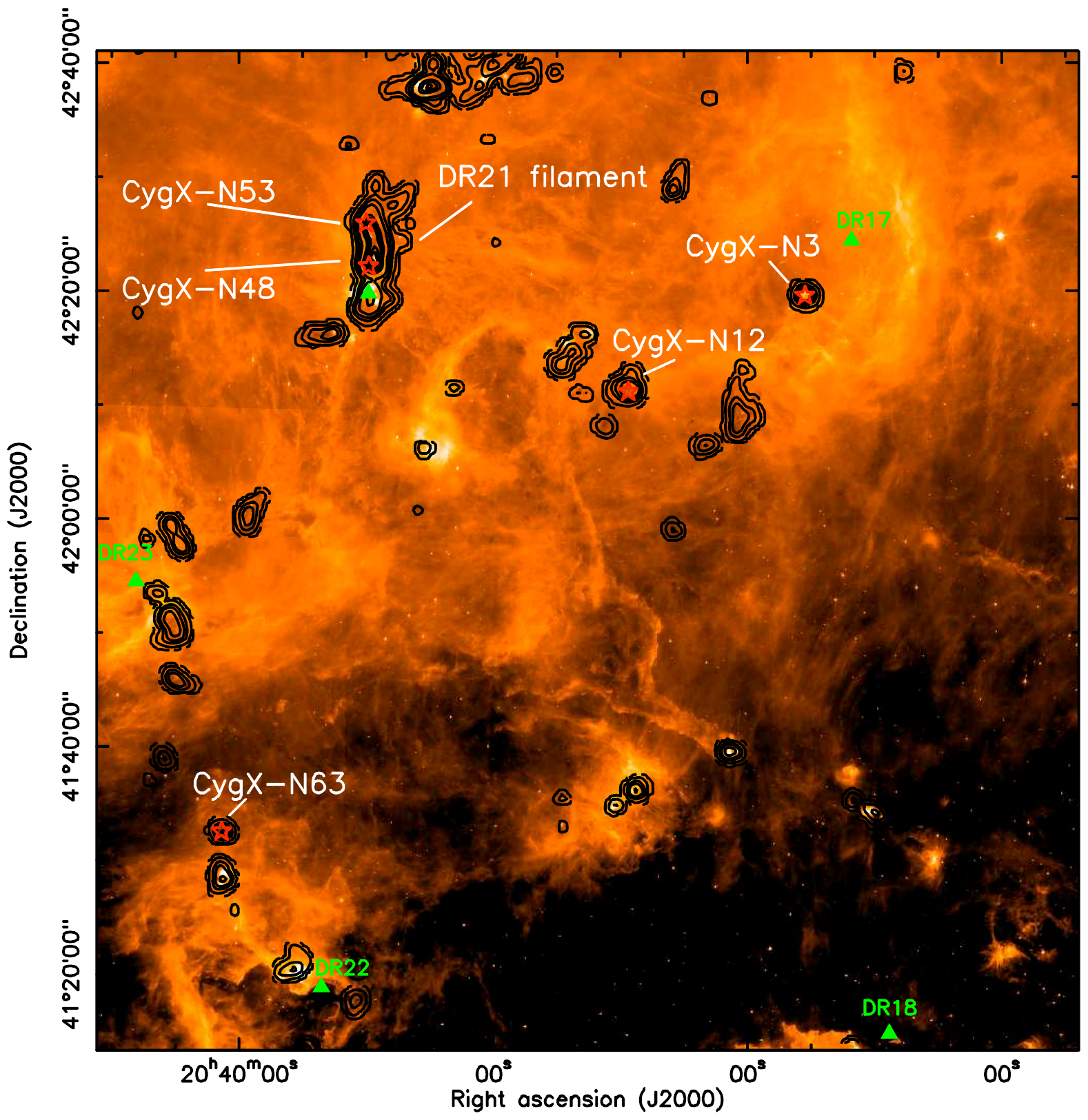}
 \caption{\textsl{Spitzer}/IRAC 8 $\mu m$ map \citep{Hora} showing a part of Cygnus-X North. Black contours show integrated intensity map of the N$_2$H$^+$ (J=1--0) line obtained with the FCRAO \citep{Schneider_prep}. Red stars indicate the position of the sample of MDCs. Green triangles indicate the position of {H\scriptsize{II}\normalsize} regions \citep{DR}.}
     \label{fig:overview}
    \end{figure}


The source selection dates back to the 1.1mm MAMBO survey of Cygnus-X \citep{M07}, from which we selected the most massive IR-quiet MDCs. All cores are compact with a similar size as nearby low-mass cores ($\sim$0.13 pc), but they are 10 times denser (with an average density of 1.9 $\times$ 10$^5$ cm$^{-3}$) and 20 times more massive ($\sim$ 60--200 M$_{\odot}$). Therefore one would expect them to be actively forming stars, which is in fact confirmed by outflow emission traced in SiO. However, the lack of strong mid-IR and free-free emission indicates that they must be in an early stage of their evolution. 

As shown in Figure \ref{fig:overview}, two of the selected cores (CygX-N48 and CygX-N53) are located in the DR21 filament which is the most massive and densest region of Cygnus-X, and is a well-known region of massive star formation (see e.g. \citet{Schneider_prep} and references therein). CygX-N48 corresponds to the submillimeter source DR21(OH)-S, while CygX-N53 is situated close to W75-FIR3 \citep{Chandler93}. The 3 other selected MDCs have been discovered by the MAMBO survey and are situated in more isolated, but still prominent molecular clumps bright in N$_2$H$^+$ line emission, which equally traces cold, dense gas. CygX-N3 in the western part is located close to DR17 which is an {H\scriptsize{II}\normalsize} region most probably excited by two OB clusters (clusters \# 12 and 14 in \citealp{LK02}). These clusters are shaping the cloud and CygX-N3 corresponds to the tip of a pillar-like cloud. CygX-N12 is also located in a cometary shaped cloud, probably influenced by the same OB clusters. However the cloud seems to be less compressed from outside than CygX-N3 \citep{Schneider06}. Finally, CygX-N63 is placed in the south of DR21 in the DR22-DR23 filament. We adopt a distance of 1.7 kpc for Cygnus-X \citep{Schneider06} so that 1{\mbox{$^{\prime\prime}$}} corresponds to 1700 AU.

\subsection{Observations}

\subsubsection{Observations with the IRAM 30m telescope}
OTF maps of spectral lines of CygX-N3, -N12 and -N63 were observed on 5 June 2007 with the IRAM\footnote{IRAM is supported by INSU/CNRS (France), MPG (Germany) and IGN (Spain).}  30m telescope using the A100 receiver with the VESPA correlator for {\hco} (J=1--0) at 89.188 GHz and {\hcop} (J=1--0) at 86.75 GHz. The average system temperature was $\sim$85 K. Sources CygX-N48 and -N53 were observed between 2-5 June 2007 as part of a large-scale OTF mapping of the DR21 filament \citep{Schneider_prep}, and have average system temperatures around 110 K. The data were corrected for the main beam efficiency of 0.78 and have a velocity resolution of 0.135 km~s$^{-1}$. The HPBW at this frequency is~$\sim$28{\mbox{$^{\prime\prime}$}}.

\subsubsection{Observations with the Plateau de Bure Interferometer}
\begin{table*}
\centering                          
\caption{Observation and data reduction parameters of the interferometric data. We give beam-sizes, position angle and noise levels of {\hco} for the combined dataset of PdBI+30m. For {\hcn}, parameters of the interferometric data are shown.}
\begin{tabular}{c c c c c c c c c }        
\hline\hline   
& & & \multicolumn{3}{c}{{\hco}}& \multicolumn{3}{c}{{\hcn}}     \\  
\hline       
Source & \multicolumn{2}{c}{Phase center} & Beam size  & PA & Obtained rms & Beam size  & PA & Obtained rms \\
& RA(J2000) & Dec(J2000) & [{\mbox{$^{\prime\prime}$}}$\times${\mbox{$^{\prime\prime}$}}] & & [mJy/beam] & [{\mbox{$^{\prime\prime}$}}$\times${\mbox{$^{\prime\prime}$}}] & & [mJy/beam] \\    
\hline                        
  CygX-N3   & 20:35:34.1 & 42:20:05.0   & 3.78 $\times$  3.09 &  63$^\circ$ & 16.3 & 3.67 $\times$  3.03 & 53$^\circ$ &  11.8 \\     
   CygX-N12 & 20:36:57.4 & 42:11:27.5  &  4.35 $\times$ 3.77 & 72$^\circ$ & 21.2 & 4.13 $\times$ 3.58 & 65$^\circ$ & 14.5\\
   CygX-N48 & 20:39:01.5 & 42:22:04.0  & 4.52 $\times$ 3.52 & 75$^\circ$ & 21.0 & 4.27 $\times$ 3.36 & 71$^\circ$ &  13.5\\
   CygX-N53 & 20:39:03.1 & 42:25:50.0  & 4.42 $\times$ 3.5   & 72$^\circ$ & 20.3 & 4.25 $\times$ 3.34 &  69$^\circ$ &  14.8\\
   CygX-N63 & 20:40:05.2 & 41:32:11.9  & 4.45 $\times$ 3.83 & 79$^\circ$ &21.3 & 4.16  $\times$ 3.58 & 66$^\circ$  & 14.3 \\ 
\hline                                  
\end{tabular}
\label{tab:obsparam}
\end{table*}

\defcitealias{B09}{Paper I}
We used the IRAM Plateau de Bure Interferometer (hereafter PdBI) to perform high angular-resolution observations at 1mm and 3mm. A more detailed description of the observations can be found in \citetalias{B09}. We obtained line emission measurements of {\hco} (J=1-- 0), and H$^{13}$CN (J=1-- 0) at  86.75 and 86.34 GHz, respectively, with a velocity resolution of 0.134 km~s$^{-1}$. The observations were done in track-sharing mode with two targets per track. The D configuration tracks were obtained between June and October 2004 with 5 antennas with baselines ranging from 24 m to 82 m. The C configuration tracks were obtained in November and December 2004 with 6 antennas in 6Cp configuration with baselines ranging from 48 m to 229 m. We mostly used as phase calibrator the bright nearby quasar 2013+370 and as flux calibrator the bright evolved star MWC349 which is located in Cygnus-X. 

\subsection{Data reduction and cleaning}
We used the GILDAS software\footnote{See http://www.iram.fr/IRAMFR/GILDAS/} for the data reduction and analysis of single-dish and interferometric data. The interferometric {\hco} dataset was reduced both with and without adding short-spacings information. The short-spacings information was obtained with the IRAM 30m telescope and implemented with the standard techniques of GILDAS.
No zero-spacings information was available for the {\hcn} dataset. For the cleaning procedure we used a naturally weighted beam to search for clean components within a circle. The resulting parameters of the data-reduction are summarized in Table \ref{tab:obsparam}. All maps are corrected for primary beam attenuation. 

%
\section{Results}
\label{sec:results}
As a continuation of the interferometric mm-continuum studies of \citetalias{B09}, we focus here on a study of the distribution and the kinematics of dense gas traced by {\hcop}, {\hco} and {\hcn} in our sample of MDCs. We first present single-dish maps of {\hcop} and {\hco}, on which we base later our radiative transfer line modeling in Sect. \ref{subsec:simline}. Then, spectra from both single-dish and interferometric maps are shown, followed by line integrated maps of {\hco} and {\hcn}.

  \begin{figure*}[]
   \centering
    \includegraphics[width=16.0cm]{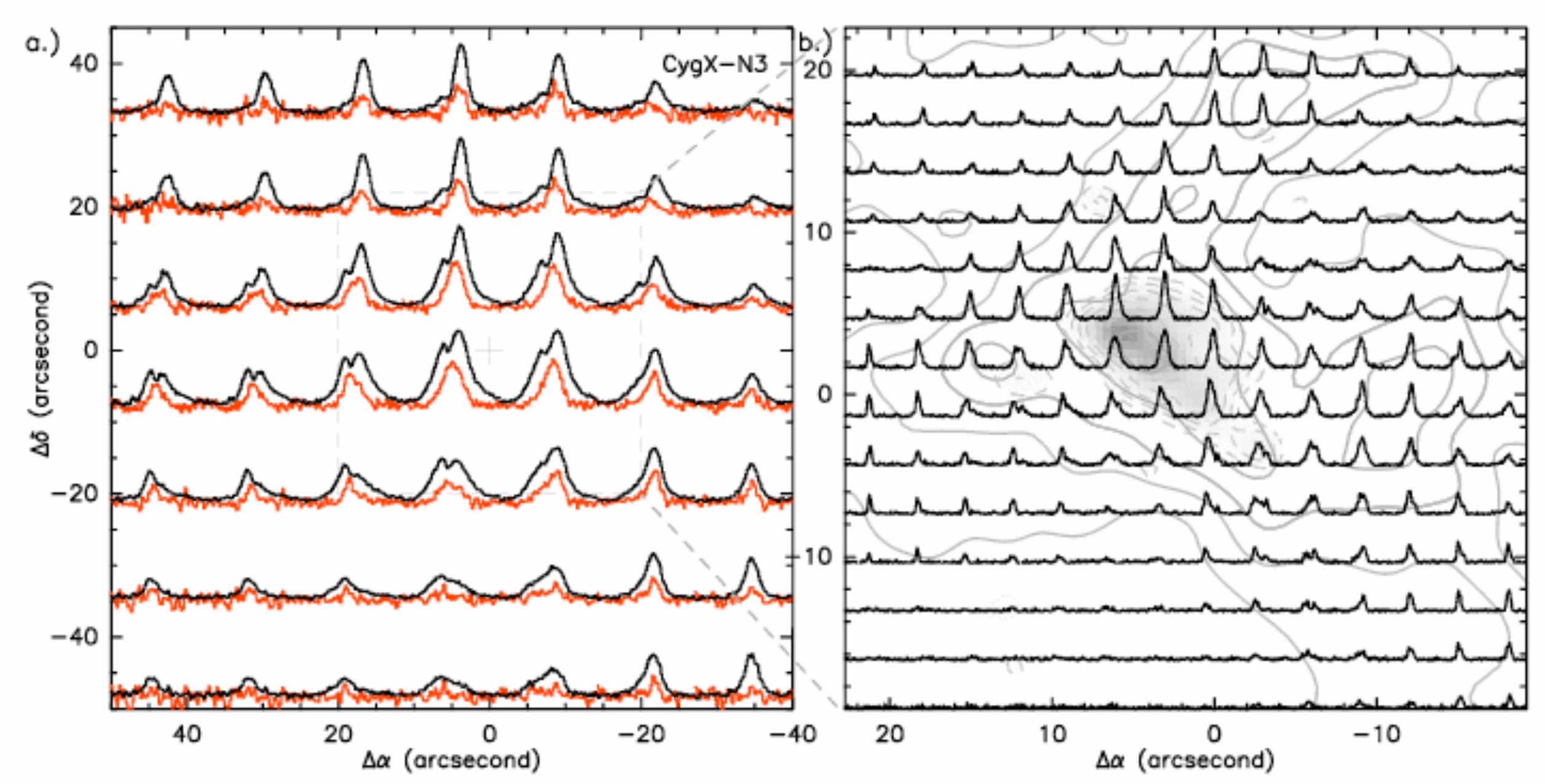}
    \includegraphics[width=16.0cm]{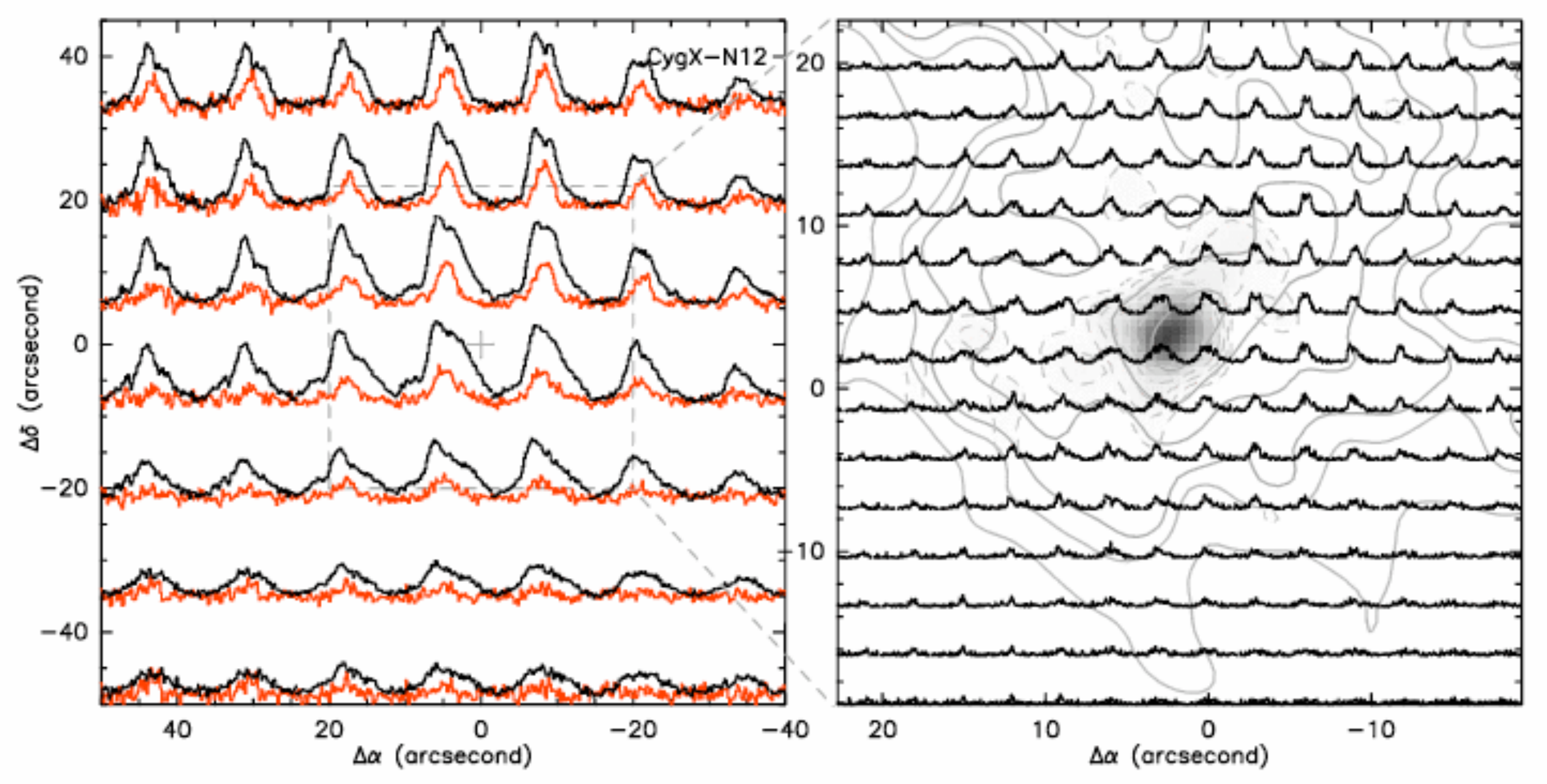}
    \includegraphics[width=16.0cm]{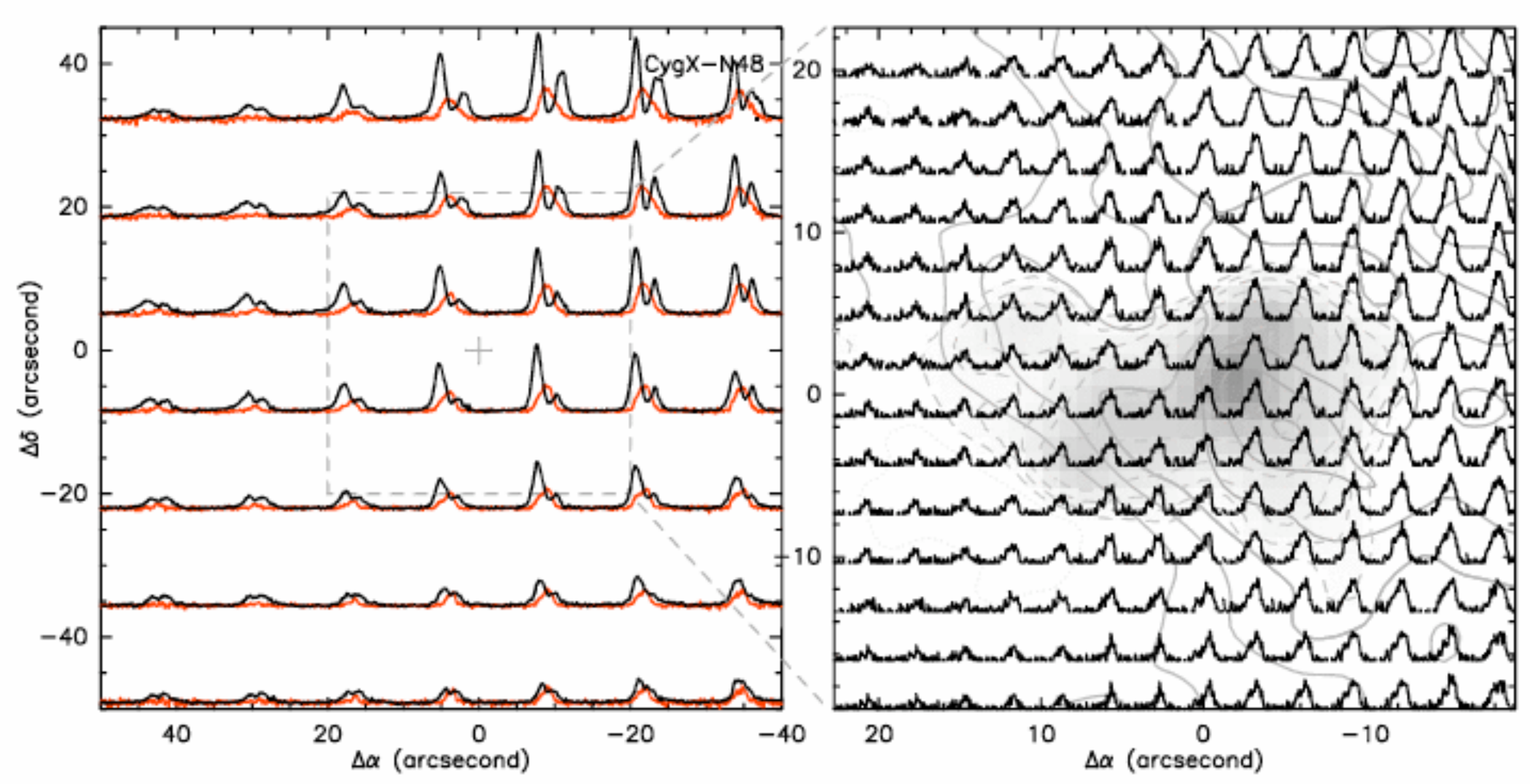}
     \caption{\textit{a)} Spectral line maps of \hcop (black) and {\hco} (red) obtained with the IRAM 30m telescope and regridded to full-sampling. The velocity range is 10 to 20~km~s$^{-1}$ for CygX-N3, -N12, -15 to +5 km~s$^{-1}$ for CygX-N48 and CygX-N53, -10~to~0~km~s$^{-1}$ for CygX-N63. The intensity of the optically thin line ({\hco}) is degraded by a factor of 2-4. Dashed squares indicate the size-scale of the right panel.
 \textit{b)} Spectral line maps of {\hco} PdBI combined with IRAM 30m, where the spectra-cubes were regridded to 3{\mbox{$^{\prime\prime}$}} resolution. Velocity ranges are 10 to 20 km~s$^{-1}$ for field CygX-N3,- N12 and -10 to 0 km~s$^{-1}$ for CygX-N48, -N53 and -N63, respectively.  The grey-scale background image is the 3mm continuum emission obtained simultaneously with the PdBI (from 3$\sigma$ to75 $\sigma$). (The grey dashed contours show the same logarithmic levels as described in \citetalias{B09} Figure 2.)  Grey solid line contours show the integrated intensity maps of {\hco} (same as in Fig. \ref{fig:pdbi} {\sl a)}).}
   \label{fig:sda}
   \end{figure*} \addtocounter{figure}{-1}
  \begin{figure*}[]
   \centering
   \includegraphics[width=16.0cm]{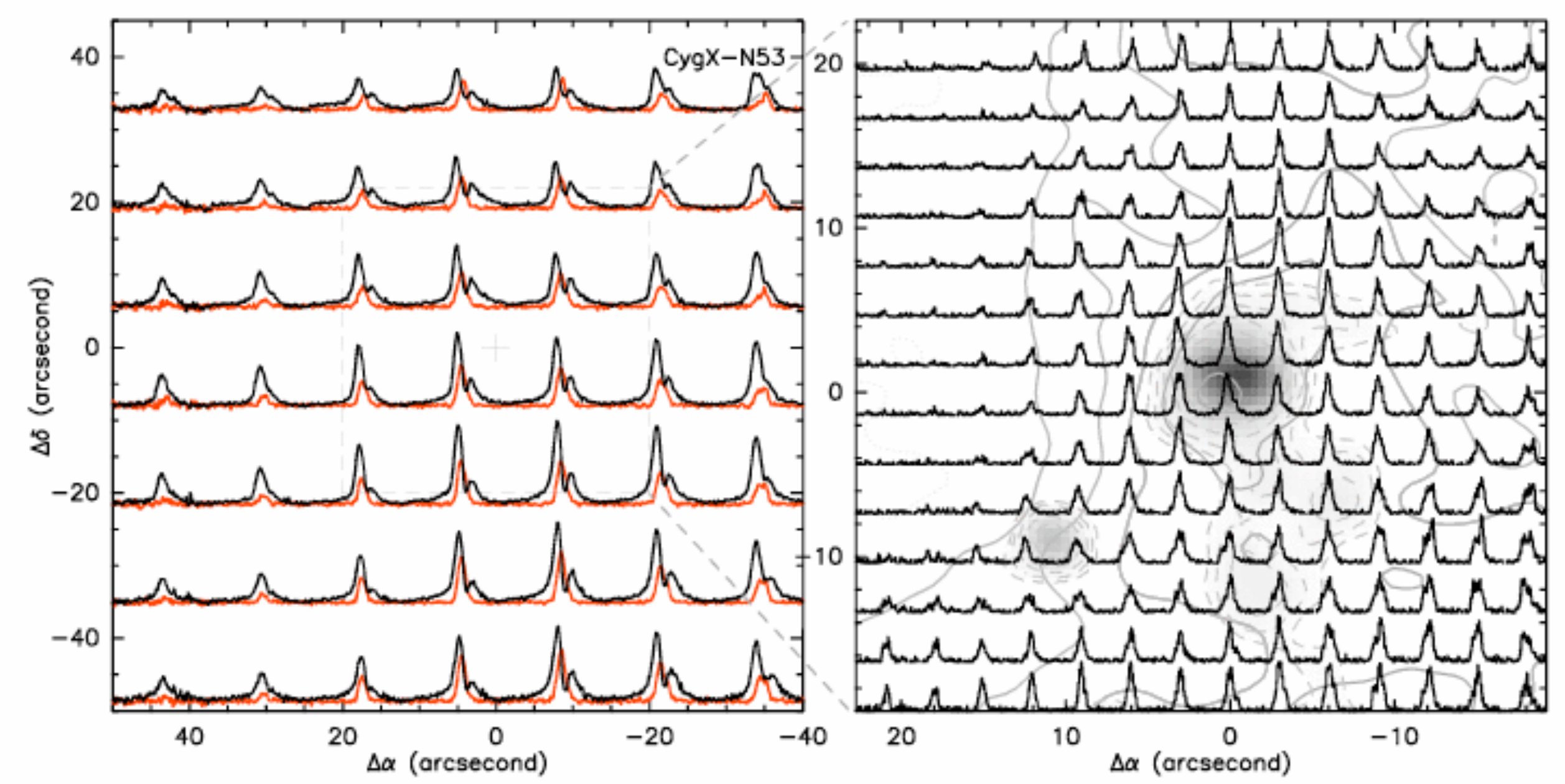}
   \includegraphics[width=16.0cm]{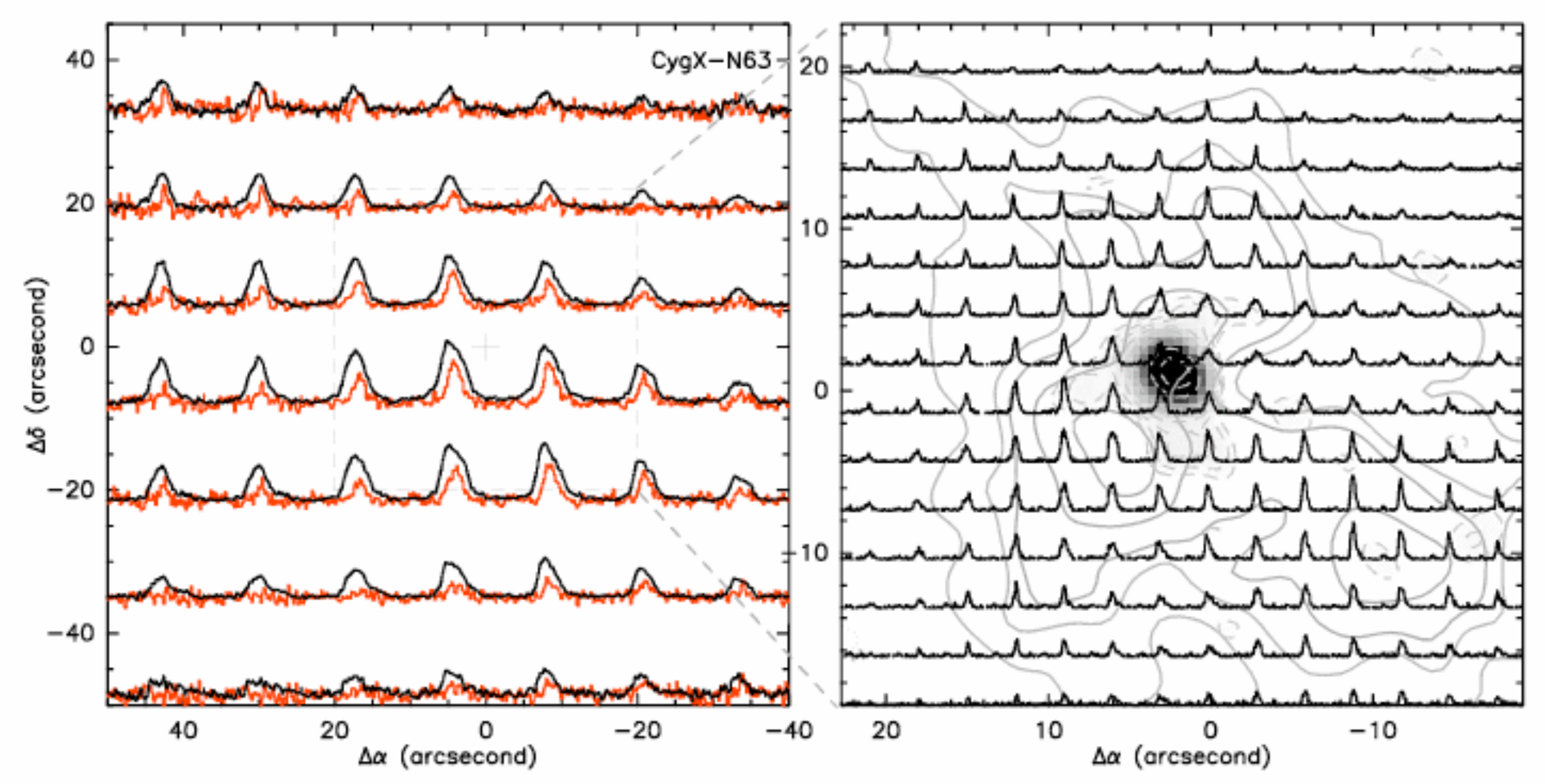}
     \label{fig:sdb}
   \caption{ -- continued.}
    \end{figure*}

   \begin{figure}
   \centering
   \includegraphics[width=4cm]{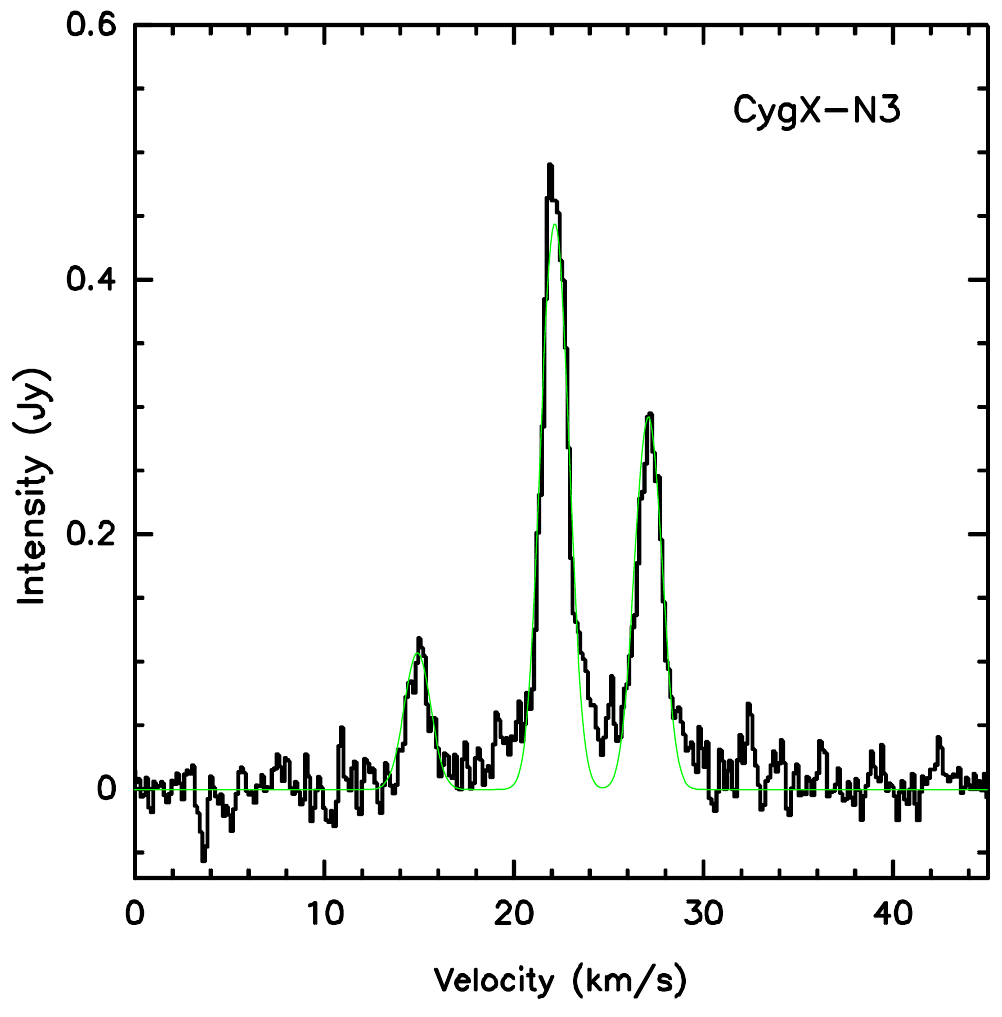}
   \includegraphics[width=4cm]{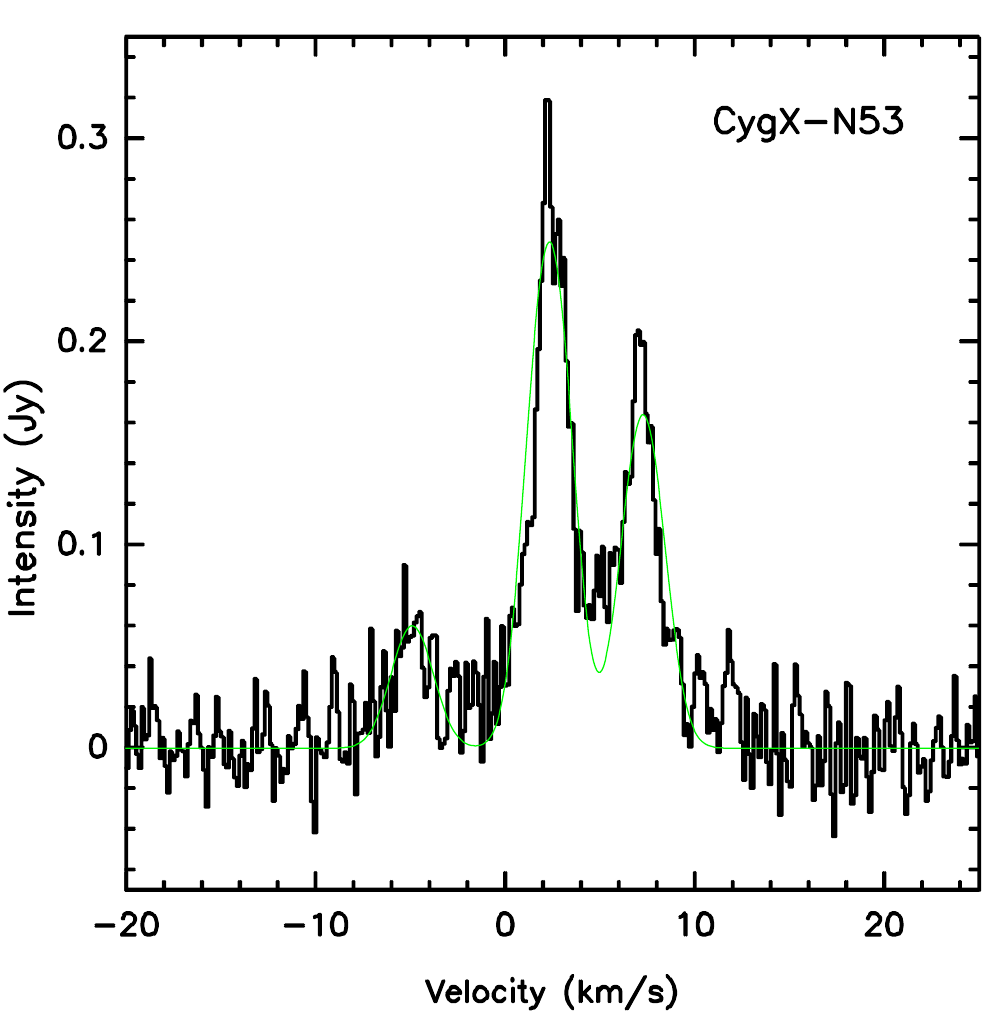}

   \includegraphics[width=4cm]{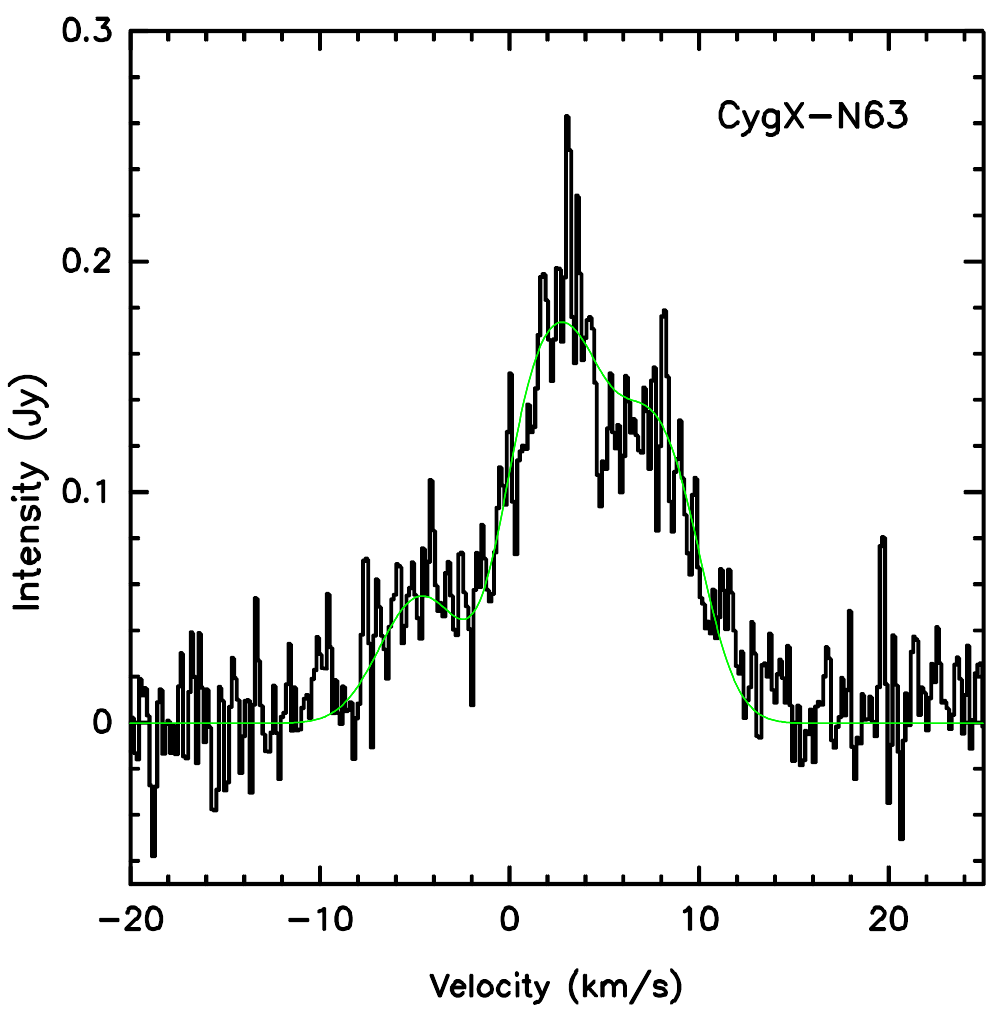}
   \includegraphics[width=4cm]{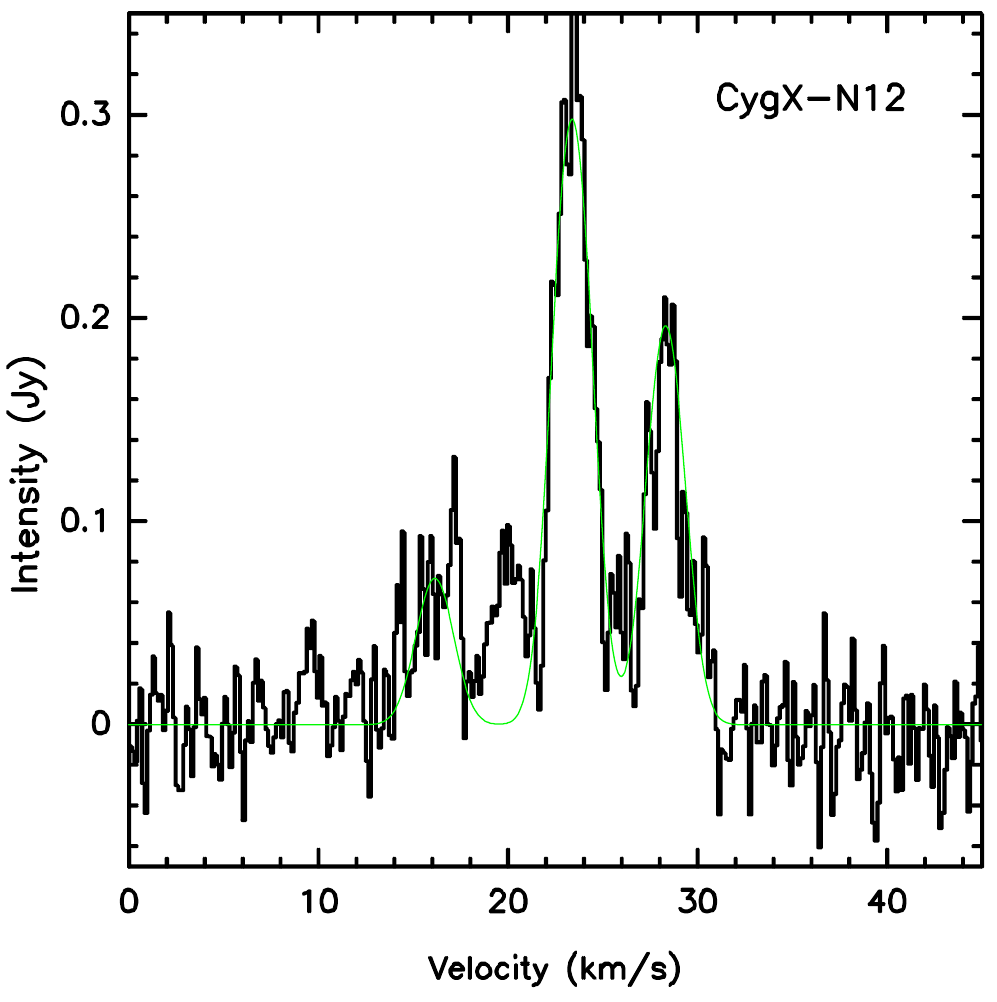}
   \includegraphics[width=4cm]{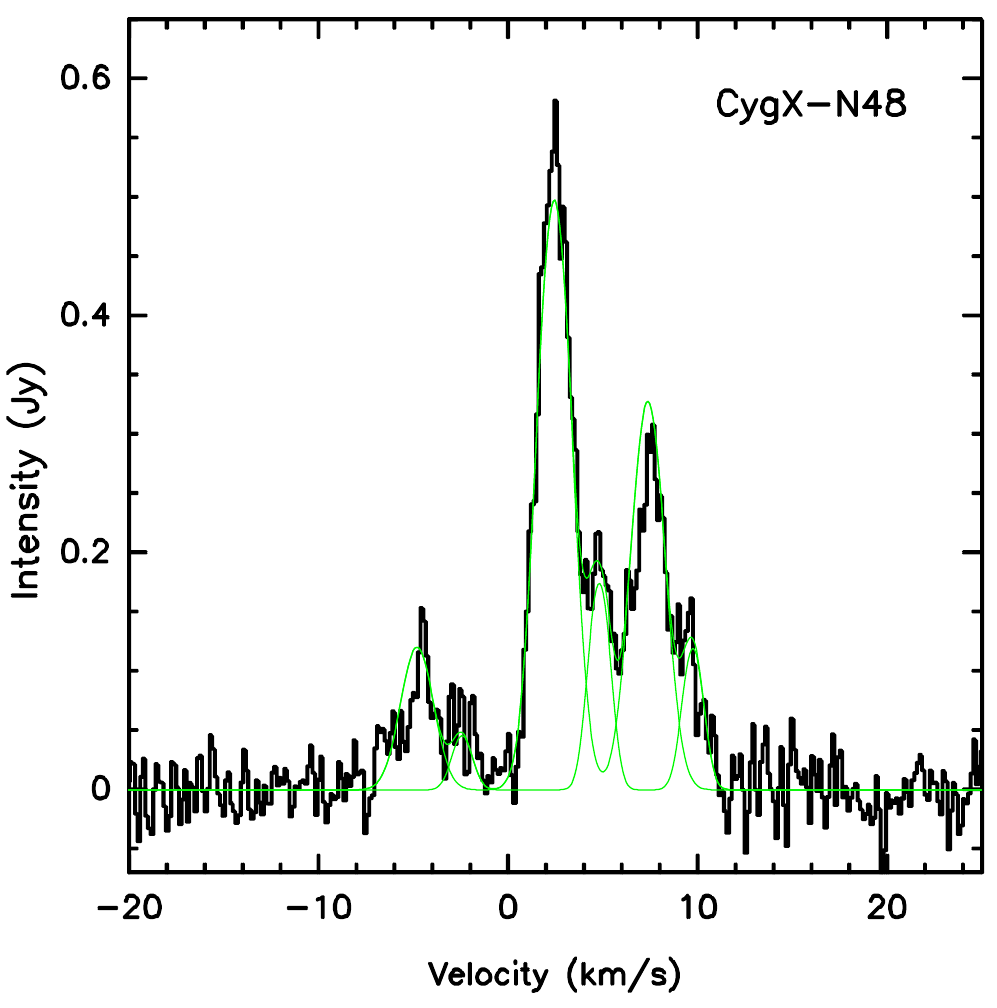}
     \caption{HFS line fitting of the {\hcn} spectra, which were averaged within 50\% contours of the peak emission. Green line indicates the result of HFS line-fitting assuming LTE conditions. All hyperfine-components are detected towards all of the cores.}
         \label{fig:h13cn_spectra}
   \end{figure}
%

  \begin{figure*}
   \begin{minipage}{18cm}
   \centering
    \includegraphics[width=5cm]{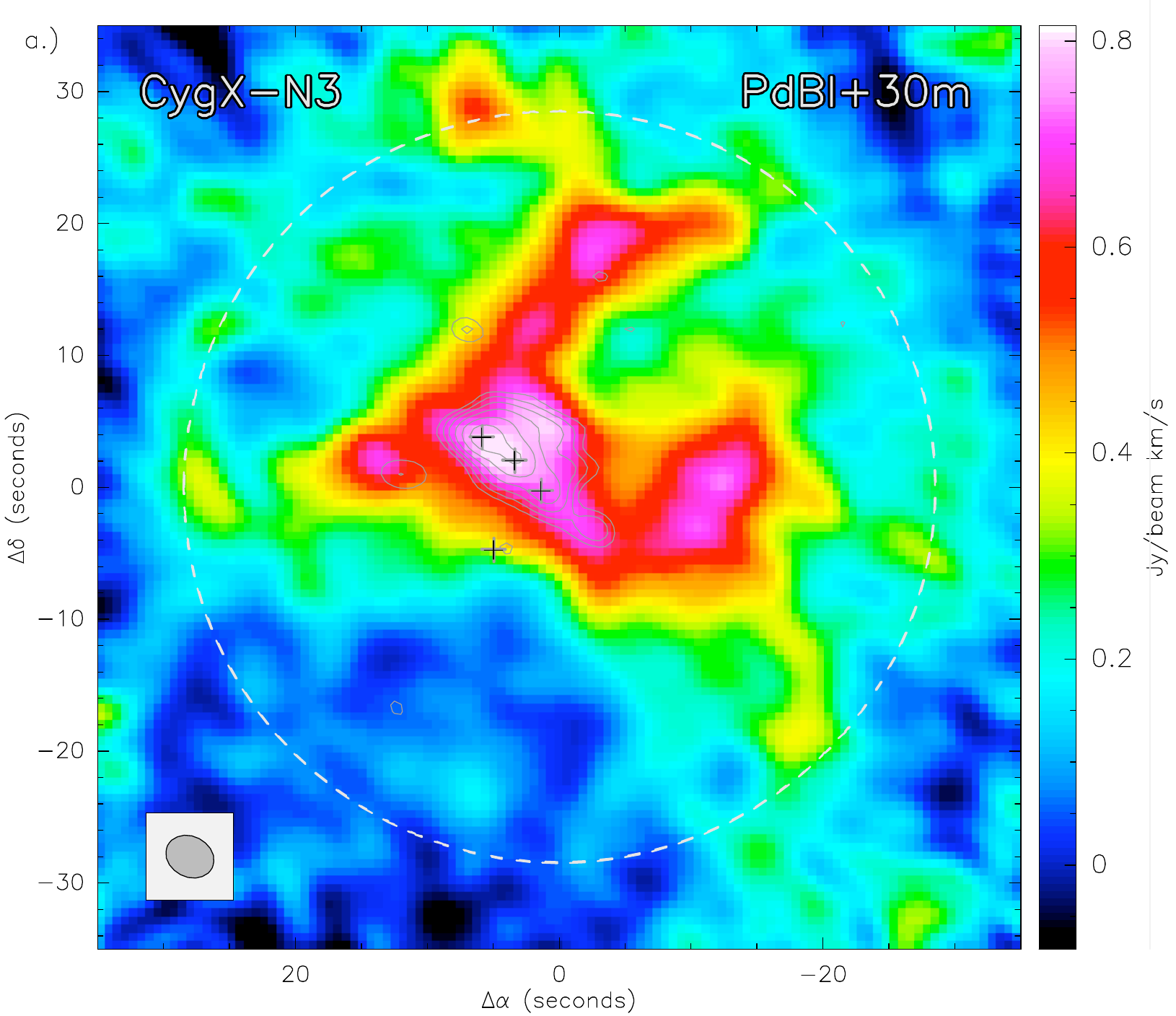}
   \includegraphics[width=5cm]{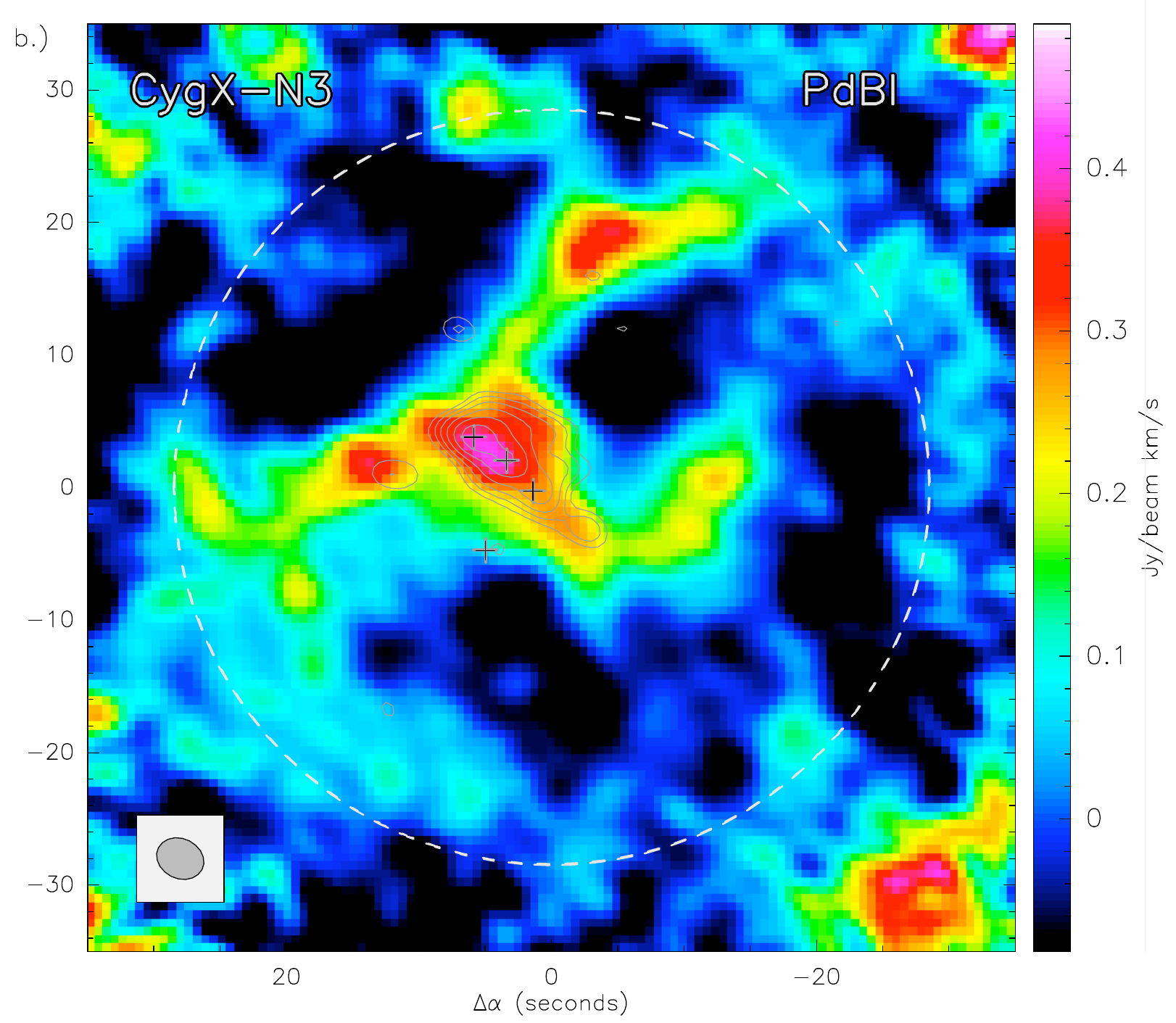}
   \includegraphics[width=5cm]{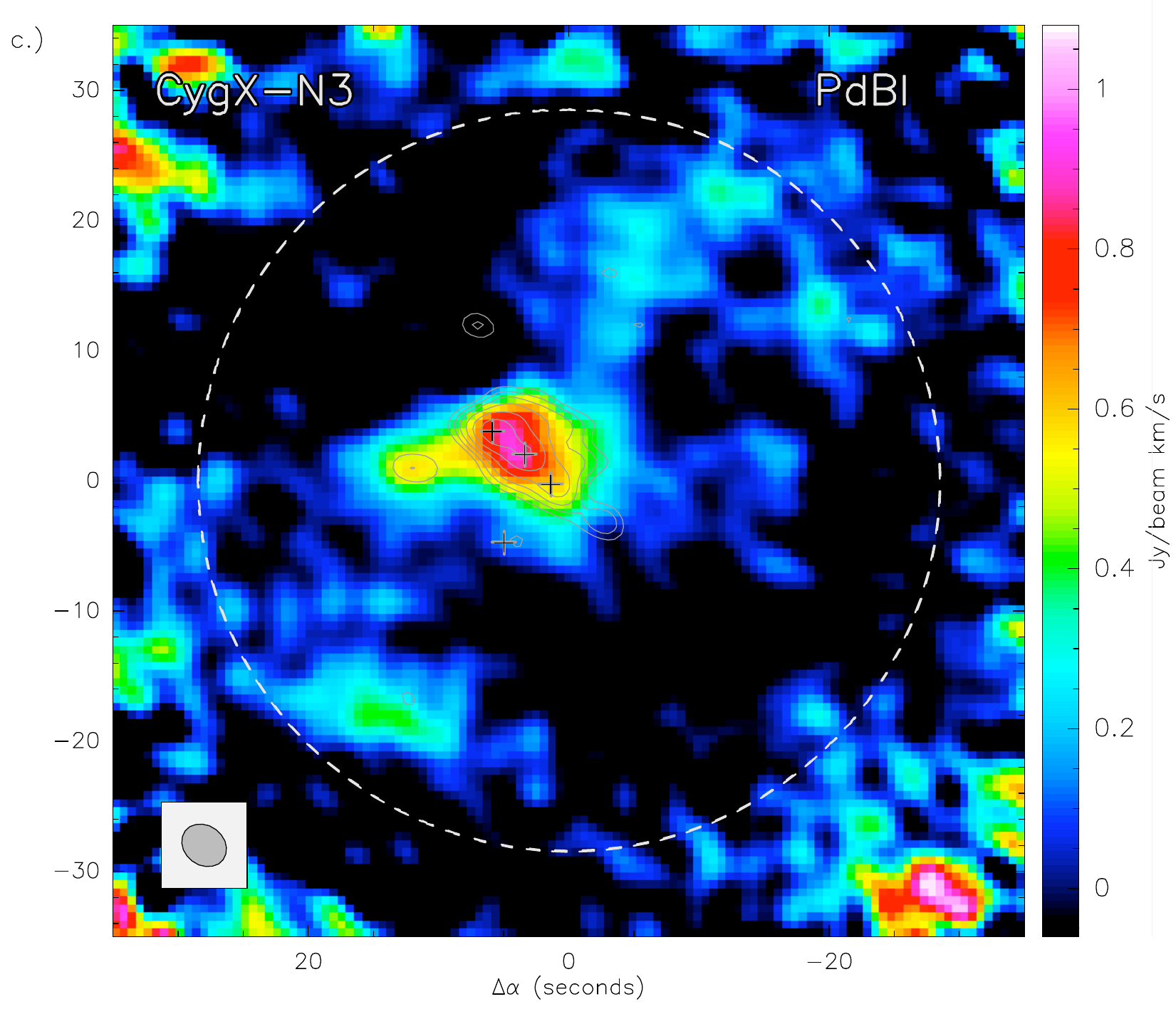} \\ 
   \includegraphics[width=5cm]{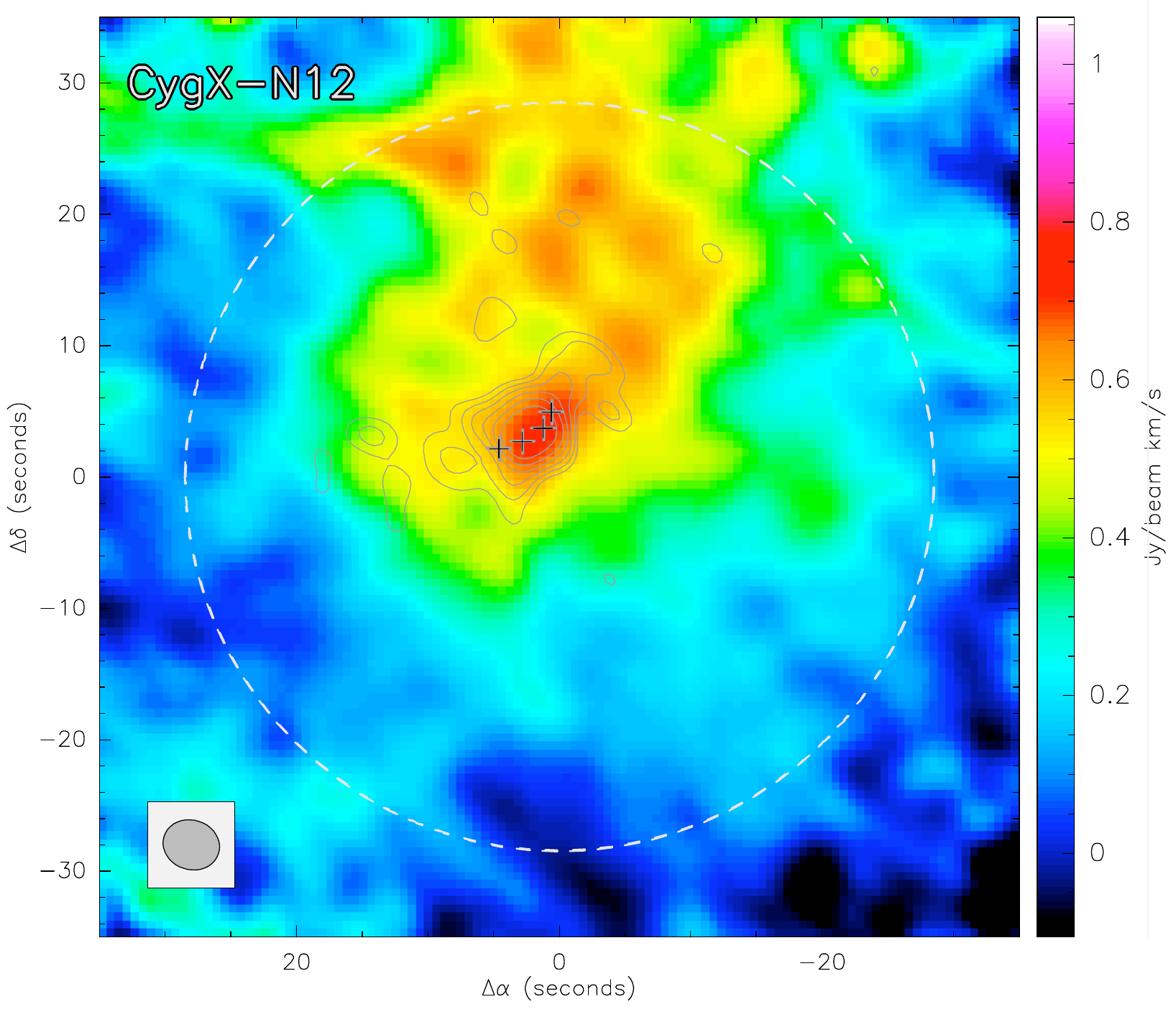}
   \includegraphics[width=5.0cm]{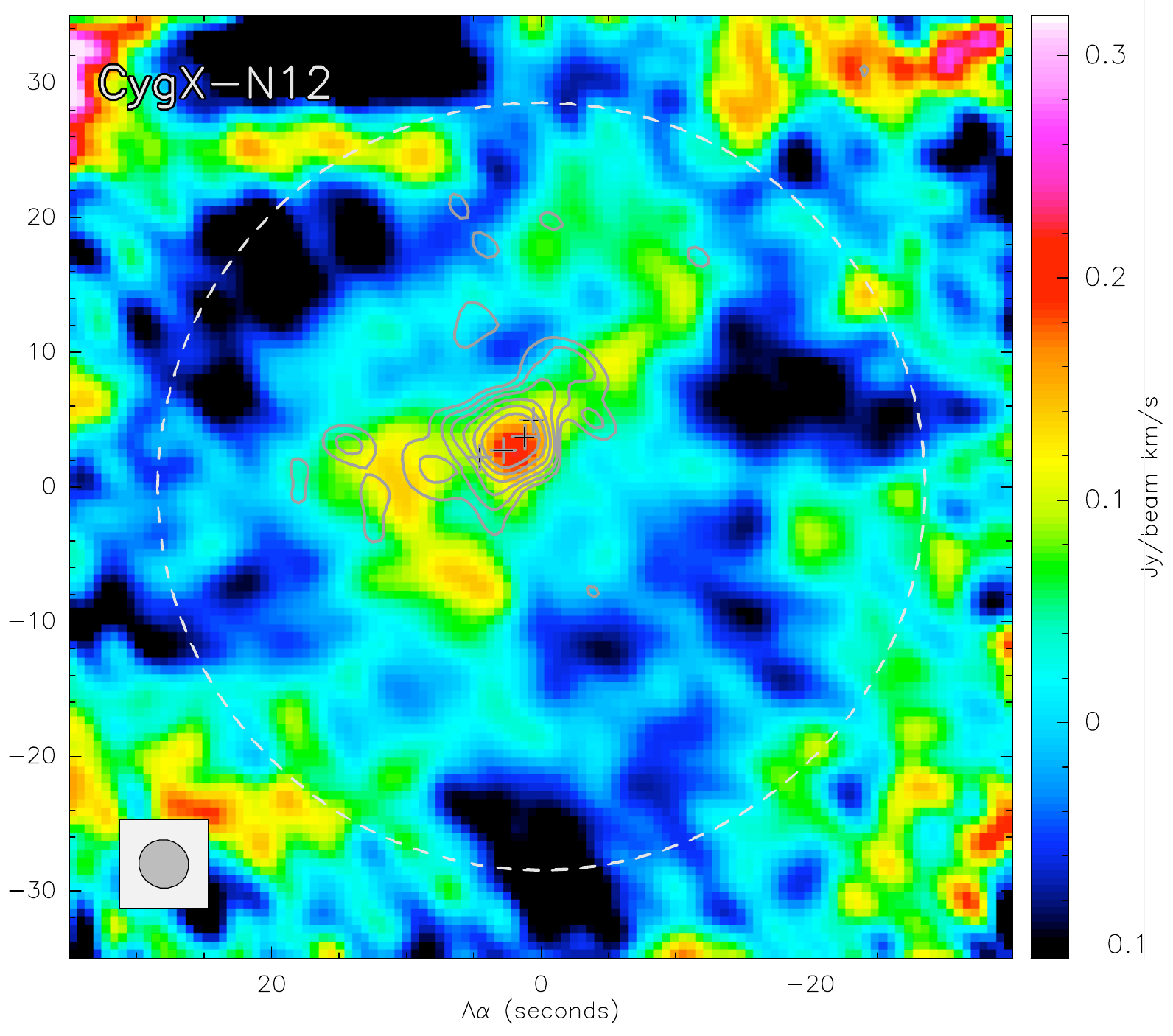}
   \includegraphics[width=5.0cm]{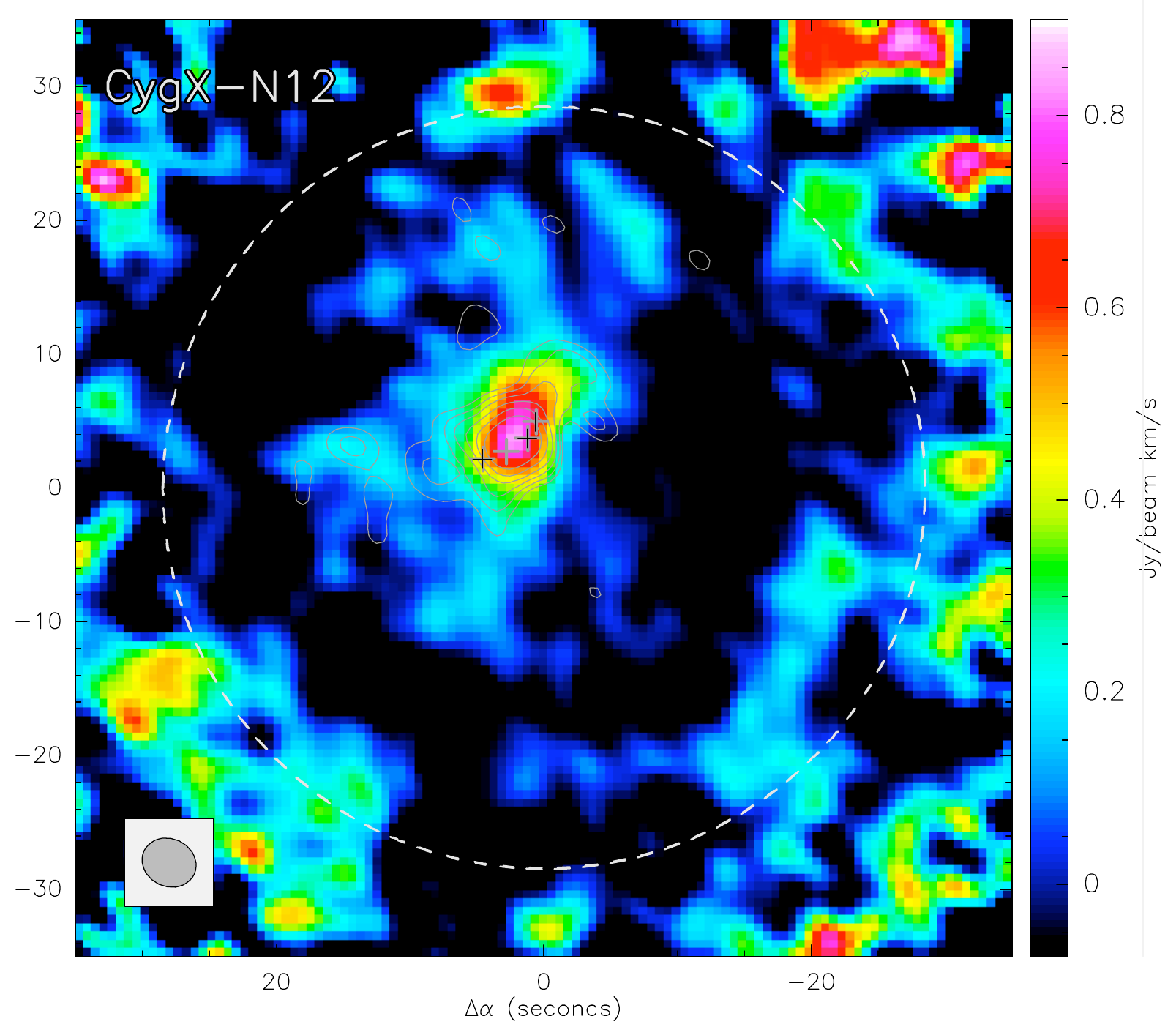}\\
   \includegraphics[width=5cm]{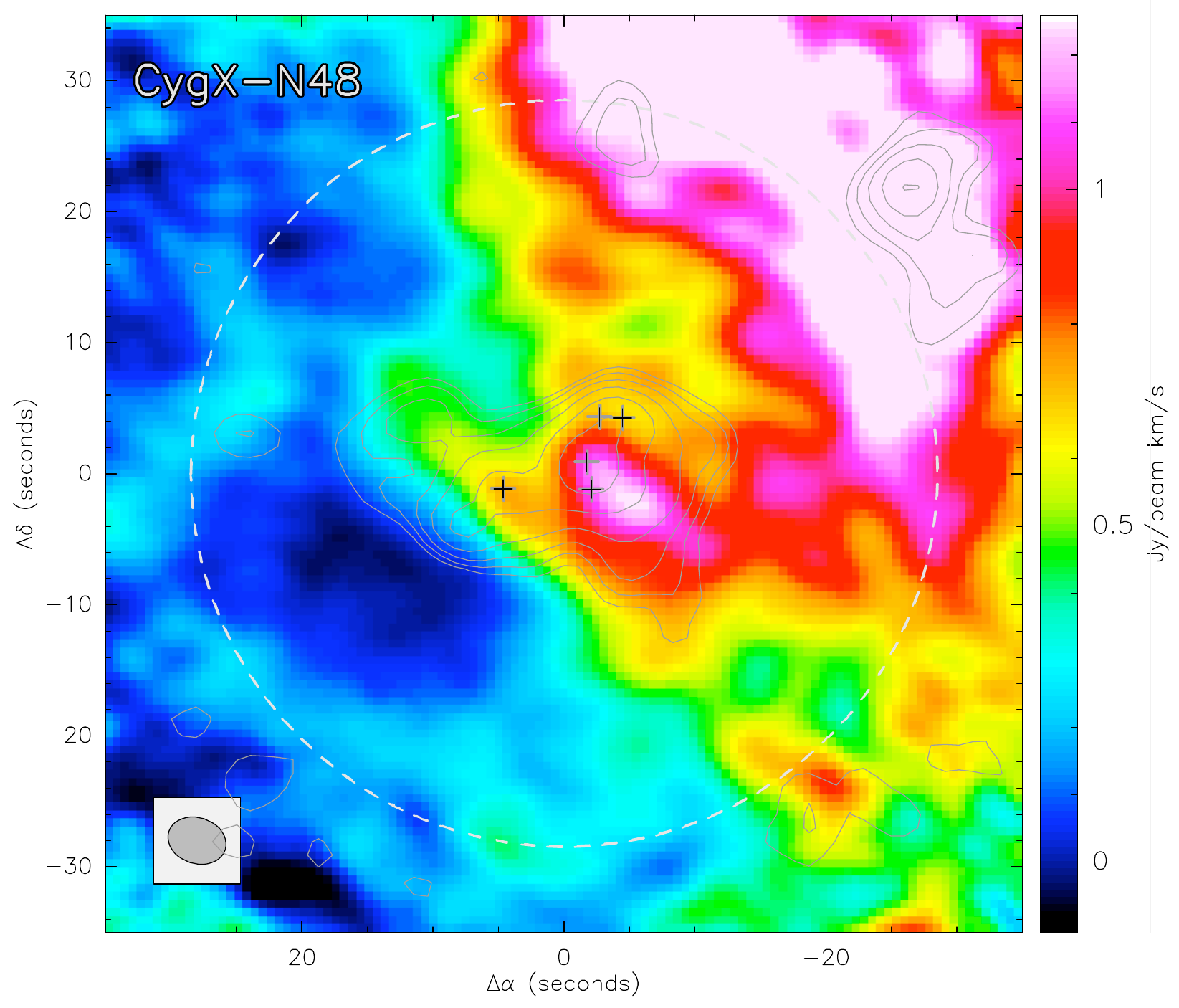}
   \includegraphics[width=5.0cm]{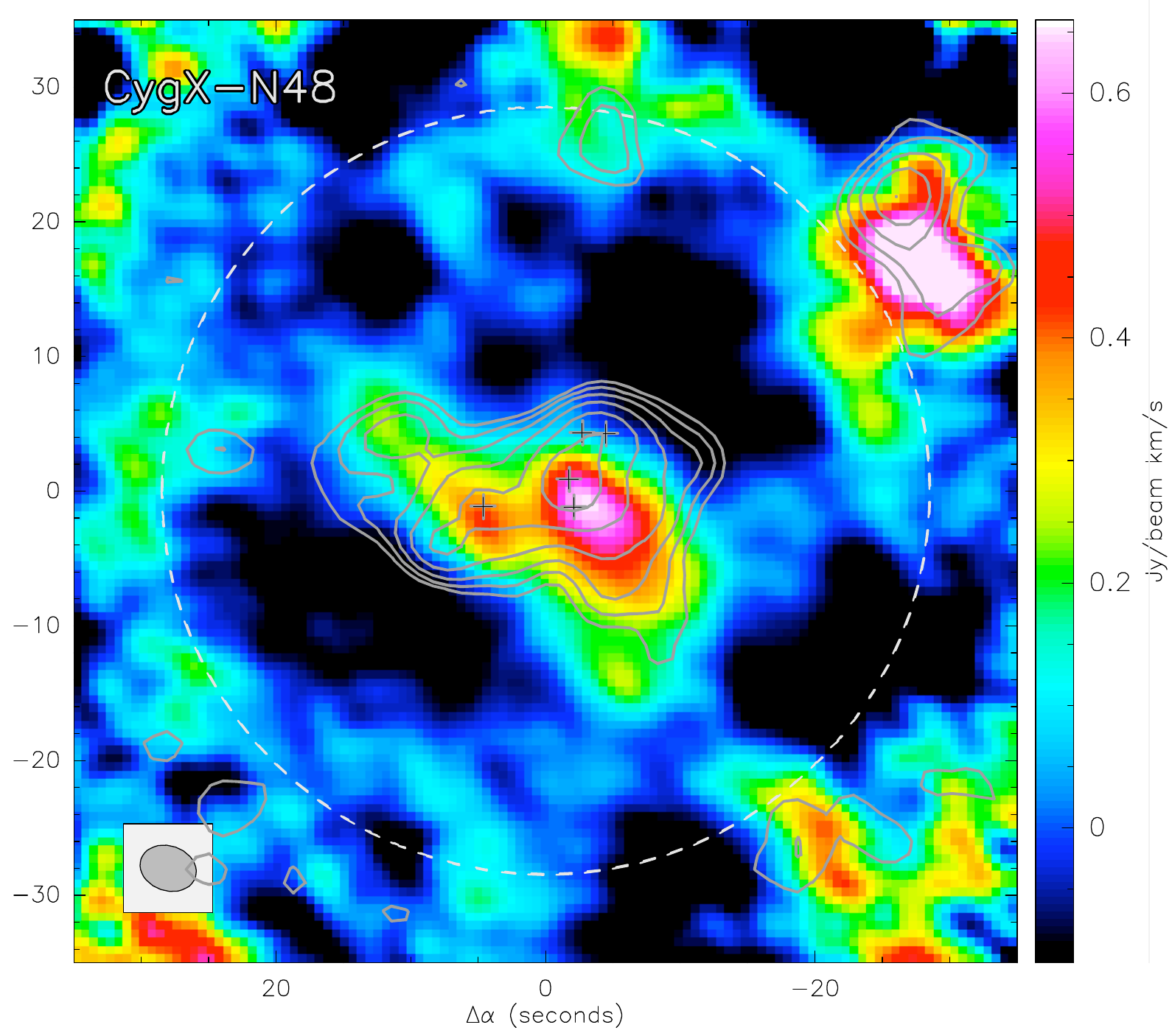}
   \includegraphics[width=5.0cm]{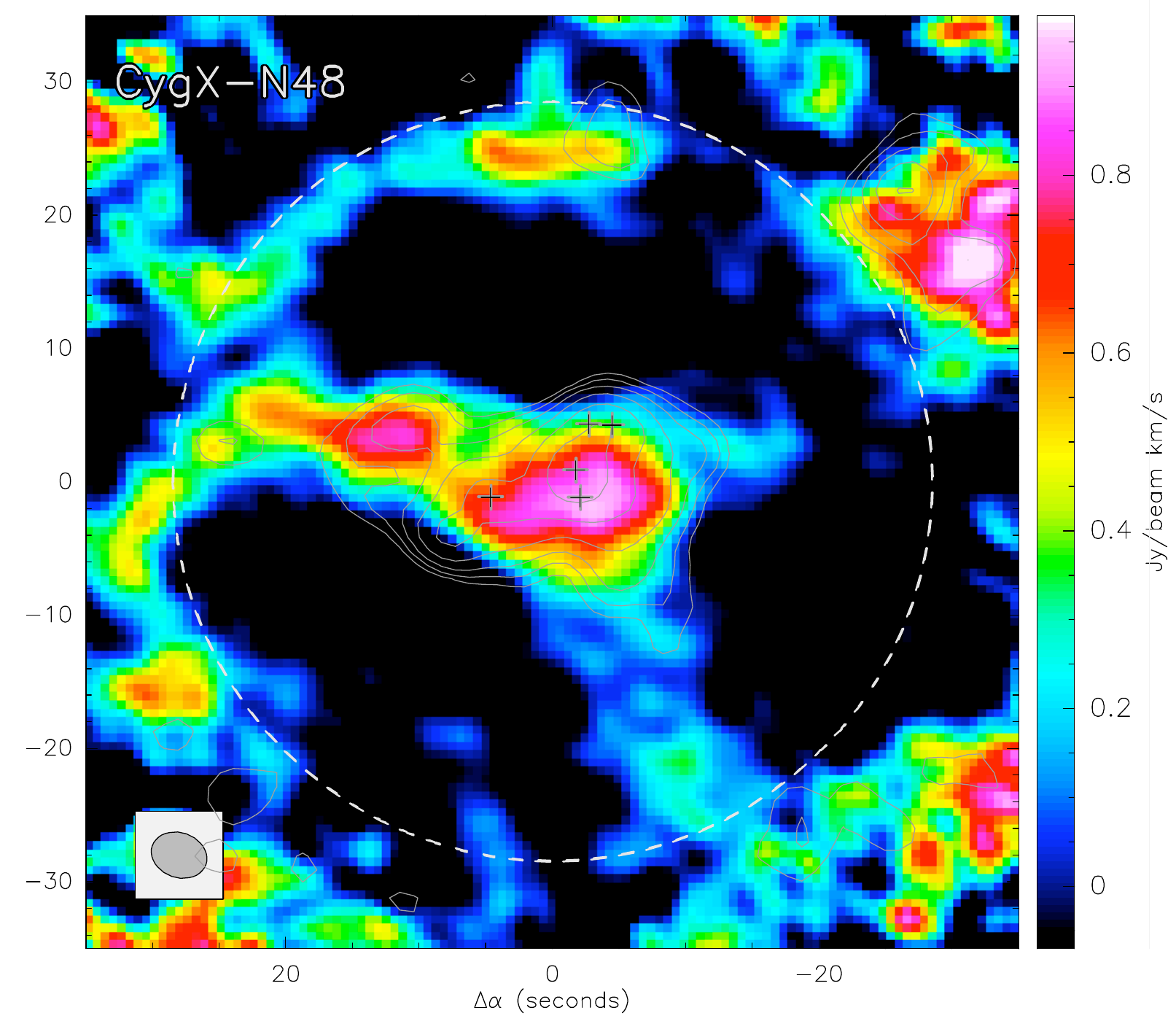}\\
    \includegraphics[width=5cm]{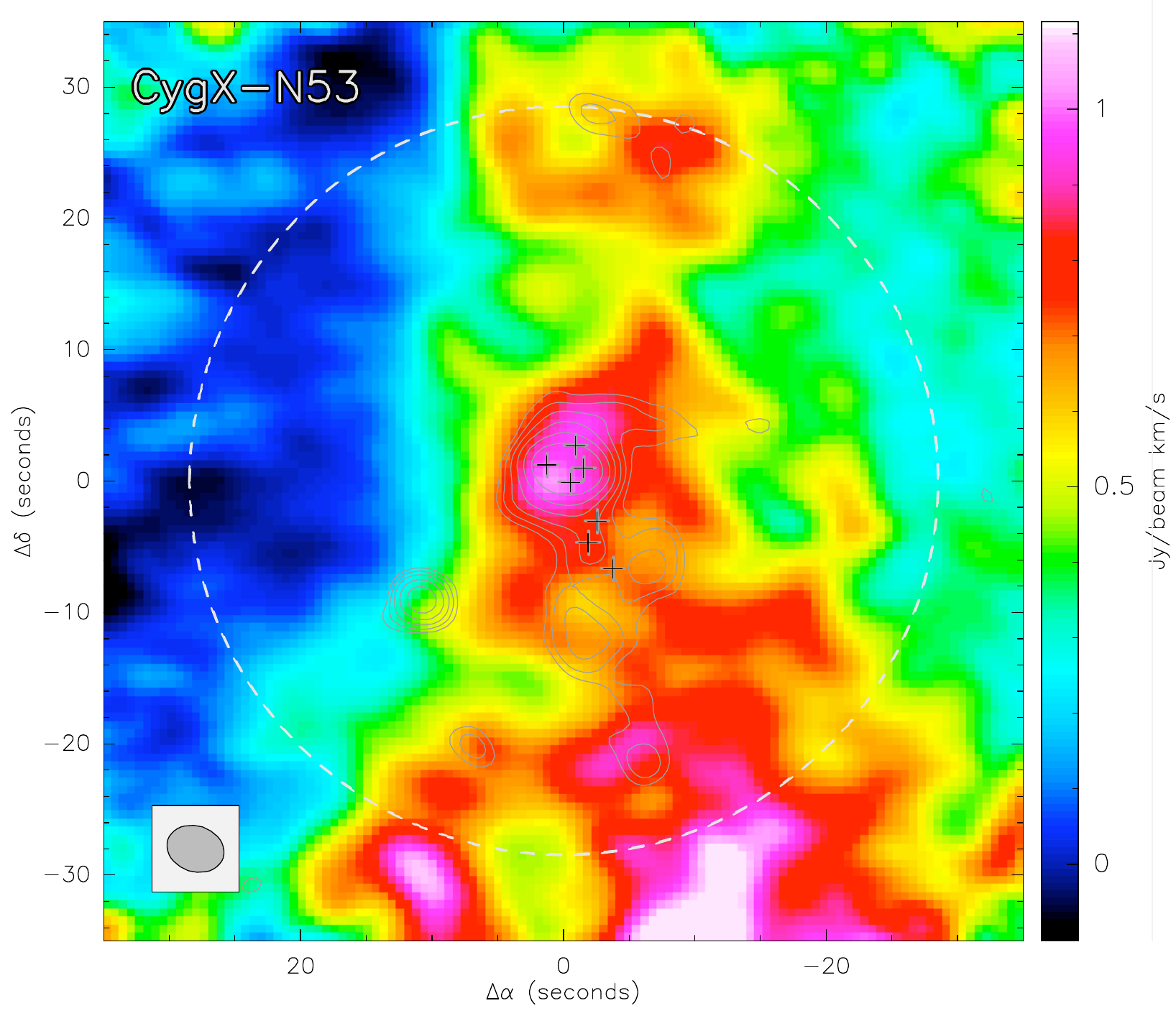}
    \includegraphics[width=5cm]{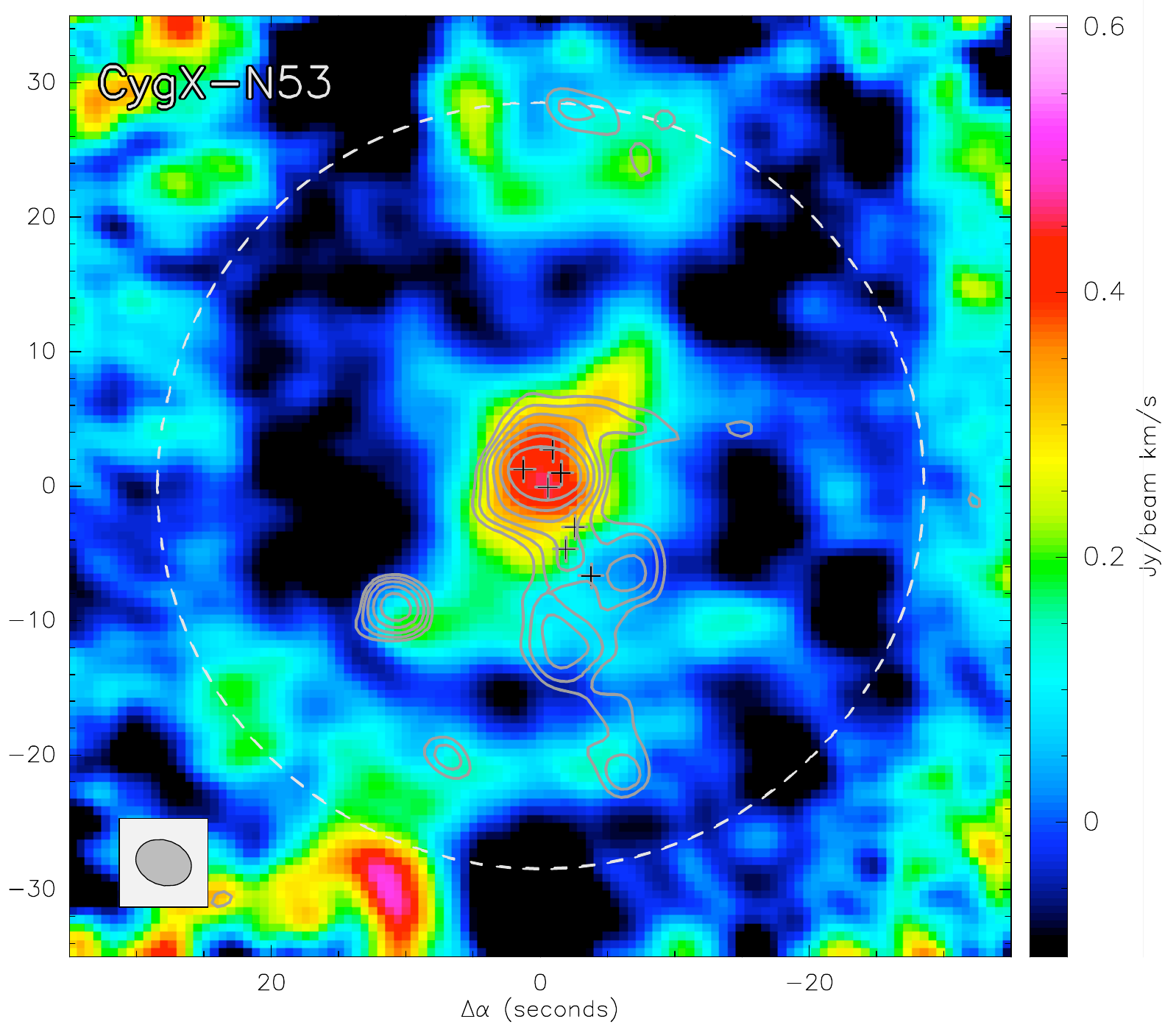}
    \includegraphics[width=5cm]{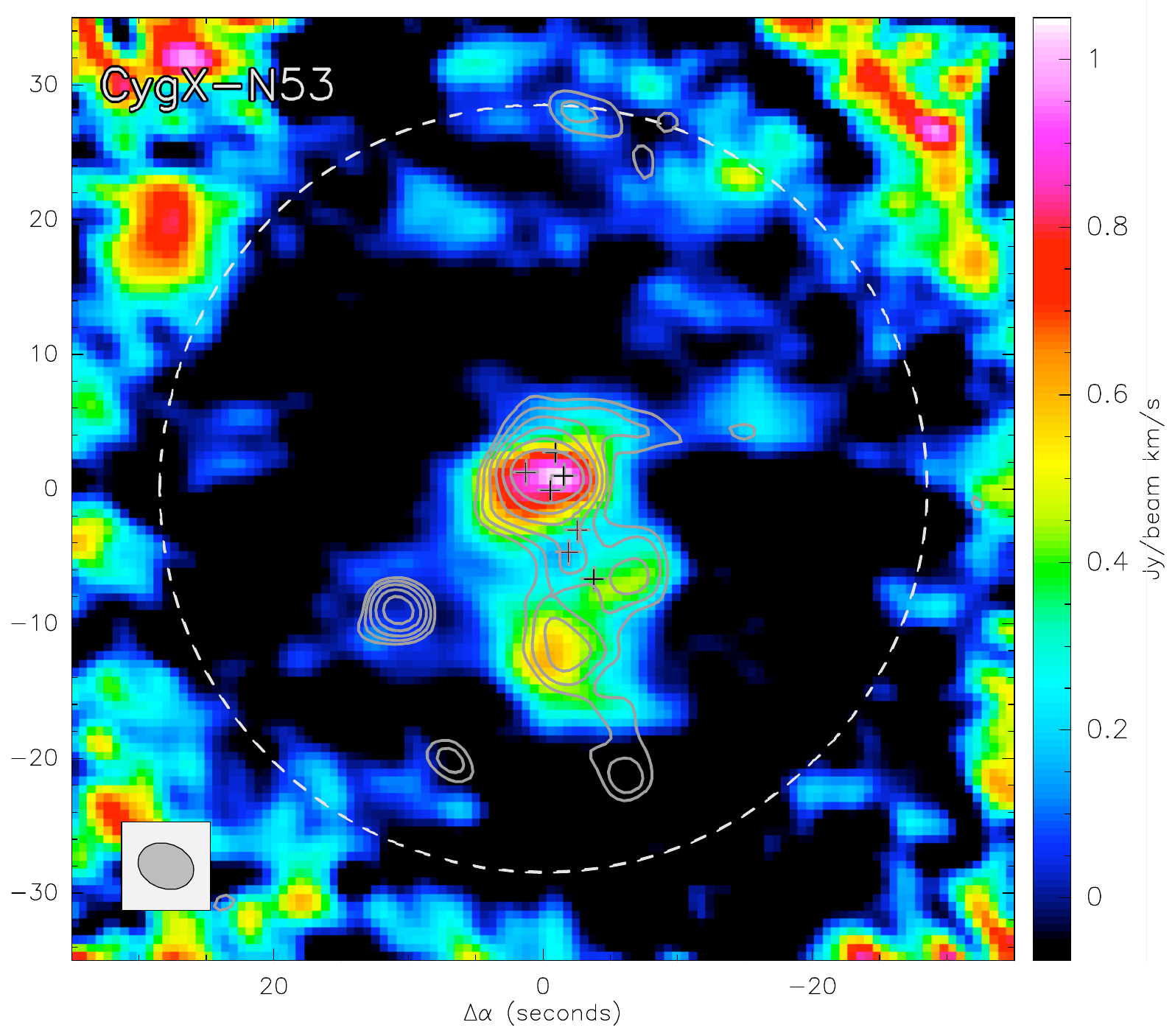}  \\
   \includegraphics[width=5cm]{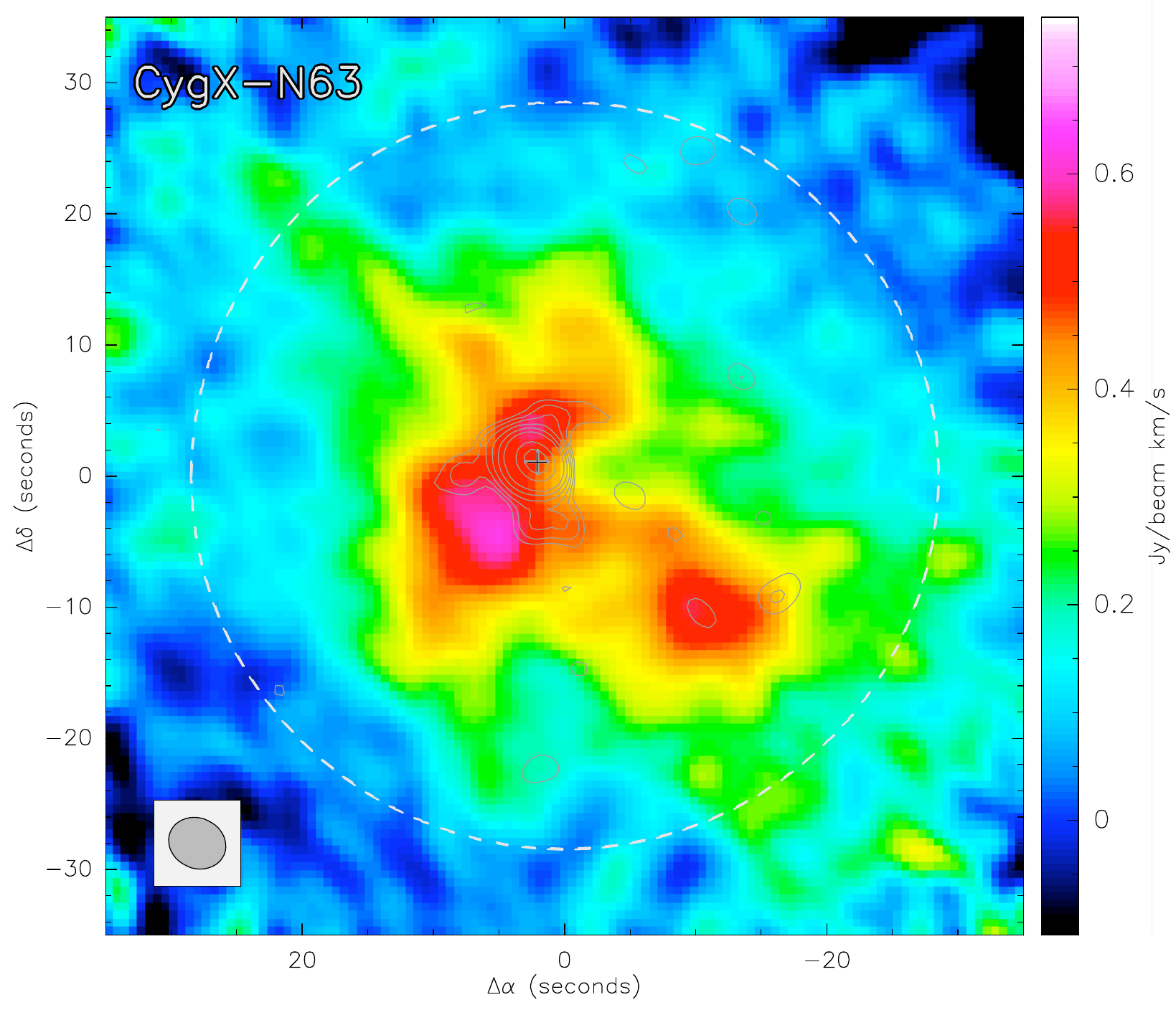}
   \includegraphics[width=5.0cm]{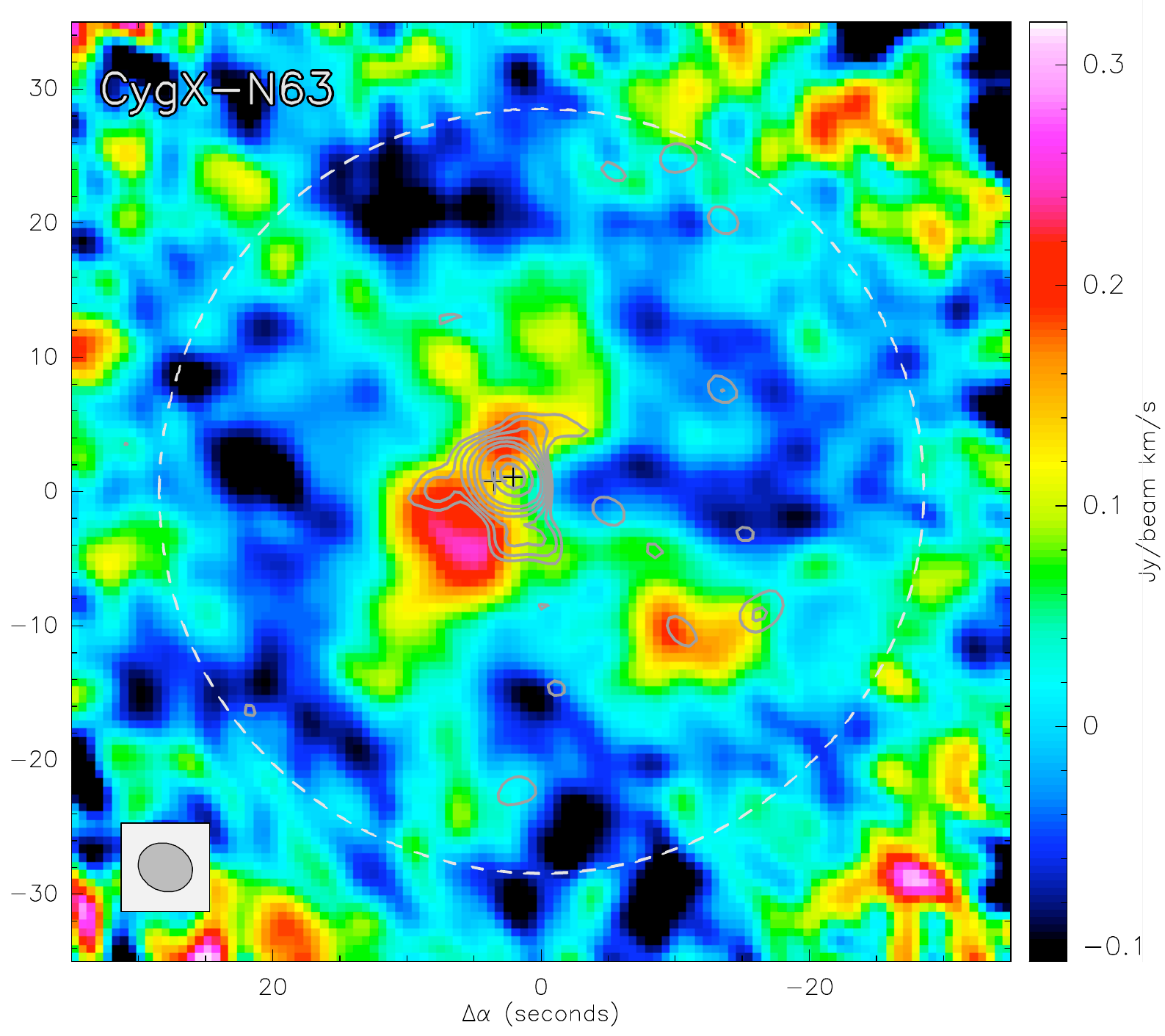}
   \includegraphics[width=5.0cm]{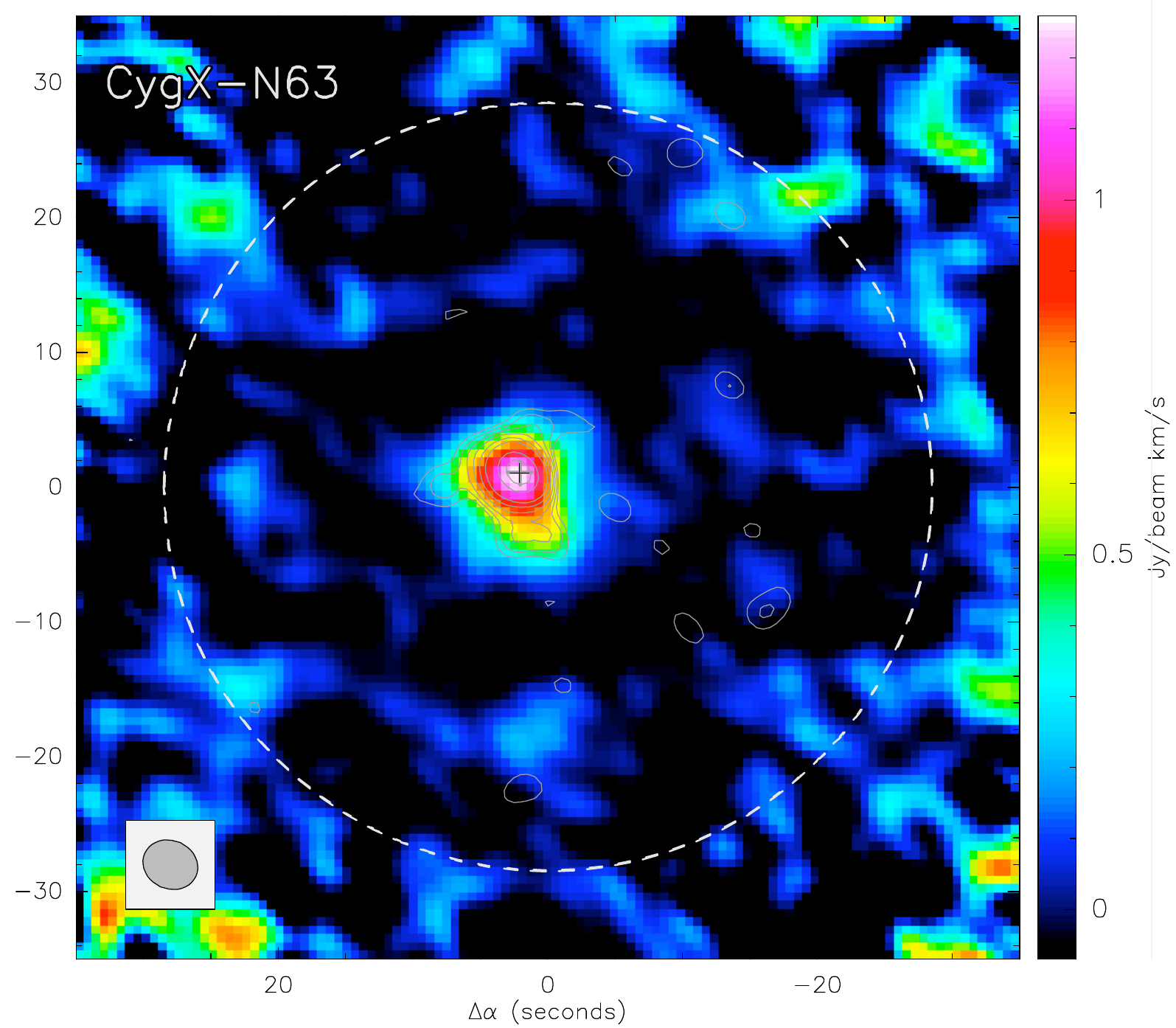} 

 \caption{{\sl a)} Integrated intensity maps of \hco (J=1--0) obtained with PdBI and combined with observations from the IRAM 30m telescope (the scaling goes from -2$\sigma$ to 35$\sigma$ for field N63 and N12, and from -2$\sigma$ to 60$\sigma$ for all other fields). Contours show the 3mm continuum emission, where levels are the same as in \citetalias{B09}. Crosses indicate the location of 1mm continuum sources reported in \citetalias{B09}. Dashed circles indicate the primary beam of the PdBI. Grey ellipses in the left lower corner show the synthesized beam. {\sl b)} Interferometric line integrated intensity maps of {\hco} (J=1--0). {\sl c)} Integrated line intensity maps of {\hcn} (J=1--0) (over all hyperfine components). }
     \label{fig:pdbi}
         \end{minipage}

    \end{figure*}
    

\subsection{IRAM 30m {\hco} and {\hcop} observations}
Spectral line maps of the IRAM 30m observations of the five MDCs are shown in Figure \ref{fig:sda} {\sl a)}. We obtained a map of line emission of $\sim$1 arcmin$^2$ of the 3 isolated MDCs (CygX-N3, {-N12}, -N63). For CygX-N48 and -N53 we cut out maps from a larger mapping of \citet{Schneider_prep} with the same dimensions. 

Both lines are well detected over the extent of MDCs at a size-scale of $\sim$0.1 pc. The large-scale distribution of emission indicates a rich gas content in all cores. For the isolated cores (CygX-N3, -N12, -N63), the distribution of {\hco} emission is centrally peaked and confirms the core as a coherent entity. For the fields located in the DR21 filament, the distribution of emission follows the main pattern of the filament and therefore shows a strong north-south elongation.


The {\hco} line shape is a rather clear gaussian in CygX-N53 and -N63, while in the other fields it shows a nearly gaussian profile. Blending due to several components and broad line-wings are remarkably present in CygX-N3 and CygX-N12. Towards CygX-N48 a red-shoulder is present in the spectra, while the stronger component is centered on the self-absorbed profiles of {\hcop} indicating two strong components in the spectra.

The optically thick tracer, {\hcop}, shows blue-shifted asymmetry in four MDCs of our sample. In CygX-N3 this profile changes into a reversed infall-profile, a red-shifted peak appears over a part of the map. The dip is at the same velocity where the optically thin line peaks, which is commonly interpreted as an expanding envelope or an outflow feature \citep{Myers96}. However, $\sim$15{\mbox{$^{\prime\prime}$}} away from the center the profiles turn into a standard infall profile. Such blue-shifted asymmetric profiles are observed towards CygX-N48 and -N53. In all these cases the dip of self-absorption is located at the peak of the optically thin line indicating infall. Interestingly, CygX-N12 and  especially CygX-N63 show only a slight asymmetry in their spectra. Since the ratio of {\hco} and {\hcop} line intensities (7, 5.5, 5.5, 1.9, 3.7 for CygX-N3, -N12, -N48, -N53, -N63, respectively) indicate that the emission of {\hcop} must be optically thick, we interpret this blue-shifted asymmetry as infall signature. 
In Section \ref{subsec:simline} we present radiative transfer modeling of these line profiles.

\subsection{IRAM PdBI {\hco}  and {\hcn}  observations}


\subsubsection{IRAM PdBI {\hco}  spectra maps}

Maps of {\hco} spectra obtained with the PdBI and processed with zero-spacings are presented in Figure \ref{fig:sda} {\sl b)}. We regridded the spectra-cubes to a fully sampled synthetic HPBW of $\sim$4{\mbox{$^{\prime\prime}$}}. These maps of spectra taken at high angular-resolution resolve several 
distinct spectral components.
  For fields CygX-N3 and -N53 individual spectral components are found with a separation of 0.5-2 km~s$^{-1}$ (4$\times$-15$\times$ the spectral resolution). They appear clearly off the main position of the continuum peaks and can be separated into two strong components each of them showing a shift in position and in velocity and tend to blend into a single line towards the continuum peaks. In fields CygX-N12 and -N48 the spectra are more complex, especially in CygX-N12 the spectra are flattened
, which makes the separation of the components difficult. Similarly in CygX-N48 the two strong components seem to be blended with other components and in the central region broad lines are seen.
(Note that for CygX-N48 a large-scale sub-filament at $0$ km$\ $s$^{-1}$ is reported by \citet{Schneider_prep} seen on the north-east, towards DR21(OH).) Note that CygX-N63 is dominated by a single component and a weak one appears separated by $\sim$0.5 km~s$^{-1}$. 

\subsubsection{IRAM PdBI {\hcn} spectra}
\label{sec:h13cn}
{\hcn} spectra were integrated within the 50\% level of the peak emission and are shown here with the corresponding hyperfine-structure (HFS) fits for each source (Figure~\ref{fig:h13cn_spectra}). The strongest lines are observed towards CygX-N3 and CygX-N48, while the other cores are weaker, but all 3 hyperfine components are detected. A larger line-width component around the main peak of the line is seen towards CygX-N3, -N53, but it is strongest in CygX-N63. The {\hcn} line-profile towards the core CygX-N63 is a single, broad component, which is significantly different from all the other fields. Two individual components can be seen towards CygX-N48. 

\subsection{High angular-resolution integrated intensity maps}

\subsubsection{{\hco} integrated intensity maps}
The integrated intensity maps  obtained for the {\hco} lines are presented in Figure \ref{fig:pdbi} {\sl a), b)}. For comparison contours of 3mm continuum emission from \citetalias{B09} are shown. Note that the two maps have a similar spatial resolution of $\sim$4{\mbox{$^{\prime\prime}$}}. 
We detect strong line-emission with the PdBI in all cores, which resolves significant substructures of molecular gas compared to the IRAM 30m observations due to the $\sim$7 times better angular resolution. 

The integrated intensity maps of {\hco} including zero-spacings are shown in Figure \ref{fig:pdbi} {\sl a)}. Including the zero-spacings allows to recover emission from larger spatial scales as well. The large-scale emission enhances the asymmetry in the emission from CygX-N48 and -N53, which are associated with a large-scale structure, the DR21 filament. CygX-N53 is located towards the northern tip of the filament, and therefore shows an intensity distribution in north-south direction, which becomes stronger towards the main part of the filament located south in the maps. For field CygX-N48 the distribution of the molecular gas emission is strongly asymmetric and drops towards the south-western edge of the core, close to a compact cluster of embedded young stars. On the other hand, towards the opposite, north-east direction the emission gets significantly stronger. This shows the same morphology as the large-scale emission of {\hco}, which is presented in Figure 5 in \citet{Schneider_prep} as well. 

Figure \ref{fig:pdbi} {\sl b)} and {\sl c)} show maps of integrated intensity of {\hco} and {\hcn} respectively, using only interferometric data (no zero-spacings added).
The distribution of dense molecular gas within the 0.1~pc-scale is highly non-uniform and shows significant substructure. The strongest line emission is generally observed towards the central regions of the cores, and coincides well with the 3mm continuum peaks (indicated by grey contours). Only CygX-N63 is standing out because there is a prominent "hole" in {\hco} emission towards the continuum emission. The two lobes surrounding the continuum peak suggest that {\hco} is highly depleted (see Section \ref{subsec:simline}) and emission is completely absent towards the central densest regions.

\subsubsection{{\hcn} integrated intensity maps}

The emission of {\hcn} (Figure \ref{fig:pdbi} {\sl c}) always peaks at the brightest continuum sources and thus the densest parts of the MDCs. Comparing the distribution of the emission of {\hco} and {\hcn}, they both trace the densest parts of the MDCs, but are locally different. In CygX-N3 all the important continuum peaks are embedded in emission of dense molecular gas, except for CygX-N3 MM4 (following the notations of \citetalias{B09}), which is outside of the {\hcn} emission. Also the north-western lobe strongly appearing in {\hco}, is weak in {\hcn} which may indicate chemical differences within the cores. CygX-N12 has a central peak embedded in an elongated, filamentary-like emission in continuum, which is similarly traced by weak {\hco} emission. Nevertheless in {\hcn} only the main peak of continuum is traced. The 3mm continuum and molecular line emission in CygX-N48 is more diffuse and follows an east-west elongation. In CygX-N53 the central continuum peak is well traced in both molecules, but while the southern lobe contains two peaks in continuum and is well-traced by {\hcn}, it is absent in {\hco}. A compact 3mm continuum peak (probably a more evolved protostar) is found south-west from the central peak, which appears in {\hco}, but is not seen in {\hcn}. As mentioned before, the main peak of CygX-N63 is absent in {\hco}, but is well traced by strong, compact emission of {\hcn}. 

Summarizing, {\hcn} is generally more centrally peaked than \hco. Such differences in the distribution of the two lines may be explained by physically different environments and/or abundance anomalies close to massive protostars. 


\section{Analysis}
\subsection{Analysis and modeling of the single-dish spectra of {\hcop} and {\hco}}
\subsubsection{Line-fitting of single-dish \hco spectra}
The single-dish maps of {\hco} (J=1--0) show a single line detected over a significant extent in each core, thus we fit them to first order with a gaussian. We use the result of the fit to derive the parameters of the spectra: line intensity, the local rest velocity (v$_{lsr}$) and the line-width (and dispersion $\sigma = \frac{FWHM}{\sqrt{8\,\rm ln(2)}}$) in each core and assume that \hco is optically thin. The values for each core at the position of the PdBI phase center are listed in Table \ref{tab:sd_params}. 

The emission associated with cores CygX-N3 and -N12 have a v$_{lsr}$ of $+15$ km$\ $s$^{-1}$,  while cores CygX-N48~and~
{$\rm -N53$}~have a v$_{lsr}$ of $-3.5, -4.2$ km$\ $s$^{-1}$, respectively. These two cores are located in the massive filament associated with DR21 and their rest velocity is consistent with that of the filament, confirming that they are physically associated. CygX-N63 is located southwards of the filament, but is associated with the south tail of the large-scale complex of DR21 and DR23 in Cygnus-X North with a v$_{lsr}$ of --4.3 km$\ $s$^{-1}$ \citep{Schneider06}. 

We derive velocity-dispersions of the orders of $\sim$1~km~s$^{-1}$ for the cores, which is similar to N$_2$H$^+$ observations from \citetalias{B09}, though 3 of the latter are smaller (by $\sim$0.1-0.2~km~s$^{-1}$) (Table \ref{tab:sd_params}).  We note that the largest line-dispersions are observed towards the most massive core, CygX-N48 (see also in Section \ref{sect:broadening}).


%
\begin{table}
\centering                          
\caption{Result of line-fitting to single-dish {\hco} lines (28{\mbox{$^{\prime\prime}$}} angular resolution): integrated intensity, local rest velocity and line-dispersion. We compare the line-dispersions obtained with single-dish N$_2$H$^+$ pointed observations towards these cores \citepalias{B09}.}
\begin{tabular}{c c c c c c}        
\hline\hline                 
Source & $\int T_{mb}\, \rm dv$ & v$_{lsr}  $ & $\sigma_{H^{13}CO^+}$  & $\sigma_{N_2H^+}$  \\    
            & [K] &[km/s]                 &   [km/s]  &   [km/s] & \\    
\hline                        
   CygX-N3  & 4.77 $\pm$ 0.1 &  15.07 $\pm$ 0.03  & 1.01 $\pm$ 0.02  & 0.98  \\     
   CygX-N12 & 3.30 $\pm$ 0.11&  15.27 $\pm$ 0.05  & 1.23 $\pm$ 0.05  & 0.89 \\
   CygX-N48 & 5.02 $\pm$ 0.14 &  -3.49 $\pm$ 0.04 & 1.26 $\pm$ 0.04 & 1.28  \\
   CygX-N53 & 5.54 $\pm$ 0.13  &  -4.18 $\pm$ 0.02 & 0.9 $\pm$ 0.03 & 0.76   \\
   CygX-N63 & 2.54 $\pm$ 0.09 &  -4.34 $\pm$ 0.03 & 0.69 $\pm$ 0.03 & 0.72  \\ 
\hline                                  
\end{tabular}
\label{tab:sd_params}      
\end{table}

%
\begin{table}
\centering                          
\caption{Result of HFS line-fitting of the PdBI {\hcn} line, which is taken from an average within the contour  of 50\% of the peak emission: integrated intensity, local rest velocity, line-dispersion and opacity. The same excitation temperature is assumed for all components.}
\begin{tabular}{c c c c c c}        
\hline\hline                 
Source & $\int I \, \rm dv $  & v$_{lsr}$  & $\sigma_{H^{13}CN}$  & $\tau$ \\    
            & [Jy/beam]     &   [km/s]  &   [km/s]         &             \\    
\hline                        
   CygX-N3   &  0.113 $\pm$ 0.002    & 14.96 $\pm$ 0.12  & 0.69 $\pm$ 0.03  & 0.1 \\     
   CygX-N12 &  0.076 $\pm$ 0.002 &  16.16 $\pm$ 0.29  & 0.98 $\pm$ 0.07 & 0.1 \\
   CygX-N48-1 &  0.05   $\pm$ 0.01 &     -2.39  $\pm$ 0.04  & 0.53 $\pm$ 0.1  & 0.15 \\
   CygX-N48-2 &  0.127 $\pm$ 0.002 &  -4.77  $\pm$ 0.02  &  0.86 $\pm$ 0.05 & 0.1 \\
  CygX-N53       &  0.063 $\pm$ 0.001 &  -4.87  $\pm$ 0.03  &  1.1   $\pm$ 0.06 &  0.1 \\
   CygX-N63      &  0.063 $\pm$ 0.006 &  -4.64  $\pm$ 0.1    &  2.05 $\pm$ 0.22 &  0.271\\ 
\hline                                  
\end{tabular}
\label{tab:hcn_fit}      
\end{table}

\subsubsection{Radiative transfer modeling}
\label{subsec:simline}
Radiative transfer modeling of the single dish spectra of {\hco} and {\hcop} was performed simultaneously using the 1D spherical, non-LTE radiative transfer code \textsl{Simline} \citep{Ossenkopf}. The spectra of {\hco} and {\hcop} were extracted for each MDC at the center position of the single-dish maps and then averaged in concentric annuli in steps of 14{\mbox{$^{\prime\prime}$}} with respect to this position. 
The physical model used for all MDCs is described in detail in Appendix \ref{app:simline}. (See also Figures~\ref{fig:sim-n3} to \ref{fig:sim-n63}). 

We used continuum data to derive the mass, thus the density profile, and NH$_3$ data \citep{Wienen} to constrain the temperature profile and varied only the density at which depletion sets in and the depletion ratio for each core within a range of 10$^5$ to 10$^6$~cm$^{-3}$ and 0-1000, respectively. The best fit results correspond to optically thick emission of {\hcop} ($\tau$$\sim$8.3 to 15.4) and optically thin {\hco} ($\tau$$\sim$0.3 to 0.5) emission in all MDCs. We find that introducing depletion at a density of $\sim$10$^6$~cm$^{-3}$ with a depletion factor of 50-100 gives generally good line intensities. Only within the coldest and densest MDC, which is CygX-N63, depletion had to be introduced at a lower density of $\sim$2$\cdot$10$^5$~cm$^{-3}$. With this modeling we show that for the optically thin cases the emission is dominated by the outer layers and not the regions close to the center, where depletion is high. This indicates that the emission is coming more from the bulk of material with densities between 10$^4$ -- 10$^6$ cm$^{-3}$, corresponding to the inter-protostar medium, the mass reservoir of the core. 


  \begin{figure*}[!htpb]
     \begin{minipage}{18cm}

   \centering
   \includegraphics[width=4.6cm]{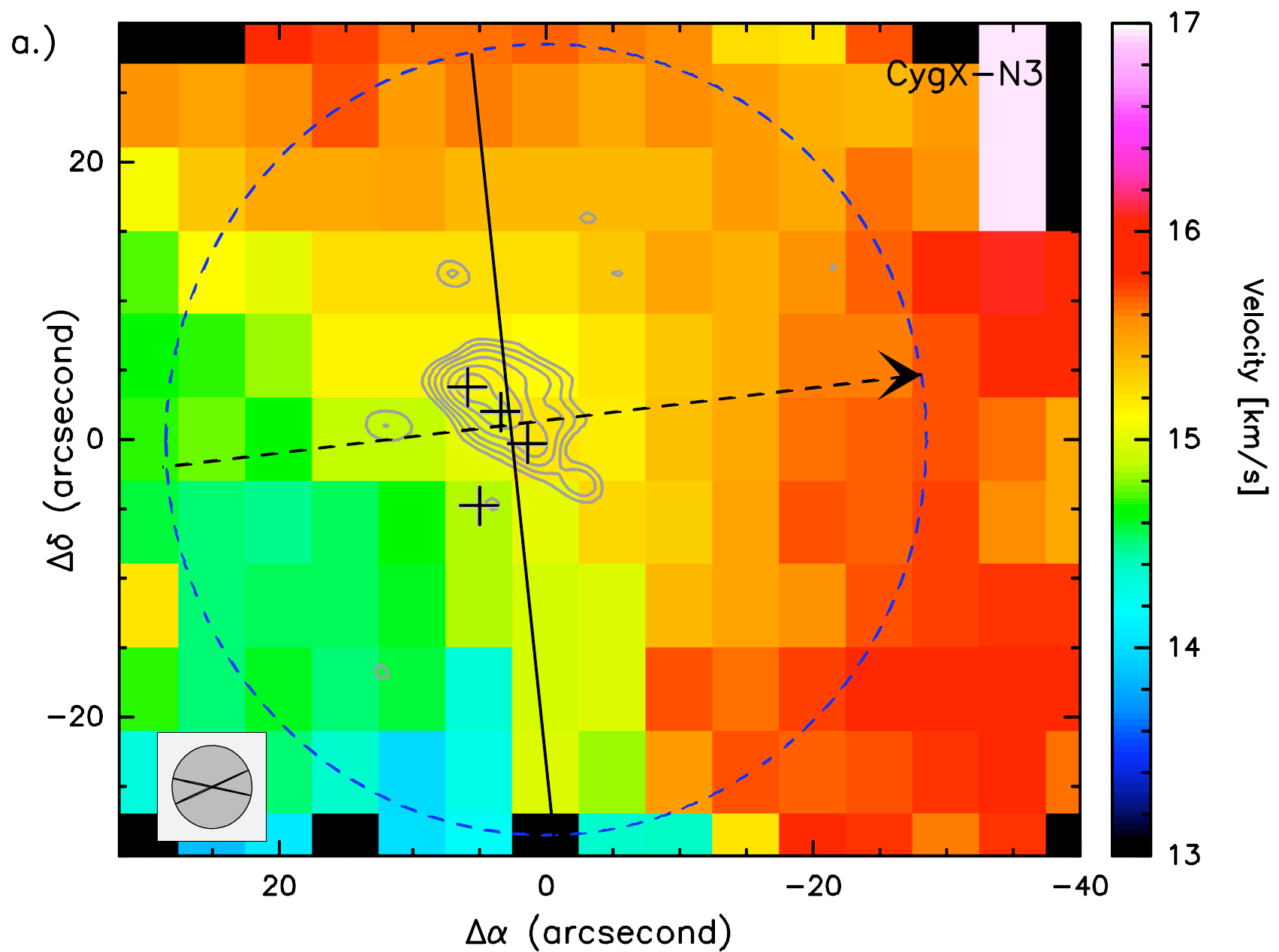}
   \includegraphics[width=4.6cm]{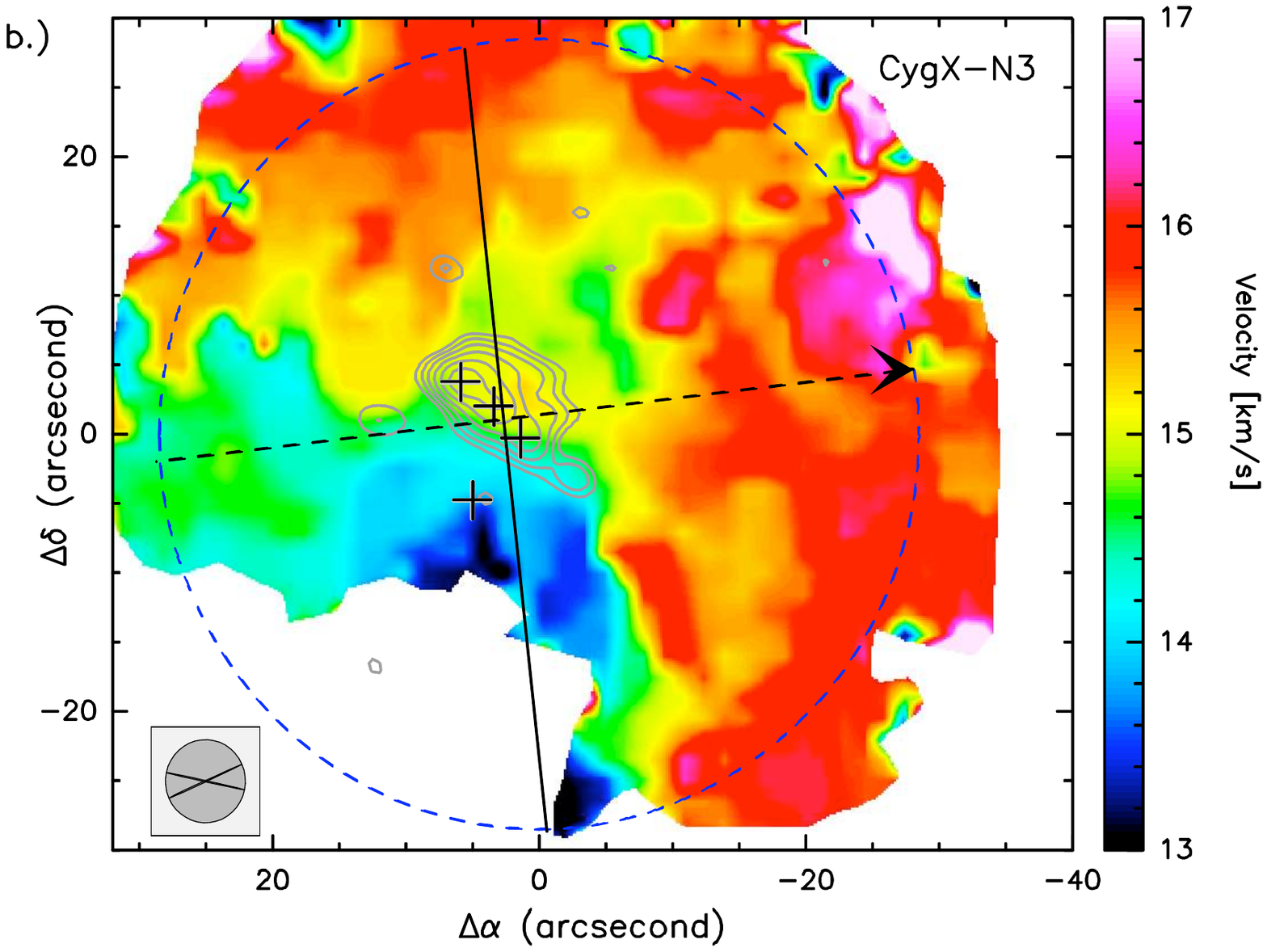}
   \includegraphics[width=4.25cm]{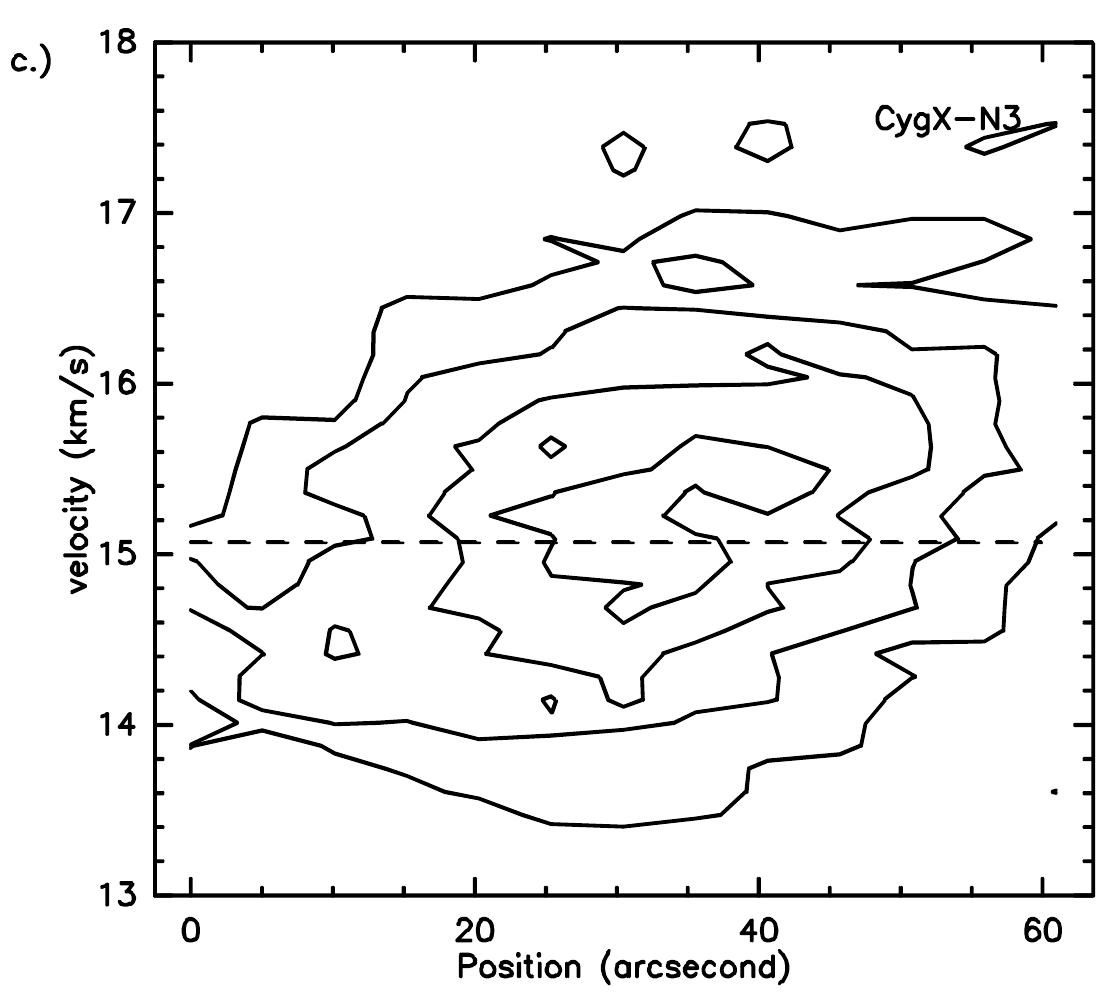}
   \includegraphics[width=4.25cm]{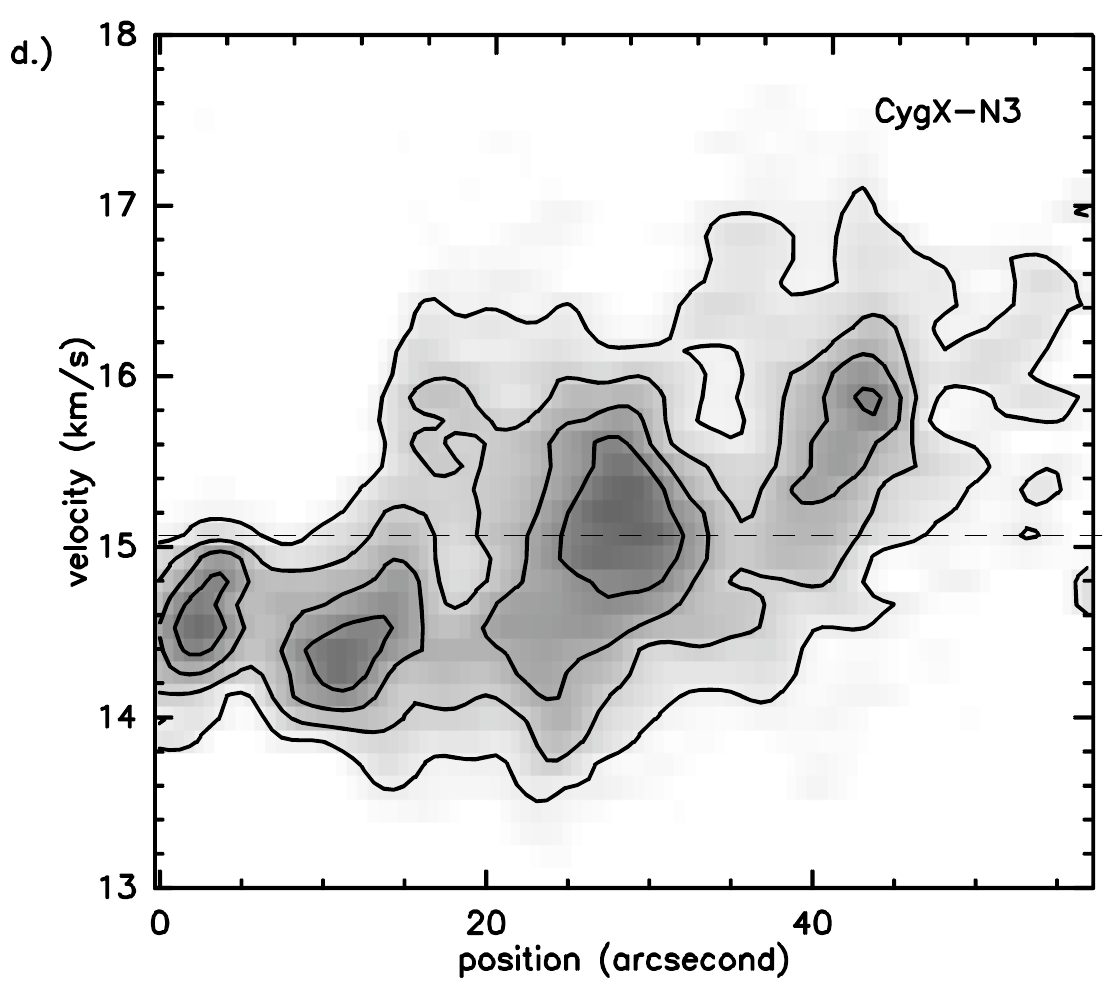}

   \includegraphics[width=4.6cm]{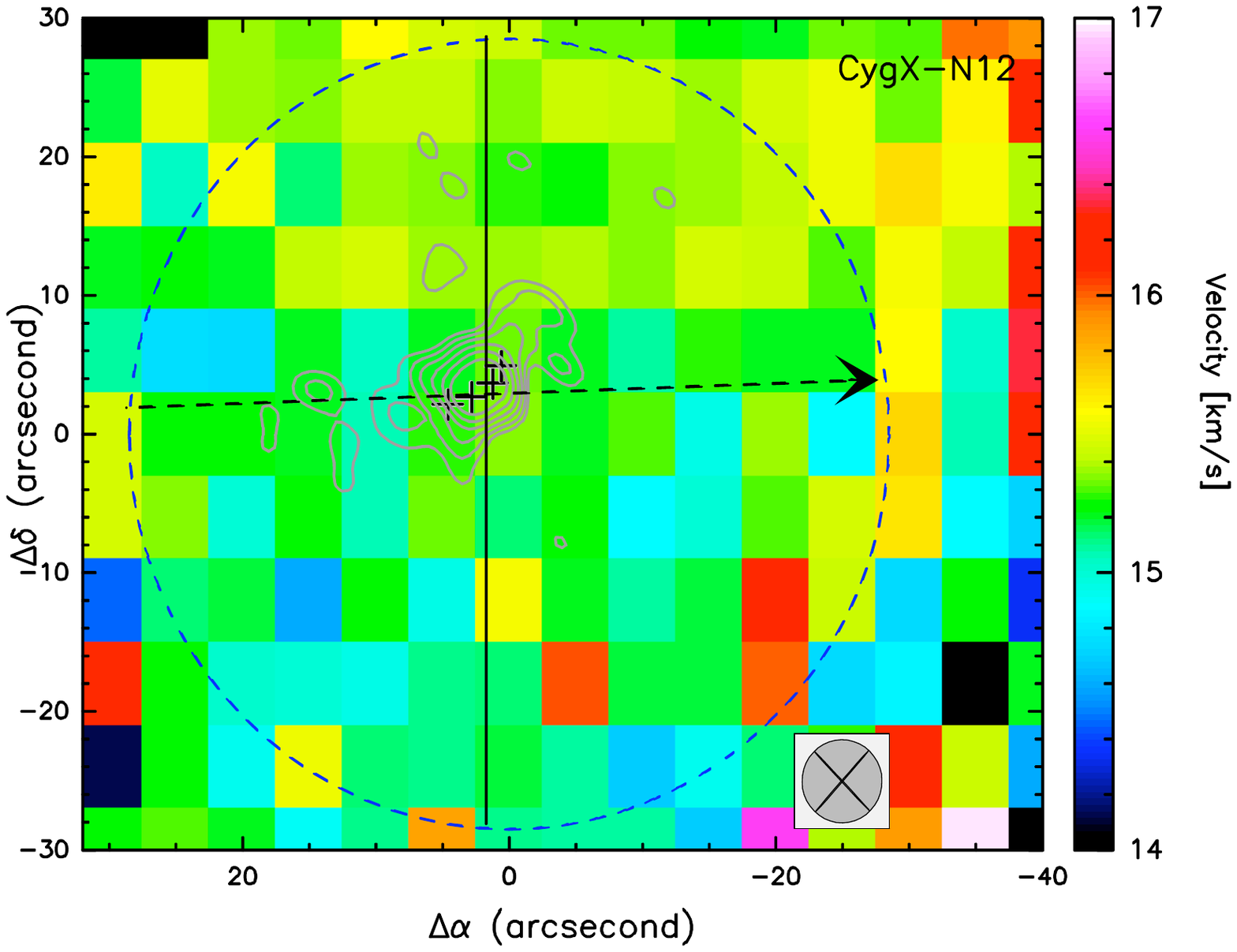}
   \includegraphics[width=4.6cm]{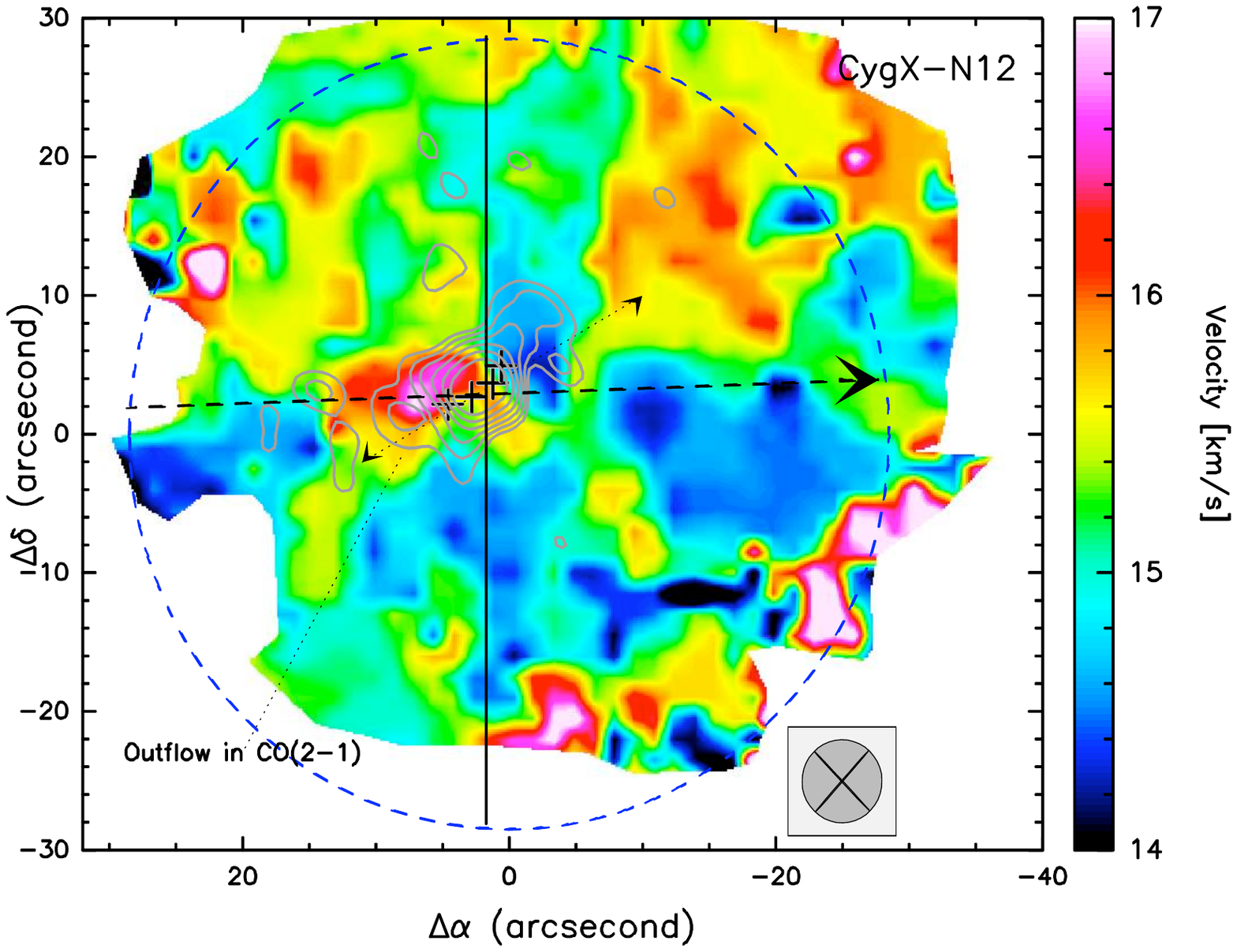}
   \includegraphics[width=4.2cm]{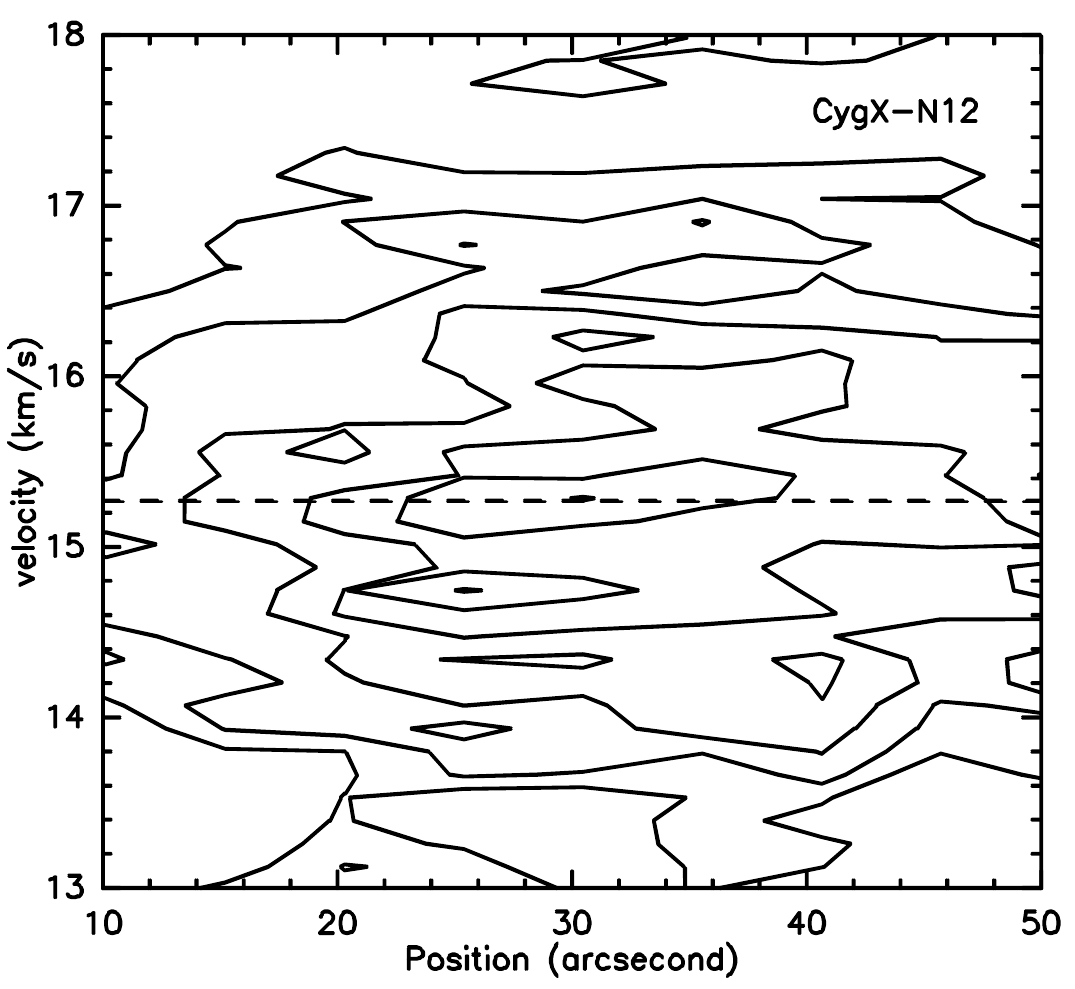}
   \includegraphics[width=4.2cm]{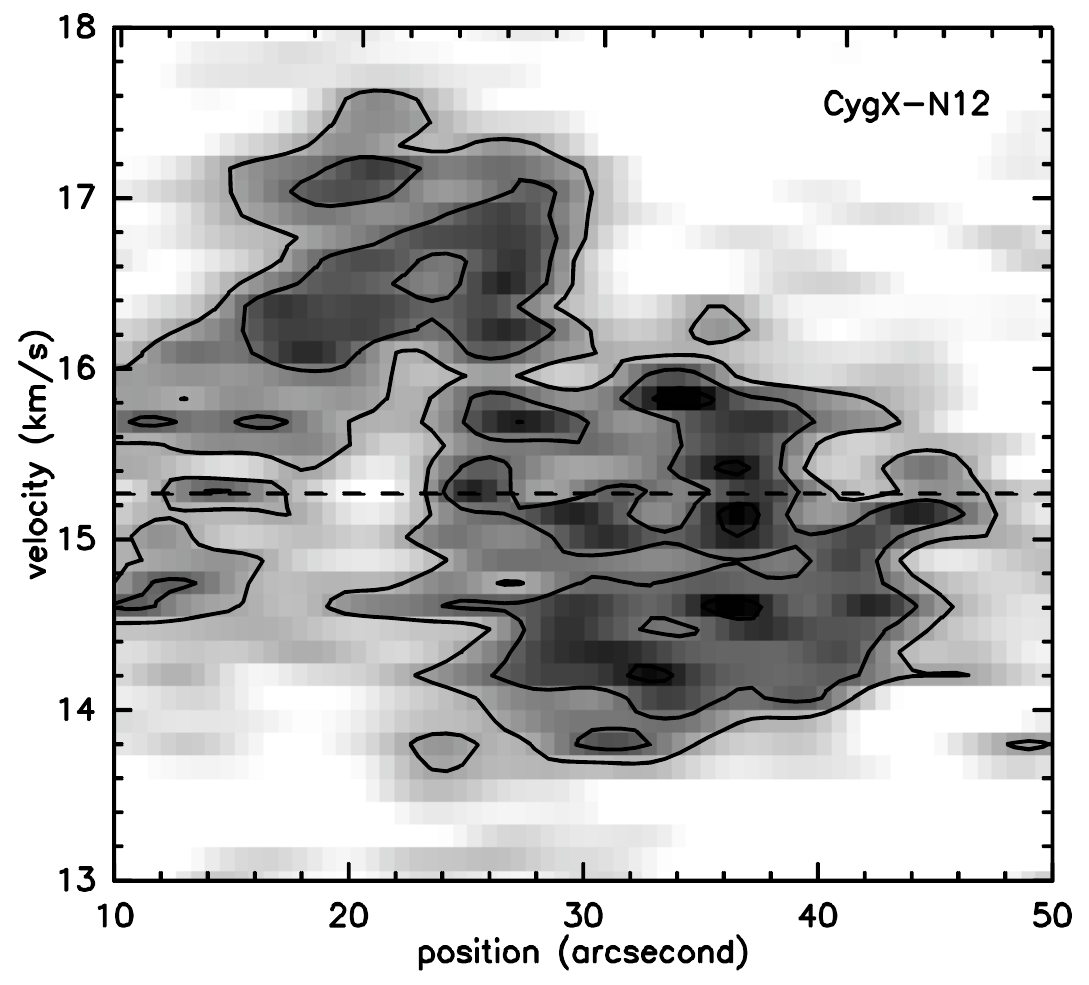}

   \includegraphics[width=4.6cm]{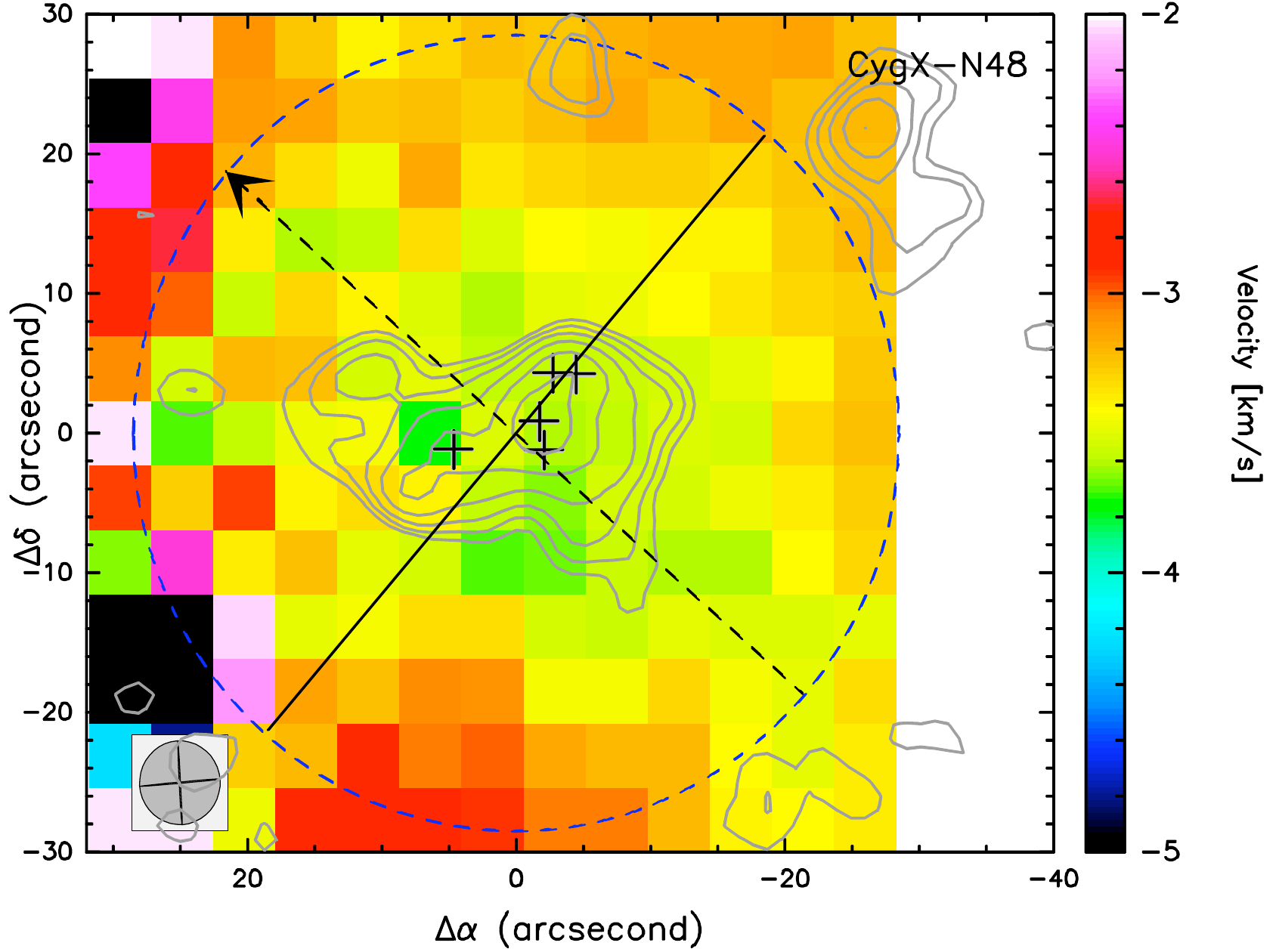}
   \includegraphics[width=4.6cm]{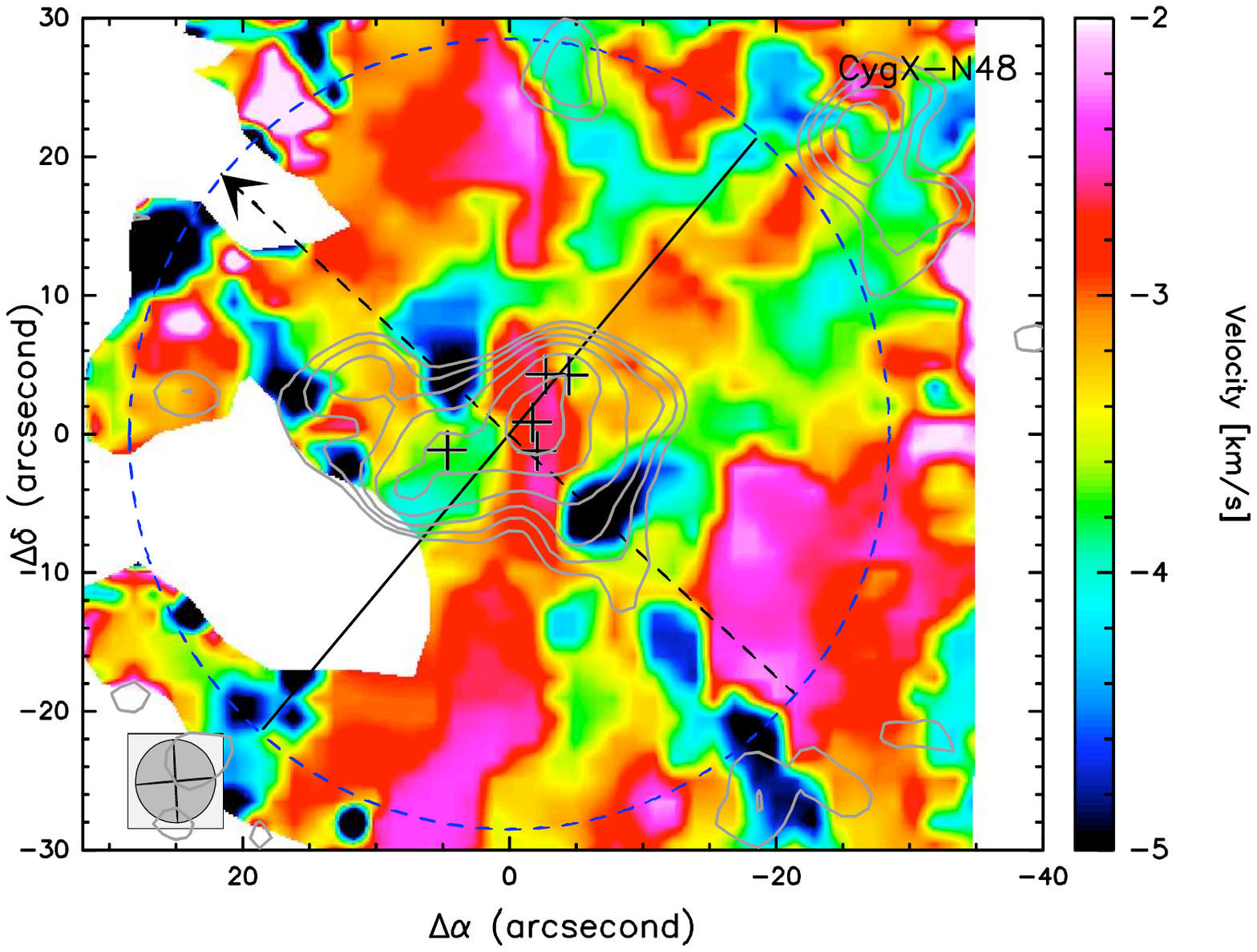}
   \includegraphics[width=4.2cm]{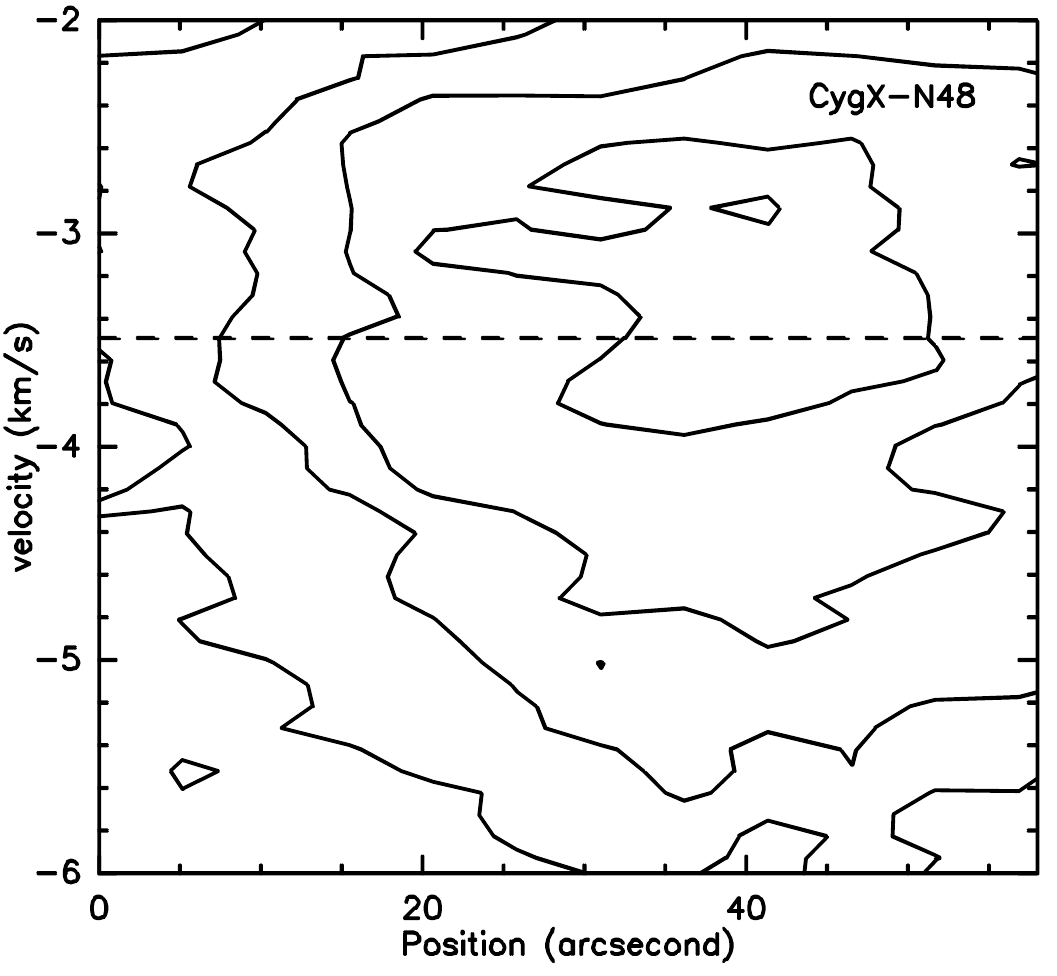}
   \includegraphics[width=4.2cm]{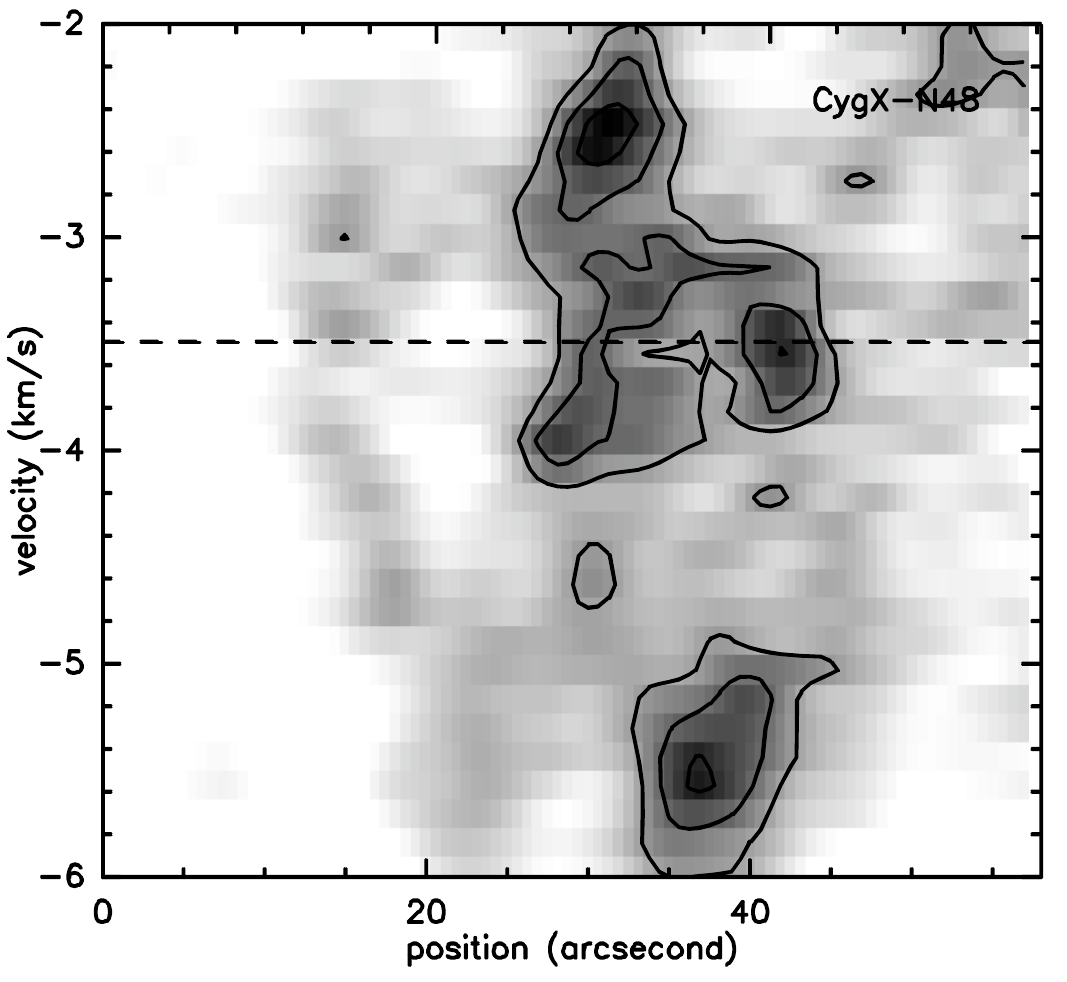}

   \includegraphics[width=4.6cm]{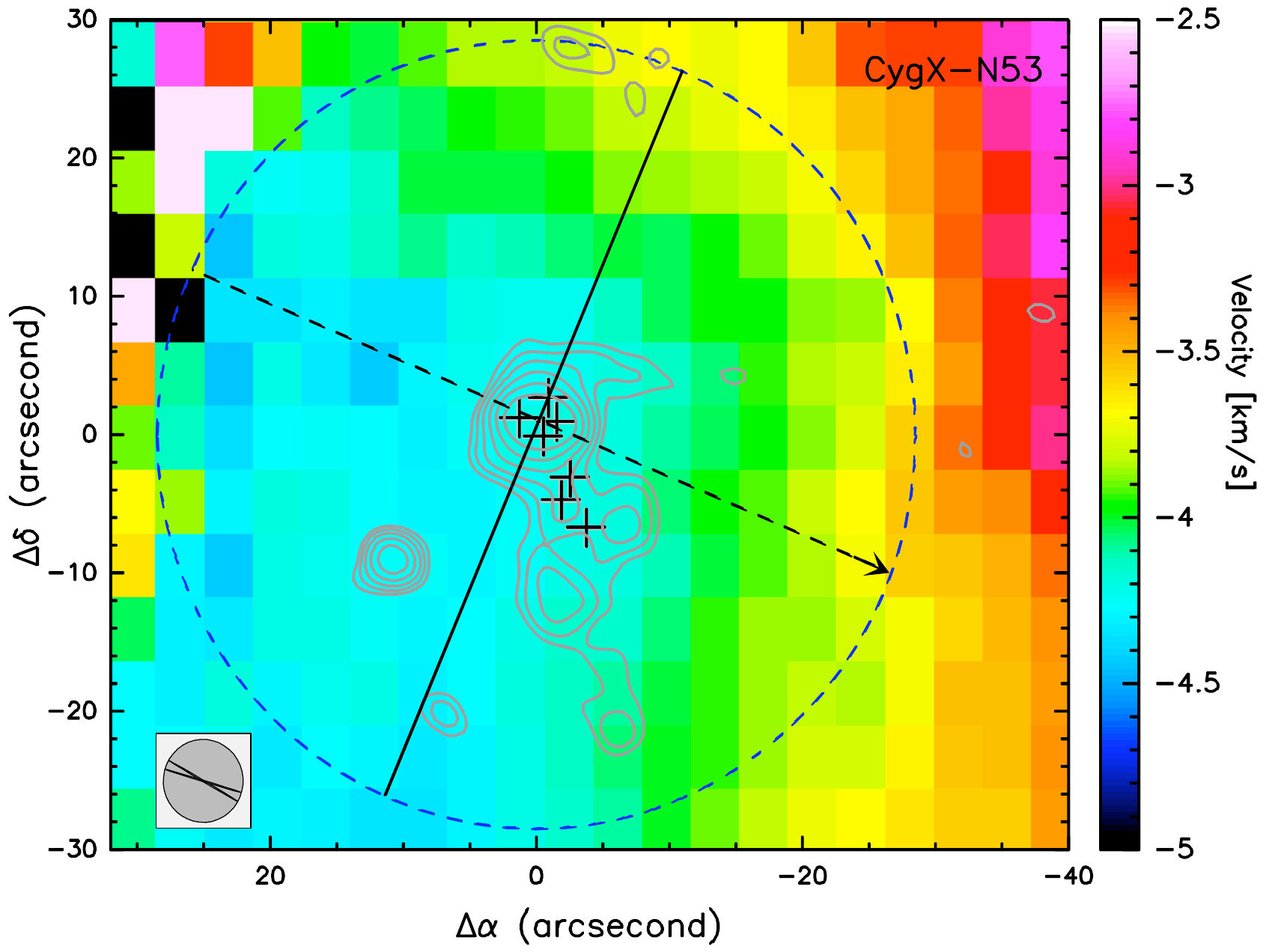}
   \includegraphics[width=4.6cm]{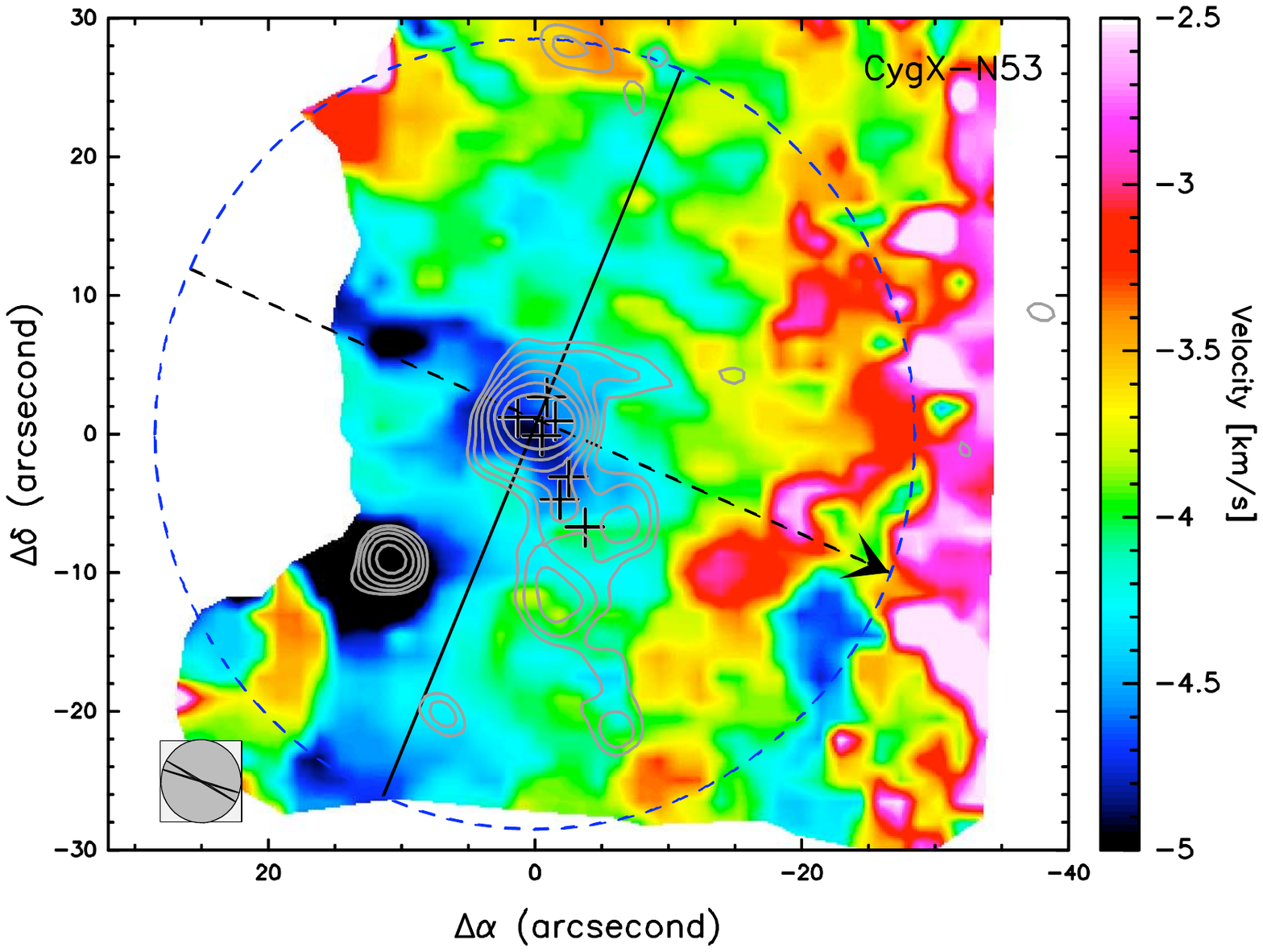}
   \includegraphics[width=4.2cm]{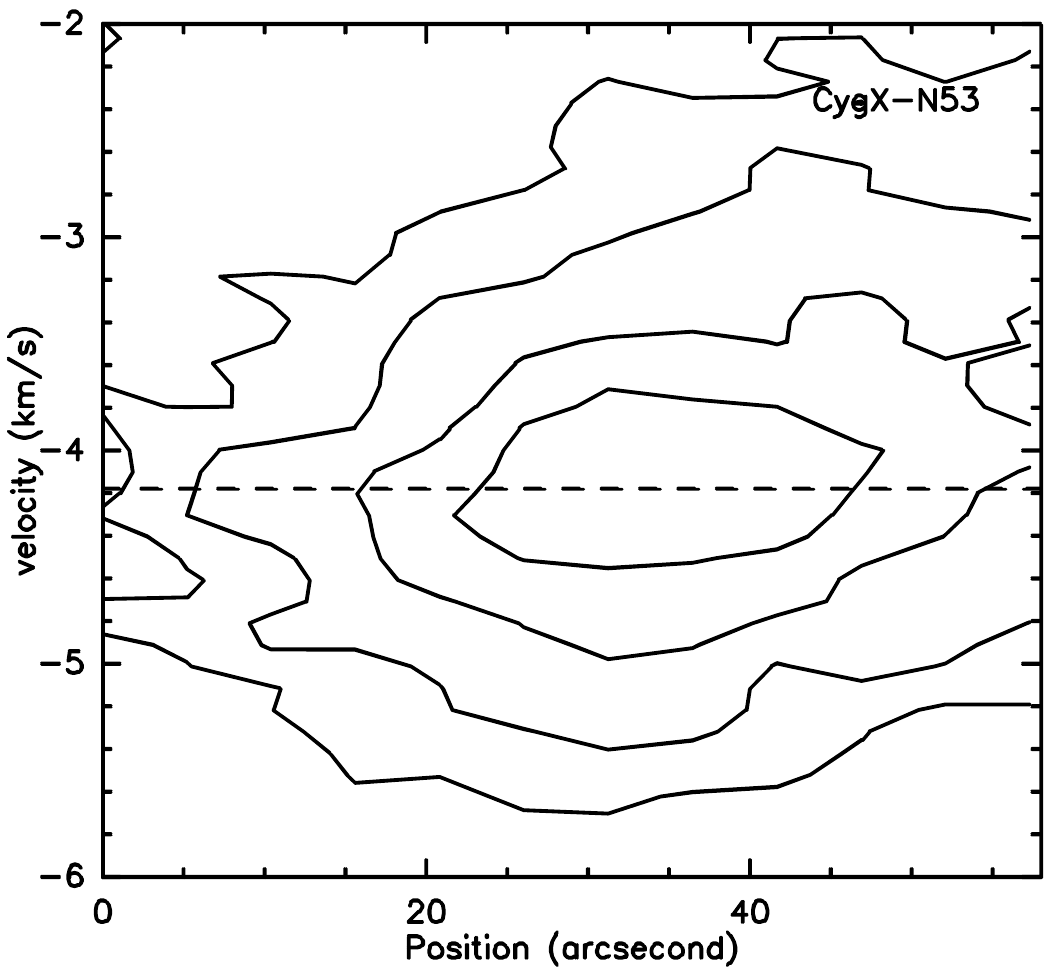}
   \includegraphics[width=4.2cm]{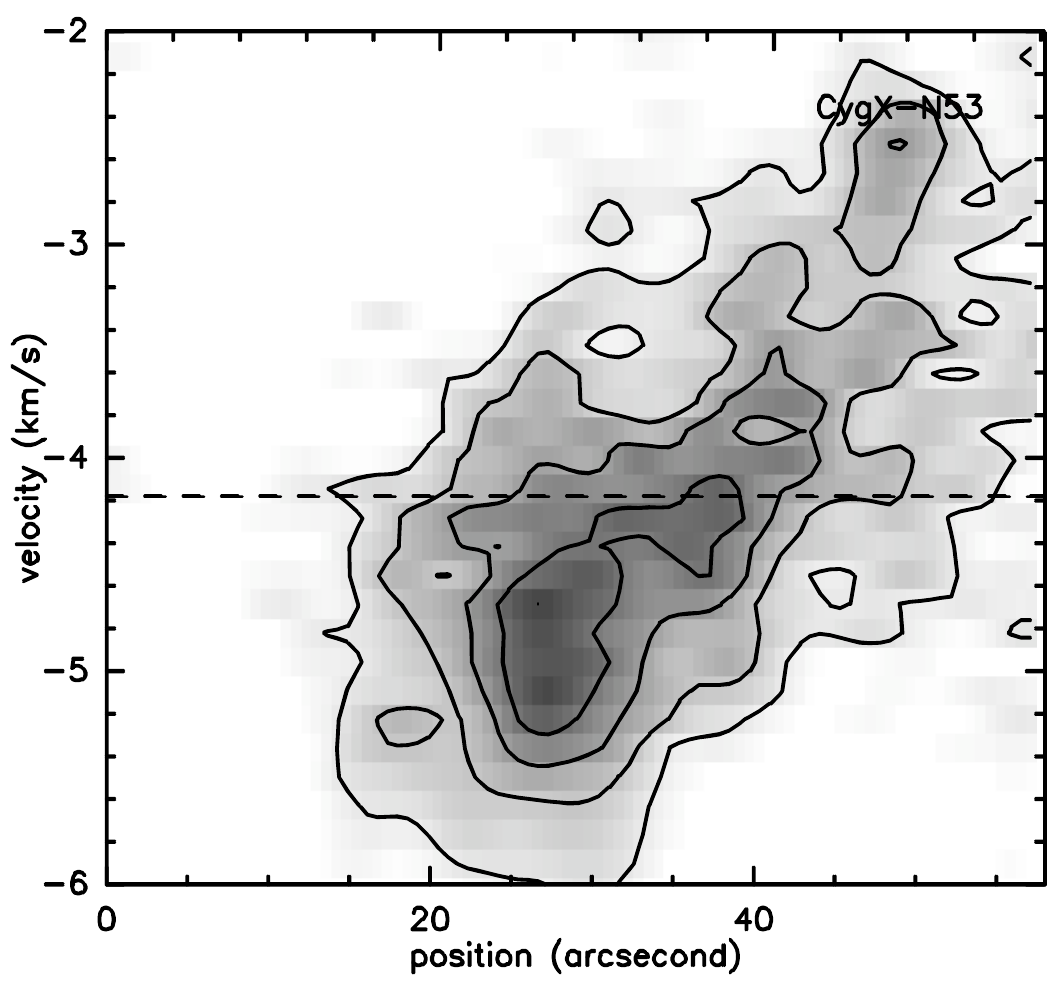}

  \includegraphics[width=4.6cm]{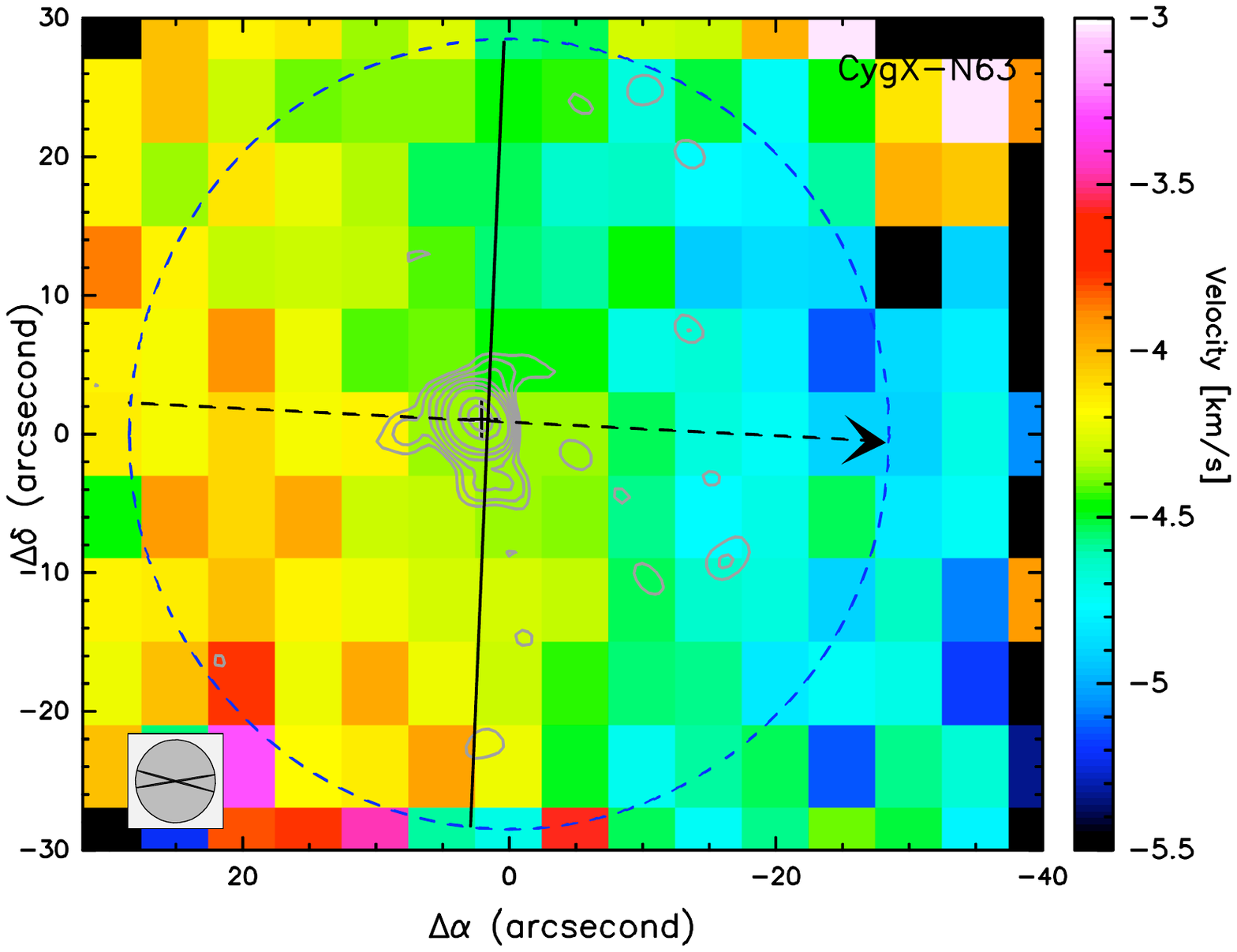}
  \includegraphics[width=4.6cm]{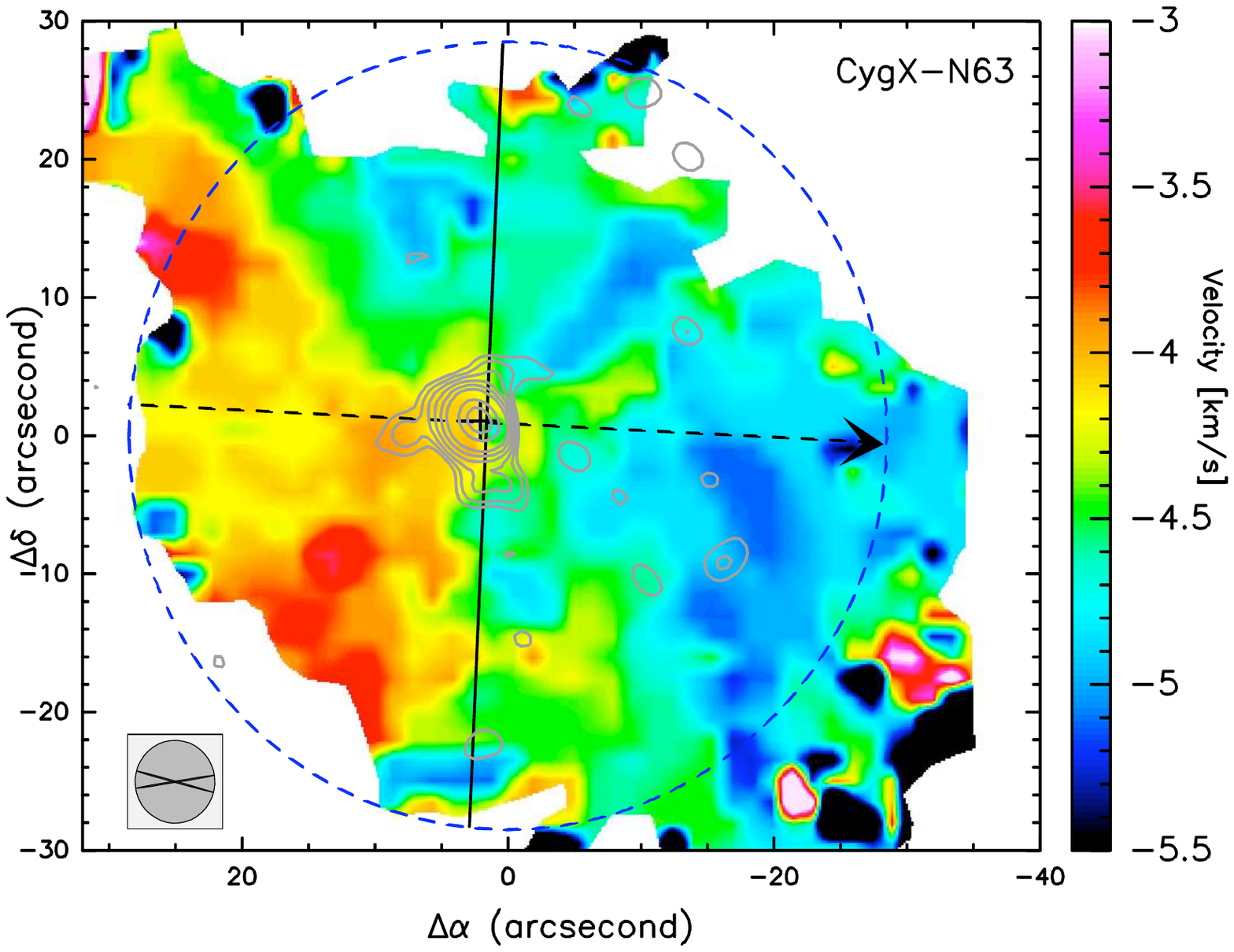}
   \includegraphics[width=4.2cm]{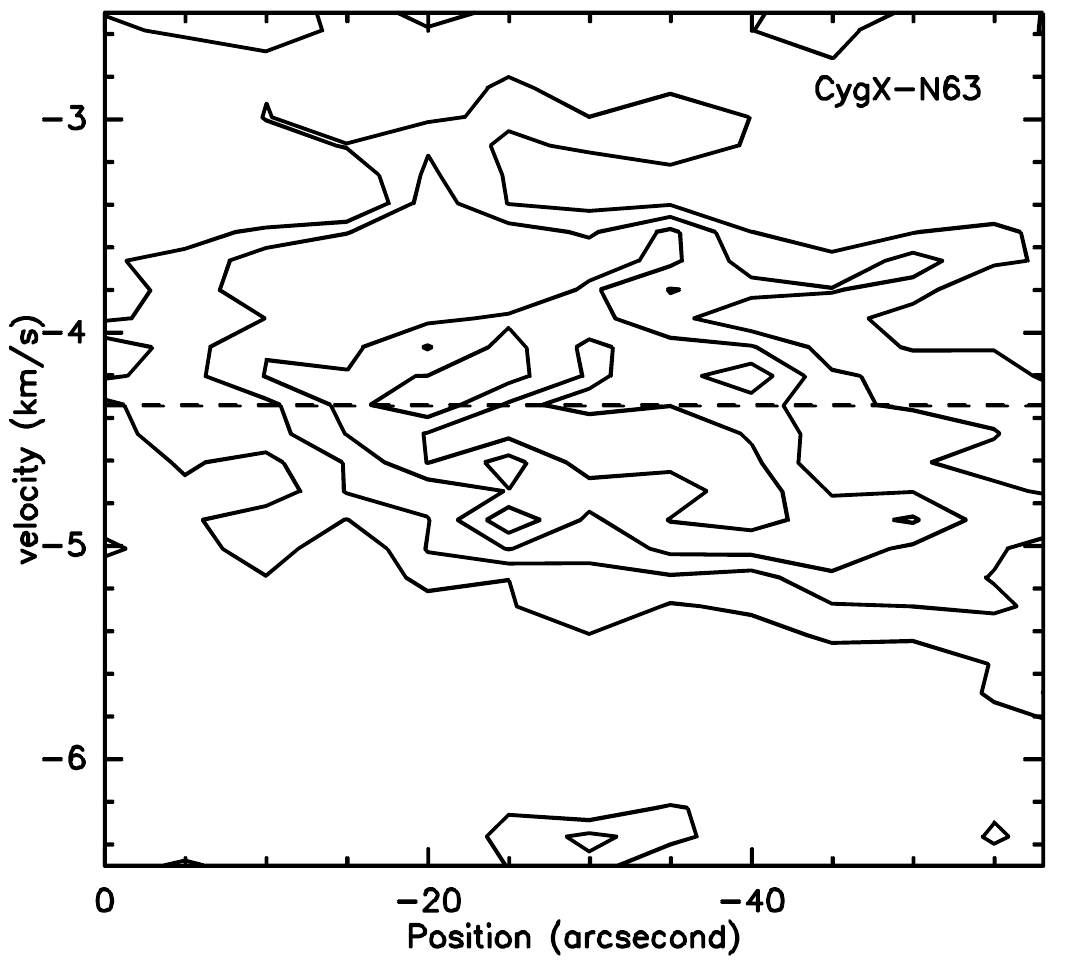}
   \includegraphics[width=4.2cm]{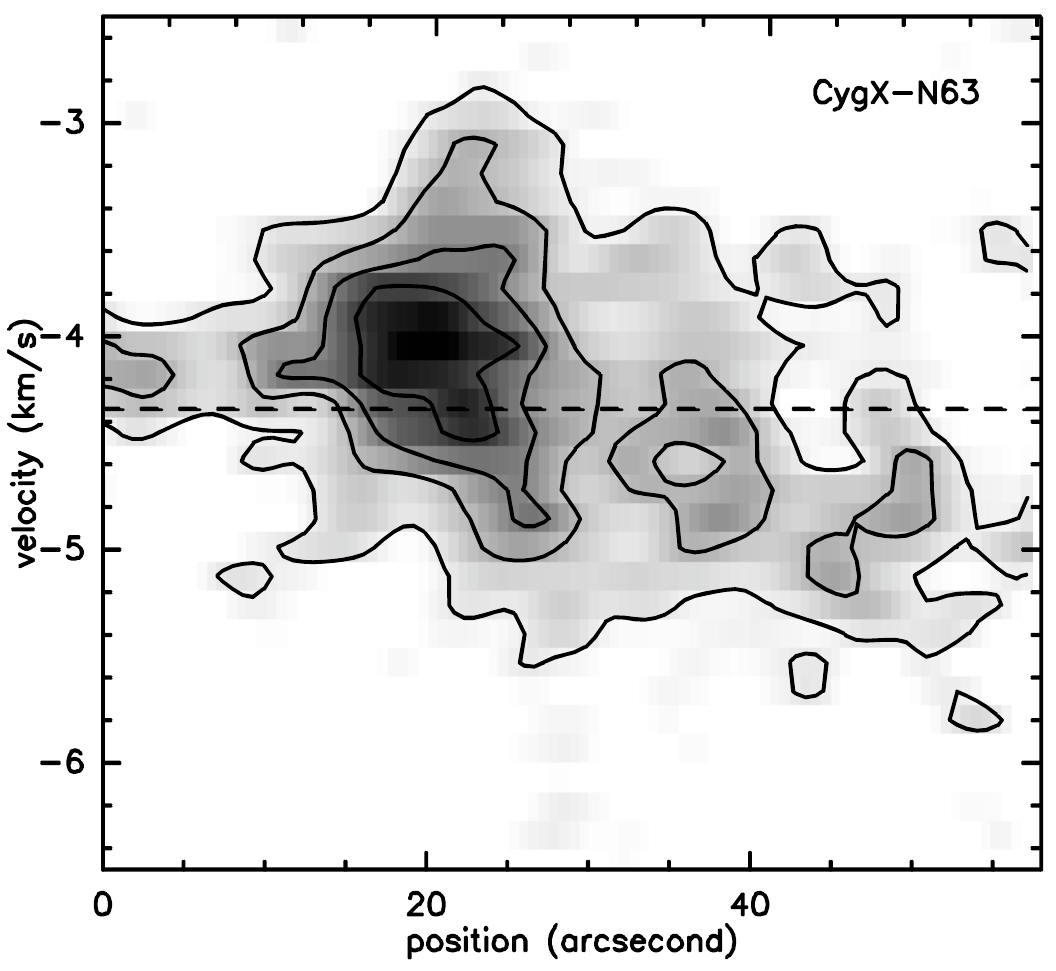}

  \caption{\textit{a)} The maps show the velocity-field of the single-dish maps of {\hco}, which is obtained by a single gaussian fit. Dashed circles show the primary beam size of the PdBI. Dashed lines show the axis of the velocity gradient, while the arrows indicate the position angle of the gradient. The error of the gradient fitting is shown in the lower corners of the images. \textit{b)} The maps show the peak velocity of the PdBI+30m {\hco} spectra. \textit{c)} Position-velocity cuts along the velocity gradient derived in Section \ref{sect:goodman} and indicated on {\sl a)} from the single-dish \hco spectra-cubes. Contour levels go from 20\% of the peak intensity by 20\% levels. Dashed lines indicate the systemic velocity. \textit{d)} IRAM 30m+PdBI position velocity cuts along the same direction. Contour levels indicate 20\% of the peak emission as well and increase by 20\% for CygX-N3,-N53,-N63, while contours start at 50\% of the peak for CygX-N12, -N48.}
  \label{fig:pdbi_velfield}
  \end{minipage}
  \end{figure*}
%

\begin{table}
\centering                          
\caption{Result of least square fitting to the velocity gradient of the IRAM 30m {\hco} maps: velocity gradients and positions angles. We calculated the $\beta$ parameter ($E_{rot}/E_{pot}$) and the specific angular momentum respectively. Note that towards CygX-N12 and -N48 no gradient was found.}
\begin{tabular}{c c c c c}        
\hline\hline                 
Source &$ \mathscr{G}$ & PA  & $\beta$ & J/M  \\    
            & [km~s$^{-1}$~pc$^{-1}$]       &   [$^\circ$] & & [km~s$^{-1}$~pc]  \\    
\hline                        
   CygX-N3   & 1.97 $\pm$ 0.7 & 96 & 30 $\times$ 10$^{-4}$  & 7.88 $\times$ 10$^{-3}$  \\     
   CygX-N12 & 0.42 $\pm$ 0.43 & -89 & 1.5 $\times$ 10$^{-4}$ & 1.68 $\times$ 10$^{-3}$\\
   CygX-N48 &  0.68 $\pm$ 0.7  & 46 & 4.2 $\times$ 10$^{-4}$ & 5.33 $\times$ 10$^{-3}$  \\
   CygX-N53 & 4.17 $\pm$ 0.7 & 67   & 250 $\times$ 10$^{-4}$ & 20.9 $\times$ 10$^{-3}$\\
   CygX-N63 & 1.17 $\pm$ 0.34 & -93 & 4 $\times$ 10$^{-4}$ &1.86 $\times$ 10$^{-3}$\\ 
\hline                                  
\end{tabular}
\label{table:vfit}      
\end{table}


\subsection{Velocity field}
\subsubsection{Global velocity field: rotation like patterns?}
\label{sect:goodman}
The systematic shifts seen in the line-positions were studied to understand the global kinematics of MDCs. We derive the velocity field with the gaussian line-fitting procedure to the {\hco} maps of the IRAM 30m telescope (see Figure \ref{fig:pdbi_velfield}). CygX-N3, -N53 and -N63 show a homogenous, but clearly not isotropic velocity field with a systematic, axi-symmetric shift in the line position (velocity). In CygX-N53 this velocity pattern follows the global velocity field of the DR21 filament \citep{Schneider_prep}. Towards CygX-N12 and CygX-N48 no clear global pattern is seen, the velocity field seems to be rather random. 
Figure \ref{fig:sda} {\sl b)} indicates that in CygX-N12 a single gaussian fit may not reproduce well the observed line profiles due to blending of more components. Similarly, for CygX-N48 a second component is seen as a weak shoulder on the spectra, therefore a single gaussian fit for these two MDCs may not be representative of the velocity field.

Altogether, for 3 out of 5 MDCs a rotation like velocity pattern is visible. To derive the axis and position angle of the velocity shift, we use the least square fit approach of \citet{Goodman93}. We applied the routine \textsl{VFIT}\footnote{IDL routine for performing least square fit to the velocity field.} \citep{Goodman93} in order to derive velocity gradients from the velocity field. The results are shown in Table \ref{table:vfit} and the position angle of the resulting gradients are indicated in Figure \ref{fig:pdbi_velfield}. The derived angle corresponds well with the observed axi-symmetry of the velocity field. The velocity gradients have the same order of magnitude of what has been observed for low-mass nearby cores \citep{Goodman93, Caselli02}, between 1.2 - 4.2 km s$^{-1}$pc$^{-1}$. The highest gradient is seen in CygX-N53, while in CygX-N12 and -N48 no gradients can be derived. 
Among the three cores, where the velocity field follows a clear symmetry, the smallest gradient is associated with the least massive, but most compact core (containing only one massive protostar), while the largest is in CygX-N53, which is embedded in the DR21 filament. That itself has a very complex kinematics with large gradients. The velocity gradient of CygX-N53 is consistent with previous estimates of \citet{Schneider_prep} indicating that this core is embedded and kinematically linked to the filament.

\subsubsection{Small-scale velocity field}
We compare the velocity field obtained with single-dish, low angular-resolution data with the high angular-resolution observations obtained with PdBI (and zero-spacings). Since the spectra show several lines dispersed within a few km~s$^{-1}$, a single gaussian fit
can not be representative of the velocity field. Therefore we extract and map the peak velocity of the spectra, which results in a velocity field tracing the bulk of motions (see Figure \ref{fig:pdbi_velfield} right panel).

At high angular-resolution the axisymmetric global velocity pattern is resolved into more substructures in velocity, indicating a much richer dynamics. Despite of the small-scale velocity fluctuations, in 3 MDCs the velocities are dominated by a large-scale gradient. CygX-N3, -N53, -N63 show a remarkably symmetrical global velocity field with systematic shifts in velocity, consistent with the single-dish maps. The velocity field in CygX-N12 and -N48 lack clear gradients and are dominated by fluctuations on small-scales, which make it very complex.
Within CygX-N12  a strong outflow is observed in $^{12}$CO (J=2-1) (Bontemps et al., in prep), which may also be apparent in the velocity field map as indicated in Figure \ref{fig:pdbi_velfield}.  
Towards CygX-N48 the velocity field is complex without any clear trends of systematic motions. The complexity of such velocity field towards CygX-N12 and CygX-N48 is due to the presence of several spectral components, which are also seen on Figure \ref{fig:sda}. 




\subsection{Position-velocity diagrams}
Position-velocity cuts were taken along a cut through the continuum peaks and perpendicular to the axis derived from the fit to the velocity fields. In Figure \ref{fig:pdbi_velfield} \textsl{c)} contours of the position-velocity maps of the single-dish observations are shown. The velocity gradients towards CygX-N3, -N53 and -N63 are clearly seen. For  CygX-N12 and -N48 a large dispersion in velocity becomes obvious, which explains the complexity of the velocity field maps. 

Figure \ref{fig:pdbi_velfield} \textsl{d)} shows the position-velocity maps along the same cuts from the high-resolution IRAM 30m+PdBI data. In all cases we resolve much richer structures confirming the existence of several line components and gradients. For CygX-N3, -N53 and -N63 these velocity cuts illustrate perhaps best the complexity of velocity features obtained with the high angular-resolution data. In CygX-N3 the pv-map shows a clumpy structure with several peaks showing a gradient from $\sim$14.5-16~km~s$^{-1}$. The peaks correspond to individual structures at different positions, but we point out that they must be physically linked, being embedded in the same flow of gas and consistent with the velocity gradient seen in the single-dish data. In CygX-N53 and -N63 the velocity shift is smooth (less clumpy) and goes from $\sim$-5.5 to -3.5~km~s$^{-1}$ and from -3.5 to -5~km~s$^{-1}$, respectively. 

For CygX-N12 and -N48 we resolve for the first time indications for velocity gradients. On these two plots (Figure \ref{fig:pdbi_velfield} {\sl d}, middle panels) we increased the contour levels to show only the 50\% level of the peak emission, since there is emission coming from broad line-wings due to outflows. For CygX-N12 the gradient goes from 17~km~s$^{-1}$ down to 14~km~s$^{-1}$, while CygX-N48 shows two components: one dispersed between -2 and -4~km~s$^{-1}$, the other component at $\sim$-5.5~km~s$^{-1}$. 

Such velocity gradients, as seen for CygX-N3, -N53 and -N63, can be interpreted in the frame of rotating structures. In case of solid body rotation, velocity scales linearly with the position, which could be consistent with the pv-maps for CygX-N3, CygX-N53 and CygX-N63. For the other fields solid body rotation is unlikely due to the complexity of the individual velocity components.
The origin of the velocity pattern is discussed in Section \ref{sect:rotation}.

\subsection{{\hcn} spectra}
We performed HFS line-fitting to the extracted {\hcn} spectra for each core and summarize the results in Table \ref{tab:hcn_fit}. For CygX-N3, -N12, -N53 and -N63 we obtain a good fit with only a single line, but we indeed identify two components for CygX-N48 at --4.77~km~s$^{-1}$ and at --2.39~km~s$^{-1}$. The line velocities of {\hcn} are in good agreement with the velocity of the \hco~ line ($\pm$0.2~km~s$^{-1}$) towards CygX-N3, and -N63, while for CygX-N12 and -N53 they are off by $\sim$0.8~km~s$^{-1}$ compared to the {\hco} line. The reason behind this is not clear, but may indicate that there are chemical differences on the projected size-scales of $\sim$ 10.000 AU and these two tracers may not trace the same gas. The HFS fitting in Figure~\ref{fig:h13cn_spectra} suggests that the assumption of LTE excitation condition is satisfactory for all MDCs.

In Section \ref{sec:h13cn} we noted already the "bump" in the spectra, peaking at the position of the main line with a large plateau on the two sides. We further investigated the possible origin of this feature by searching for complex molecules at these frequencies but found no lines that could result in such a strong contribution. 
This is the most prominent feature in CygX-N63, where the line-dispersion is surprisingly large ($\sim$2~km~s$^{-1}$) compared to the other MDCs. We discuss in detail the possible origin of that in Section \ref{sect:tracer2}.


\section{Discussion}


From a theoretical point of view, there is either a high level of turbulence in the MDCs, capable to provide a significant support against gravity to allow a quasi-static evolution towards the
monolithic collapse of the cores, or the cores are not in equilibrium and their formation and evolution is mostly driven by dynamical processes.
Specific and different kinematics is expected in the two scenarios (e.g. supersonic micro-turbulence versus complex, but organized flows).

We first discuss the origin of the observed line emission (Sect.~\ref{sect:tracer}), then we analyze the physical origin of the velocity dispersion at large ($\sim$0.1 pc) and small scales ($\sim$0.03 pc, Sect.~\ref{sect:broadening}) to ultimately discuss the amount of micro-turbulence in the cores which could stabilize and regulate their evolution. 
We finally review the obtained results in the context of the two scenarios in competition (Sect.~\ref{sect:scale-turbulence}). 

\subsection{Molecular lines to trace high-density material}
\label{sect:tracer}

\subsubsection{\hco ~traces the inter-protostar medium}

{\hco } has often been detected around low-mass protostars following an elongated structure perpendicular to their outflow. Having a high critical density, it is acknowledged as a good tracer of low-mass protostellar envelopes (e.g. \citealp{Saito01};  \citealp{Jorgensen04}). The average density in our sample of MDCs is high ($>$10$^5$ cm$^{-3}$) \citep{M07}, while the peak density in protostars estimated in \citetalias{B09} is reaching up to $\sim$10$^7$-10$^8$ cm$^{-3}$. These MDCs lack significant emission at infrared wavelengths, and recent NH$_3$ measurements \citep{Wienen} confirm that they are cold with kinetic temperatures between 18--28 K. Such high densities at low temperatures should lead to a high level of depletion of virtually all C-bearing molecules as a consequence of CO depletion onto grains (e.g. \citealp{Tafalla02}; \citealp{Bergin07}). On the other hand, the desorption temperature for CO is of the order of 20~K, i.e. similar to the observed temperatures here, considering also 
that if protostars have formed already, they might 
heat-up their cold environment, releasing and restarting the formation of C-containing molecules. The precise level of depletion is therefore uncertain and the abundances of different molecules (here \hco and \hcn) as a function of density and temperature is even more difficult to predict.


From the radiative transfer modeling (Section \ref{subsec:simline}, Appendix \ref{app:simline} and in Figures \ref{fig:sim-n3} -- \ref{fig:sim-n63}) we conclude that {\hco} is always optically thin 
and the emission is dominated by the gas between the protostars and can not probe their highest densities 
due to significant depletion in their central regions. We find indications of the highest depletion towards CygX-N63
, where in its central regions a significantly higher depletion ratio and lower depletion density was used to reproduce the correct line intensities.
Summarizing this picture, we base the following discussion on considering that the emission of {\hco} comes from the dense material within the MDC, but rather representing the bulk of material in which the protostars are embedded, than their inner envelope as for low-mass protostars. 

\subsubsection{\hcn, a better tracer of the highest density regions?}
\label{sect:tracer2}

Modeling of e.g. \citet{Lee04} has shown that HCN may also be affected by depletion, but to a somewhat lower level than {\hcop}. In surveys for low-mass pre-stellar cores \citet{Sohn07} observed that HCN traces better the inner part of the cores than {\hcop} and shows a similar distribution as N$_2$H$^+$, which is known to trace the inner (dense, cold) part of the cores (Tafalla et al. 2006; Di Francesco et al. 2007). 
Nevertheless, \citet{Lee04} also pointed out that HCN is part of a complex chemical network, whereas {\hcop} depends directly on the CO abundance. Based on the morphology of the distribution of emission presented in Figure \ref{fig:pdbi}, we show here that {\hcn} has a better correspondence with the central higher density regions (continuum peaks), than {\hco}. The most prominent example is CygX-N63, where {\hco} is completely depleted while {\hcn} peaks at the strong continuum source. The origin of this may be due to two reasons: {\hcn} may be simply less depleted - as it has already been pointed out in examples of low-mass cores, and/or {\hcn} may be affected by different chemistry, with for instance overabundance due to strong shocks emerging from the close vicinity of the protostars.


\subsection{Micro-turbulence and bulk motions within the MDCs}
\label{sect:broadening}

Analysis of line-widths reveals the contribution of thermal versus non thermal motions. The non thermal velocity dispersion is usually interpreted as isotropic turbulence down to small scales where it is often referred to as micro-turbulence. 
However it may have contribution from bulk motions, such as rotation, infall and coherent flows or streamings.
It is of high importance to measure the true level of micro-turbulence since this leads to a local support of similar nature as thermal motions and thus contribute to the global support of the cores against gravity. 

We first discuss the global velocity dispersion as compared to previous studies (Sect.~\ref{sect:virial}).
The analysis of the single dish observations showed that the line widths of \hco ~for most of the observed MDCs have contributions of systematic, global motions which are not micro-turbulence, namely rotation (Sect.~\ref{sect:rotation}) and infall (Sect.~\ref{sect:infall}). The high resolution PdBI data confirm these motions but, in addition, disentangle several velocity components at small scales (see Sect.~\ref{sect:flows}  below). Finally, it is only after these different contributions to the line-widths are recognized, that we will be able to discuss the possible remaining level of micro-turbulence to support the cores (Sect.~\ref{sect:scale-turbulence}).

\subsubsection{Global velocity dispersion and virial equilibrium}
\label{sect:virial}

A one-dimensional thermal velocity dispersion for \hco and \hcn ~gives $\sigma_{T} = \sqrt{\frac{k_B\ T}{\mu_N\ m_H}}$, where $k_B$ is the Boltzmann constant, T is the temperature, $\mu_n$ is the molecular weight equal to 30 and 28 for \hco and \hcn ~respectively and $m_H$ is the mass of a hydrogen atom. Following \cite{M07} and \citetalias{B09} we assume a temperature of 20~K, which gives a thermal velocity-dispersion of 0.074 and 0.077~km~s$^{-1}$ for \hco and \hcn ~respectively, while the observed dispersion is $\sim$1~km~s$^{-1}$ (Table \ref{tab:sd_params}). This is much ($\sim$15$\times$) larger than the thermal velocity-dispersion, indicating that non-thermal motions significantly contribute to the line broadening.

\begin{figure}
\includegraphics[width=9cm]{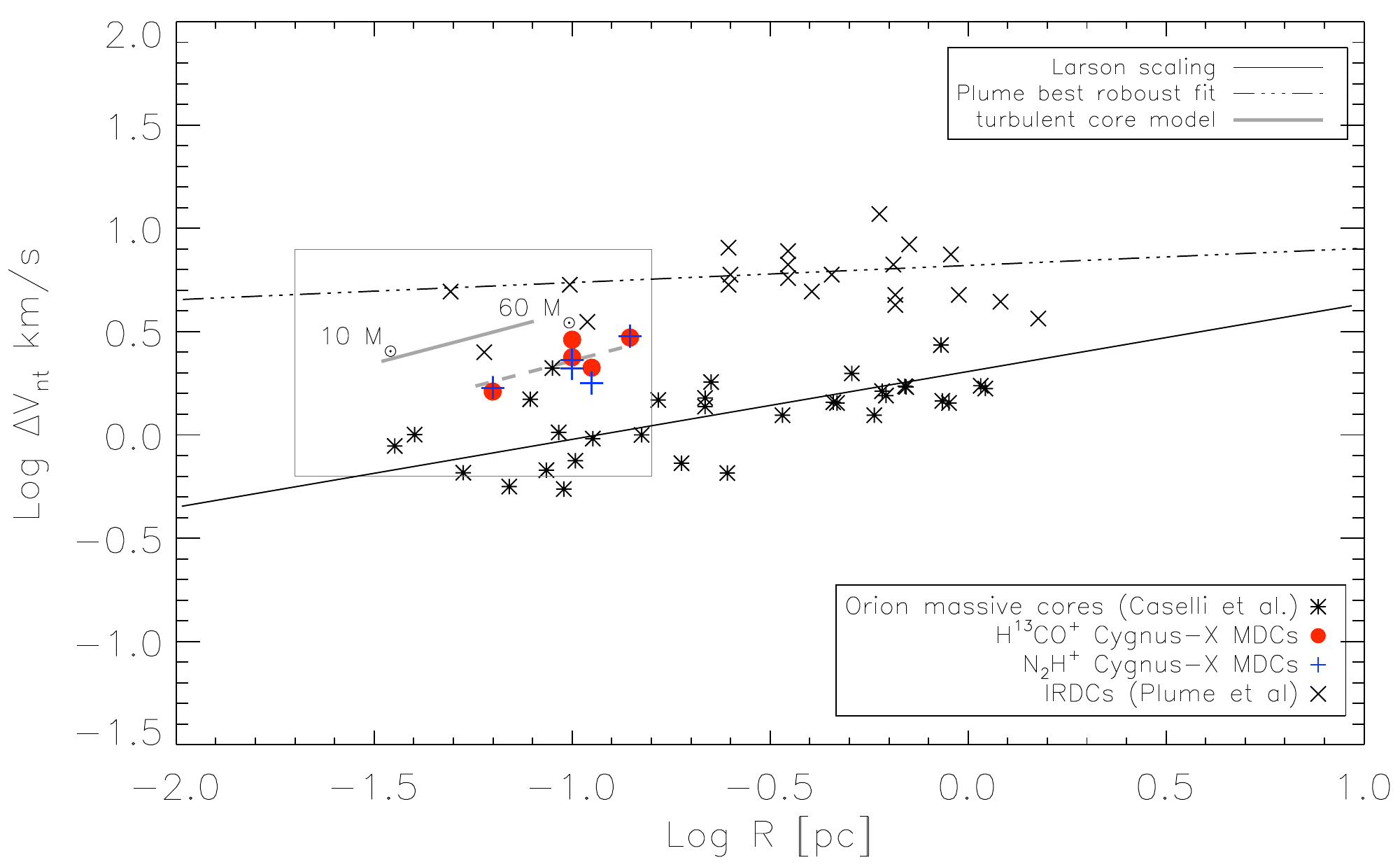}  
\caption{We plot line-widths (FWHM) of the Cygnus-X MDCs (red large dots and blue crosses) versus the core size. Black symbols indicate line-widths taken from the literature \citep{Caselli95}. Black line indicates a Larson like line-widths size-relation, while the dashed-ditted line is adopted from \citet{Plume97} representing their best robust fit to their observed line-width size-relation of IRDCs. Grey line indicates the fiducial model of \citet{MT03} with a canonical surface density of $\sim$~1g~cm$^{-2}$ and a final stellar mass of 10 to 60 M$_{\odot}$. Dashed grey line indicates the same model with a clump surface density 3 times lower. A small box indicates the zoom presented in Figure \ref{fig:linewidth}.}
\label{fig:caselli}
\end{figure}

Figure~\ref{fig:caselli} shows the line-widths of the single dish {\hco} (this work) and N$_2$H$^+$~lines (from \citetalias{B09}) for the 5 MDCs in comparison with samples of single-dish observations of massive cores in Orion \citep{Caselli95} and Infrared Dark Clouds (IRDCs) \citep{Plume97}. The Cygnus X MDCs appear {to be intermediate between} the Orion cores of \citet{Caselli95} and the larger distance IRDCs of \citet{Plume97}.



We use Eq. 18 and 20 with the fiducial model assumptions of \citet{MT03} to predict the required turbulent velocity dispersions at a corresponding core radius to keep the MDCs in equilibrium: 

\begin{equation}
	\sigma_s = 1.27 \Big( \frac{m_{*f}}{30\, \rm M_{\odot}} \Big)^{1/4} \Sigma_{cl}^{1/4}
\end{equation}

\begin{equation}
  R_{core} = 0.057 \Big( \frac{m_{*f}}{30\, \rm M_{\odot}} \Big)^{1/2} \Sigma_{cl}^{-1/2} 
\end{equation}

where the parameter $m_{*f}$ corresponds to the final mass of the star and $\Sigma_{cl}$ denotes the average clump surface density. We calculate the turbulent dispersion and the core size adopting $\Sigma_{cl}$ = 1g cm$^{-2}$ (like \citealp{MT03}) for a 10 to 60~M$_{\odot}$ final star. This adopted surface density is comparable or at least a factor of 2 lower than the observed surface density of MDCs \citepalias{B09}.

Our measurements at the scale of the MDCs (0.1 pc scale), which themselves indicate upper limits for the turbulent support (due to other contributions to the line-width), are smaller than the model predictions of \citet{MT03}. In fact the maximum level of micro turbulence, which is observed, would only support cores with surface density of a factor of 3 lower than $\Sigma_{cl}$ = 1g cm$^{-2}$
(see dashed line in Figure~\ref{fig:caselli}).

\subsubsection{Angular momentum and rotation of the MDCs}
\label{sect:rotation}

The velocity field derived from 
{\hco} maps at the size-scale of the cores ($\sim$0.1 pc) ~indicates that non-thermal, organized motions such as rotation also 
contribute to the 
observed line-widths. According to \citet{BB00} if the cores were dominated by purely isotropic micro-turbulence, the velocity field should be completely random.  From Figure \ref{fig:pdbi_velfield} it is clear that CygX-N3, -N53 and -N63 show indications for organized motions, while  for CygX-N12 and N48 this is less obvious. For the before mentioned MDCs the velocity maps are axisymmetric and the axis of the symmetry is close to be aligned with the mini-filament containing the protostellar fragments. As a first order assumption we interpret this phenomenon as rotation, which is in fact a natural consequence of preserving the initial angular momentum of the collapsing cloud. We note that recent numerical simulations by \citet{Dib10b} do report a variety of dynamical patterns for cores formed in a turbulent, magnetized, and self-gravitating molecular clouds ranging from easily recognizable rotational features such as the ones observed in CygX-N3,-N53, and -N63 to more complex features, such as the ones seen in CygX-N12 and -N48.

\begin{figure}
\includegraphics[width=9cm]{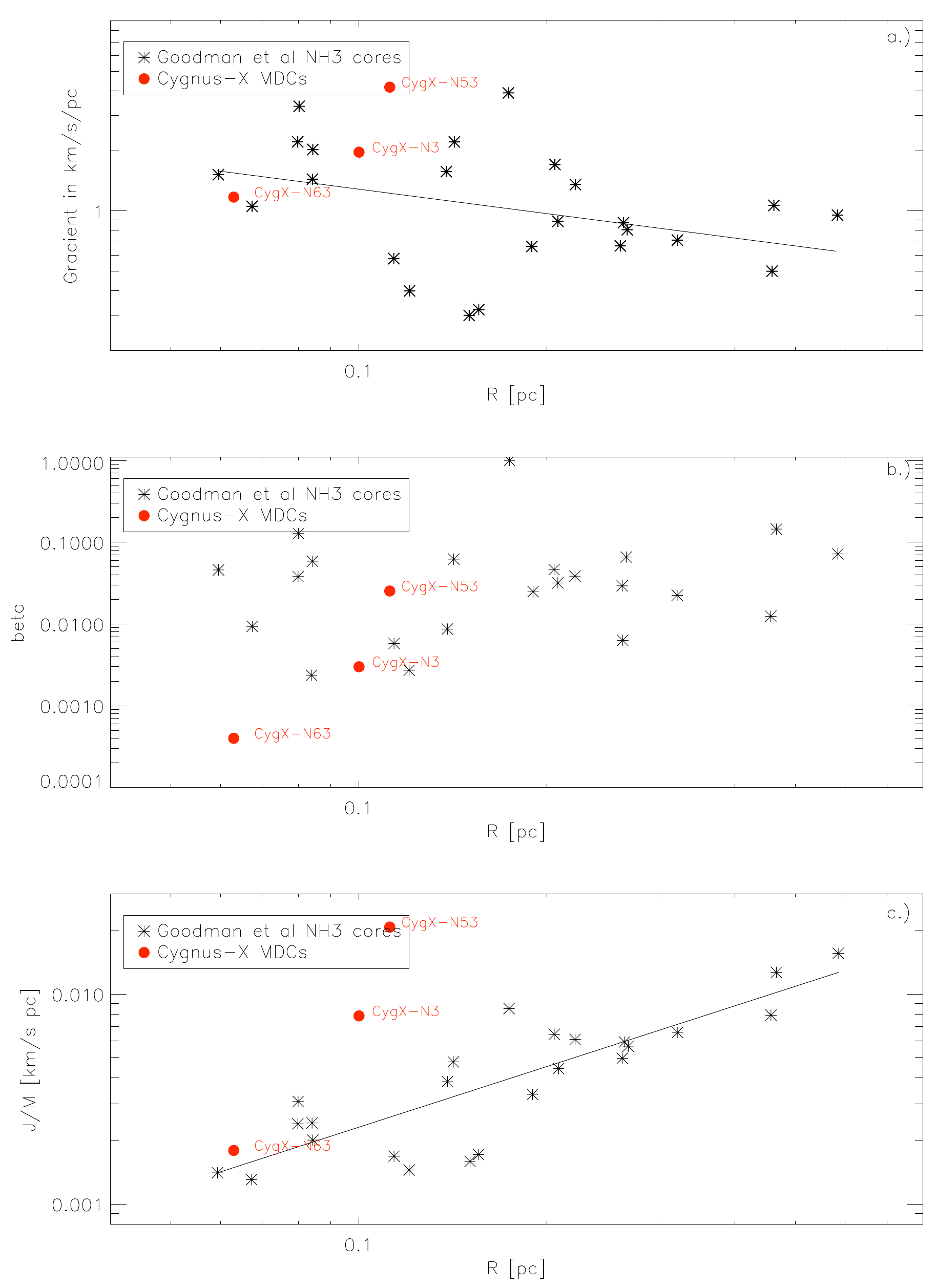}  
\caption{\textsl{a)} Plot of size versus velocity gradient comparing a sample of low-mass cores \citep{Goodman93} with the Cygnus-X MDC sample. Red symbols represent the Cygnus-X MDCs, while black stars show the sample of \citet{Goodman93}. Black line shows the best fit from \citet{Goodman93} with $ \mathscr{G}$ = 10$^{-0.3 \pm 0.2}$R$^{-0.4 \pm 0.2}$.} \textsl{b)} Plot of $\beta$ versus size relation shown for the same sample. \textsl{c)} Plot of the specific angular momentum versus size. Black line shows the best-fit line from \citet{Goodman93} with J/M = 10$^{-0.7 \pm 0.2}$R$^{1.6 \pm 0.2}$0.
\label{fig:goodman}
\end{figure}

To further discuss the observed angular momentum we rely on the systematically derived velocity gradients. Our values of 1.2-4.2 km s$^{-1}$pc$^{-1}$ are consistent with those reported in the literature of both low and high-mass objects.
We also calculated the $\beta$ parameter (see Table \ref{table:vfit}), which shows the ratio between the rotational and potential energy ($E_{rot}/E_{pot}$). Values between 2~$\times$~10$^{-3}$ and 1.4 were reported in a sample of low-mass cores by \citet{Goodman93}. Our $\beta$ values are low (4~$\times$~10$^{-4}$ - 2.5~$\times$~10$^{-2}$) (see Figure \ref{fig:goodman} {\sl b}) indicating that gravity significantly dominates over the rotational energy, especially in CygX-N3 and -N63. The most extreme example is the most compact core, CygX-N63 ($\beta\sim$~4$\times$10$^{-4}$), which hosts a single or a close binary massive protostar. A priori one could expect this core to behave similarly to low-mass cores, where rotating envelopes around Class 0 objects are reported. Its low velocity gradient corresponds to a very low specific angular momentum (Figure \ref{fig:goodman} {\sl c}) fitting on the trend found for low mass cores by \citet{Goodman93}. Nevertheless the other MDCs have higher specific angular momentum. We suggest that the origin of this diversity is related to the small-scale dynamics, and that the origin of the velocity field may be different for CygX-N63 and the other cores (see in Section \ref{sect:flows}).

In contrast, towards CygX-N12 and -N48 no clear rotation-like gradient are evident in the large-scale ($\sim$0.1 pc) velocity field maps with complex velocity patterns (see left panels of Figure~\ref{fig:pdbi_velfield}). This suggests that these 2 MDCs do not have a large resulting rotation in projection on the sky. Their main rotation axis may be close to the line of sight or these MDCs have a truly smaller angular momentum than the others. In the high angular-resolution maps, indeed some velocity gradients are observed at small scales with similar velocity drifts like in the other MDCs, indicating that the angular momentum may be redistributed at smaller scales. 


The position-velocity cuts along the derived velocity gradient (Figure \ref{fig:pdbi_velfield} right panel) show a centrally peaked elongated pattern for CygX-N3, -N53, -N63, which may correspond to large rotational toroids up to 40{\mbox{$^{\prime\prime}$}}(68.000 AU), 45{\mbox{$^{\prime\prime}$}} (76.500 AU) and 35{\mbox{$^{\prime\prime}$}} (60.000 AU) respectively, which may serve as a common and coherent reservoir of material for forming protostars. 
Such rotating structures with similar size-scales were observed towards low-mass Class 0/Class I objects (e.g. \citealp{Goodman93}; \citealp{Caselli02}) and towards massive star-forming cores as well (e.g. \citealp{Beltran04}; \citealp{Cesaroni05}; \citealp{F09}), but the nature and origin of these rotating toroids are not well established yet, especially in cores hosting the earliest stages of massive star-formation.

\subsubsection{Importance of global infall}
\label{sect:infall}
The Cygnus-X MDCs have larger masses for the same sizes and slightly larger velocity dispersions compared to the Orion cores. The virial-masses are determined in \citetalias{B09} and comparing the virial-mass to mass ratios (1.16, 0.87, 1.06, 0.75, 0.51 for CygX-N3,-N12,-N48,-N53,-N63 respectively), we find that most of the Cygnus-X MDCs are gravitationally unstable and are therefore expected to be in global collapse. Four out of the five MDCs indeed display a typical blue-shifted asymmetric \hcop ~line profiles suggesting infall motions. As showed in Sect.~\ref{subsec:simline}, the line profiles of both \hcop~and \hco ~can be reproduced with radiative transfer modeling with typical infall velocities between 0.1-0.6~km~s$^{-1}$. This indicates that in addition to the rotation-like organized motions discussed above, the global velocity dispersions in the MDCs have also a contribution from infall motions.
The strongest infall is seen towards the most massive core CygX-N48 ($\sim$0.6~km~s$^{-1}$) and is similarly large towards CygX-N53, i.e. the 2 MDCs located in the DR21 filament studied by \citet{Schneider_prep}. CygX-N3 shows a red-shifted asymmetric profile, which is indicative of expanding motions. However the stronger peak observed at $\sim$16.5~km~s$^{-1}$ is coinciding with a second velocity component identified only from the high-resolution spectra map in Section \ref{sec:spec_decomp} (see also Table \ref{table:sp_fit}) and therefore we argue that this peak has contribution from an additional velocity component, rather than expansion. {\hcop} emission towards CygX-N63 is significantly weaker than in the other cores and radiative transfer models show that depletion is the strongest in this core. Nevertheless the {\hcop} line shows a slight, blue-shifted shoulder, which also points to infall-motions. Our best radiative transfer model gives an optically thick line for \hcop~with an infall speed of $\sim$0.1~km~s$^{-1}$. The high depletion, as shown by \cite{Rawlings01} based on numerical simulations, might explain this rather weak indication of infall. 



\subsubsection{Coherent flows seen at high spatial resolution}
\label{sect:flows}

   \begin{figure*}[!]
   \centering
   \includegraphics[width=\linewidth]{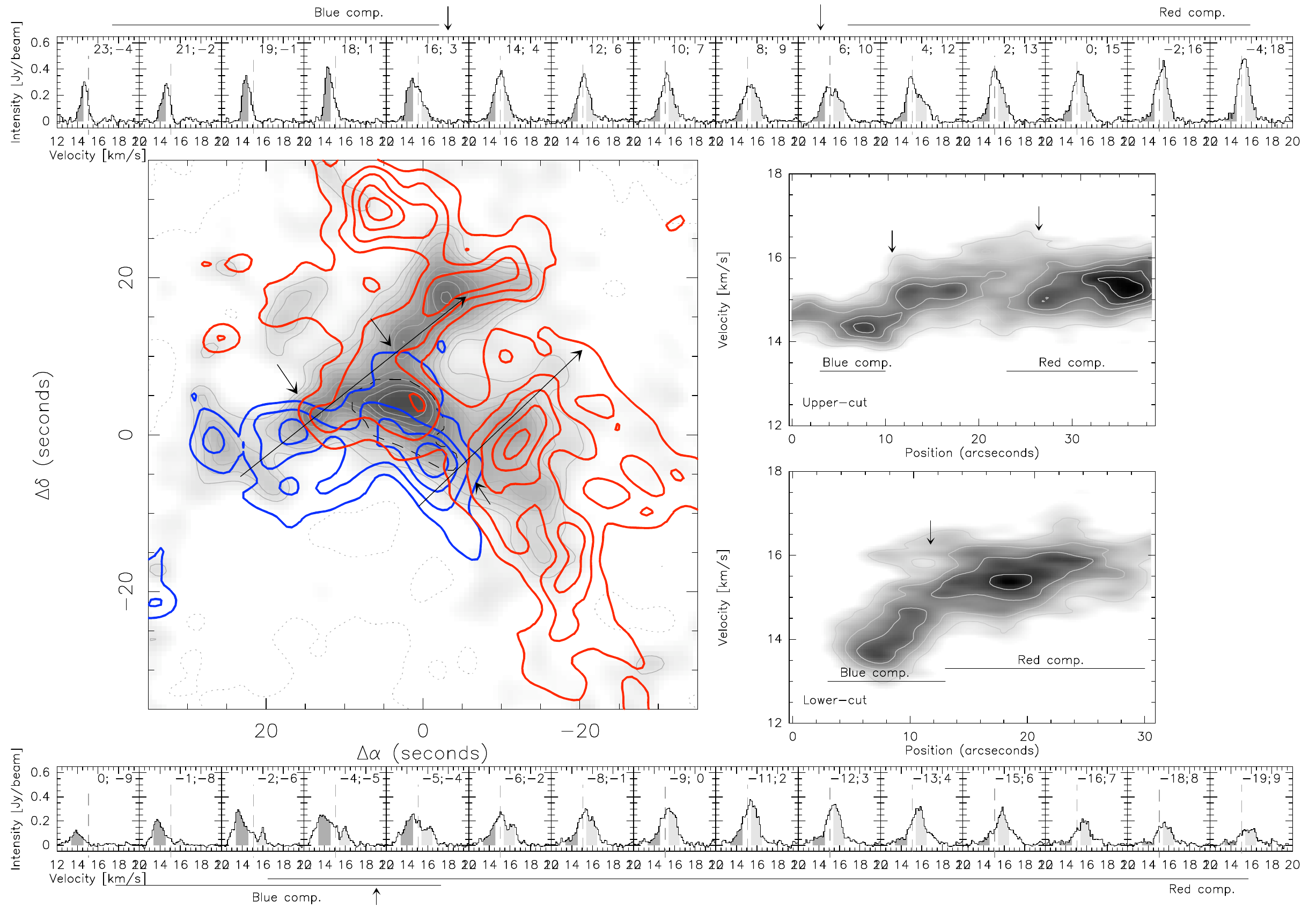}
  \caption{ 
   Map of {\hco} emission obtained with the PdBI and zero-spacings. The grey scale shows integrated intensity around the v$_{lsr}$, between 14.5-15.5~{\kms}. Blue contours correspond to integrated intensity between 13.5-14.5~{\kms} and red contours show integrated intensity between 15.5-16.5~{\kms}. Contours go from 5$\times$rms noise and increase in steps of 3$\times$rms noise. The black dashed contour shows the 30\% of the peak intensity of the 3mm continuum map. Fully sampled spectra are extracted in two cuts following two filamentary structures, where the position corresponding to each spectra are shown in the upper right corner. The integration ranges from the blue shifted component are in dark grey and white grey shows the red component. The v$_{lsr}$ is shown by a dashed grey line. The right panels show position-velocity diagrams for both cuts respectively. }

         \label{fig:N3}
   \end{figure*}
       \begin{figure*}
   \centering
   \includegraphics[width=\linewidth]{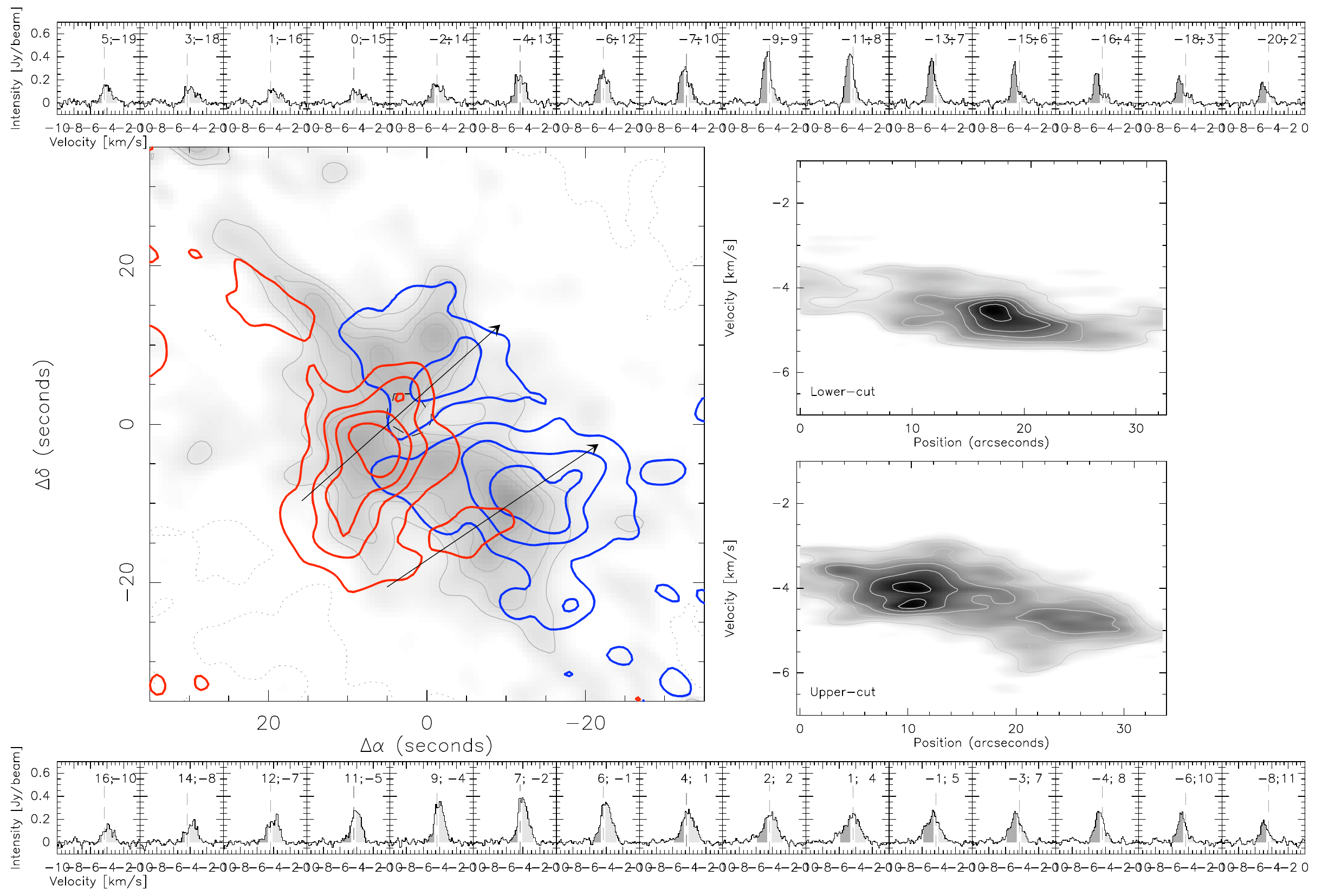}
     \caption{CygX-N63 presented similarly as Figure \ref{fig:N3}. Integration ranges for the grey scale is -4.0 to -4.7~{\kms}, red contours are integrated between -2.7 to -4.0~{\kms}, blue contours are integrated between -4.7 to -5.7~{\kms}.
Contour levels go from 5$\times$rms noise by steps of 3$\times$rms noise.}
         \label{fig:N63}
   \end{figure*}



 The high spatial resolution PdBI velocity maps in Figure~\ref{fig:pdbi_velfield} (second column of panels) display velocity fields which are similar to what is seen in the purely single dish data (first column of panels in Figure~\ref{fig:pdbi_velfield}) in particular the global rotation-like gradients discussed above in Sect.~\ref{sect:rotation}. On the other hand these global motions are found to
have a more complex morphology at small scales with stronger local gradients,  
and more importantly with the {\hco} line profiles, which are not single gaussians and clearly split into several individual velocity components.

These individual components may correspond to coherent features in velocity. To trace the bulk emission from the mass reservoir we integrate around the rest velocity of each MDC and then separate the emission from the other components by integrating in the blue and the red side of the velocity range. This representation allows to study the distribution of gas associated with the bulk emission and traces the morphology of the potential coherent velocity features of these individual components.


Figure \ref{fig:N3} shows CygX-N3, where the bulk emission at a rest velocity of 15~{\kms} (grey scale) is coinciding with the main continuum peaks (dashed black contours) and shows an elongated, filamentary structure centered on it as well as hints for sub-filaments in the north-east and south-west edge, perpendicular to it. The blue and red-shifted spectral components show a bipolar structure indicative of velocity gradients or shears along these sub-filaments. In order to further investigate the velocity structure of these sub-filaments, we show position-velocity cuts along them. The upper cut shows a global smooth shift in velocity, but with hints for two sharp transitions or jumps at offset 16{\mbox{$^{\prime\prime}$}}, 3{\mbox{$^{\prime\prime}$}} and 6{\mbox{$^{\prime\prime}$}}, 10{\mbox{$^{\prime\prime}$}}. 
The lower cut shows also a strong red component with a smooth shift in velocity and at $\sim$15~{\kms} a sharp change in velocity between offsets $-4${\mbox{$^{\prime\prime}$}}, $-5${\mbox{$^{\prime\prime}$}} and $-5${\mbox{$^{\prime\prime}$}},$-4${\mbox{$^{\prime\prime}$}}. Such velocity shears may be similar to the velocity jump revealed in NGC2264-C by \citet{Peretto06}.


We show spectra across the cut on the sub-filaments. In the upper one, until offset 16{\mbox{$^{\prime\prime}$}}, 3{\mbox{$^{\prime\prime}$}} the blue component dominates and seems to stay at the same velocity, then the spectra suddenly get broader peaking at the rest velocity. After 8{\mbox{$^{\prime\prime}$}}, 9{\mbox{$^{\prime\prime}$}} the red component appears and becomes more dominant at 0{\mbox{$^{\prime\prime}$}}, 15{\mbox{$^{\prime\prime}$}}, even if the separation between the main and the red component is less clear. The position of the red component is also compatible with a fixed velocity along the sub-filament. The separation between the blue and the bulk emission is $\sim$0.75~{\kms} and $\sim$0.55~{\kms} between the red and the bulk emission. The lower cut shows both the blue and red spectral components from -1{\mbox{$^{\prime\prime}$}},-8{\mbox{$^{\prime\prime}$}} to -5{\mbox{$^{\prime\prime}$}},4 and a velocity jump from the blue to the main component is seen at -6{\mbox{$^{\prime\prime}$}},-2{\mbox{$^{\prime\prime}$}} offset. 
It is apparent, that the velocity separations are much larger in the lower, than in the upper cut up to a total velocity difference of $\sim$2.5~{\kms} between the red and the blue component in the lower cut. Towards the continuum peak the spectral components merge into a single, broader line one suggesting \textsl{physical interaction} between the two components. Summarizing, velocity shifts along organized gas structures are seen but a more careful analysis of the line profiles suggests rather local shears than smooth velocity gradients between the individual velocity components. Interestingly these velocity jumps correspond to weak continuum sources (see Figure~\ref{fig:shear}), which may be at an early stage of their formation.

In contrast CygX-N63 is shown in Figure~\ref{fig:N63}, where the bulk emission at a rest velocity of $-4.3${\kms} is similarly centered close to the continuum peaks and shows hints of a north-west to south-east organized elongated structure. The red and blue shifted emission shows a bipolar structure perpendicular to the main filament. The position-velocity cuts along these structures show though a much smoother shift in velocity or a weaker velocity shear. Cuts of spectra also show that the line-profiles are less complex as for CygX-N3 and could be compatible with a single line component, which shows a shift in velocity of $\pm$ 0.6 km~s$^{-1}$.  
Altogether, CygX-N63 seems to be more ordered, than the example above. 

Following this systematic way, we find similar trends in the other fields, but due to a larger complexity of individual velocity components they are less clear for CygX-N12, -N48 and -N53 respectively (see Figures \ref{fig:N12}, \ref{fig:N48}, \ref{fig:N53} in Appendix~\ref{app:filaments}). Cuts of spectra and position-velocity maps are presented therefore centered on the continuum peaks and perpendicular to each other in order to give a glimpse of the spatial distribution of the velocity components. Towards CygX-N12 the main component is dominated by spherical extended emission with a large number of individual components and no clear morphology can be found for the blue and red components. As the spectra show, the two components are both visible over a large extent of the map with a velocity of $\sim$14.3~km~s$^{-1}$ and $\sim$15.7~km~s$^{-1}$. These lines show either a shift in velocity over $\sim$$\pm$1 km~s$^{-1}$ with respect to the systemic velocity of 15.2 km~s$^{-1}$ or these may be additional components. 

Similarly, towards CygX-N48 several line components are detected within a velocity range of 6~km~s$^{-1}$ corresponding to the most complex dynamics in the sample. Two main components can still be separated, one around -5.5~km~s$^{-1}$ and one around -3.8~km~s$^{-1}$, both showing shifts in velocity and additional components appear as well indicating complex kinematics.

In CygX-N53 the bulk emission at a rest velocity of $-4.2$~{\kms} shows a north-south elongation, however the determination of the rest velocity may be biased due to the surrounding large-scale filament, which may have a redder bulk emission. Bipolar blue and red-shifted emission is seen close to the continuum peak, but the tendency is less clear as for CygX-N3.

We conclude that, except perhaps CygX-N63, all MDCs of the sample display a mixture of several velocity components separated by up to several km~s$^{-1}$, 
and show drifts and discontinuities in velocity. The spatial distributions of these components suggest that they are associated with the MDCs and indicate that they do not show up as only individual fragments which could be associated with the dust continuum fragments discussed in \citetalias{B09}. Instead, they appear as filamentary structures, showing systematic velocity differences leading to local shears. For three of the five MDCs, these velocity features averaged at the scale of the MDC result in a rotation-like motion, and for four of them, there is also an indication of global infall. This is like if the global motions (rotation and infall) split into small scale individual streams or flows when observed at high spatial resolution (see Sect.~\ref{sect:converging-flows} for a further discussion of the nature of these flows).



\subsection{Importance of the turbulent support at small-scales}
\label{sect:scale-turbulence}


\begin{figure}
\includegraphics[width=9cm]{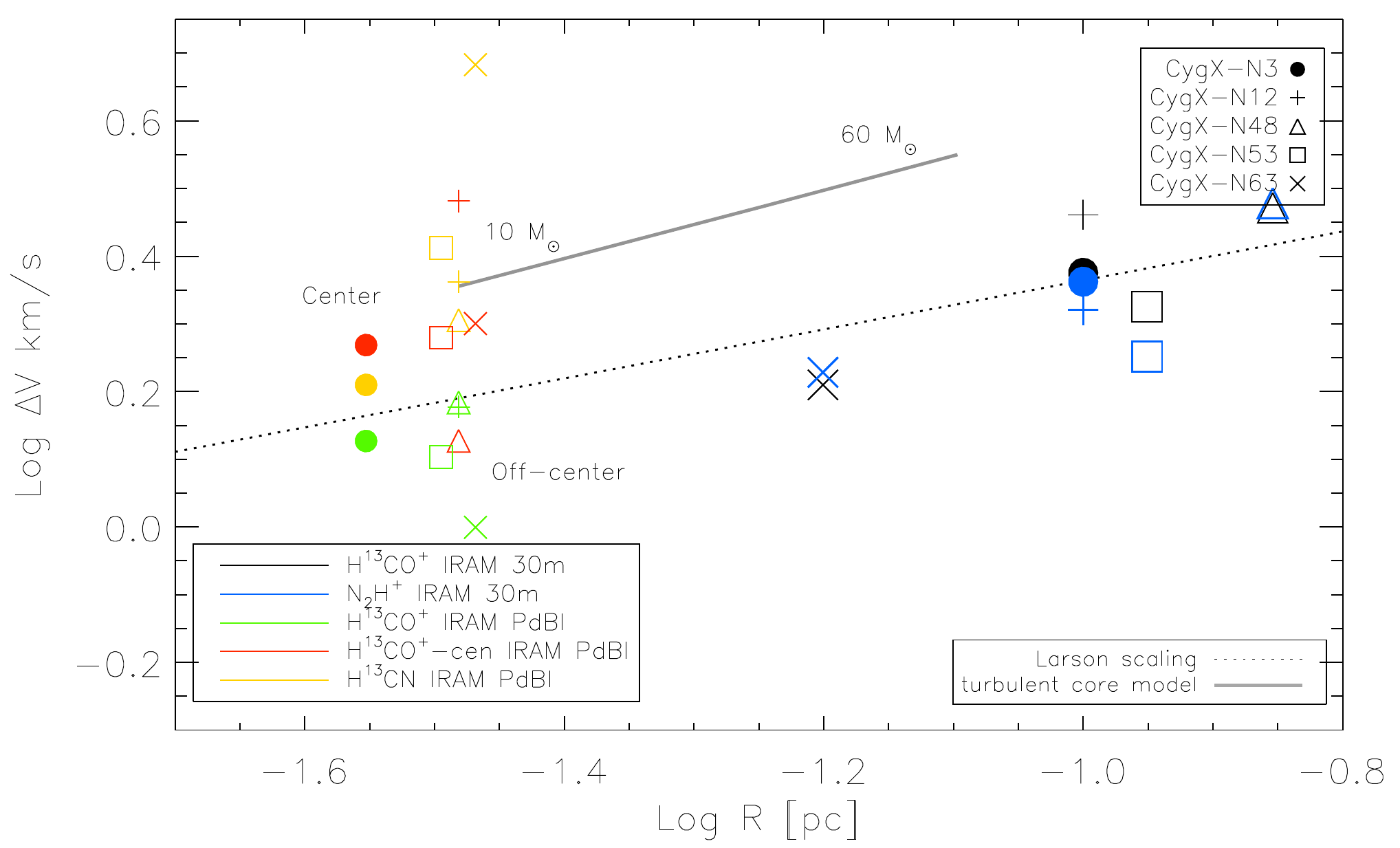}  
\caption{We plot line-widths of the MDCs taken from single-dish and high angular-resolution data. The corresponding sizes are taken from \citet{M07} for the single-dish data, while for the high angular-resolution data sizes show the synthesized beam of the PdBI. The Larson-scaling is indicated by a dotted line. Grey line indicates the fiducial model of \citet{MT03} with a canonical surface density of $\sim$~1g~cm$^{-2}$ and a final stellar mass of 10-60 M$_{\odot}$. Different colors indicate the different tracers observed with only single-dish or with the interferometer.}
\label{fig:linewidth}
\end{figure}
\label{sect:micro-turbulence}

A detailed understanding of global organized motions both at low and high angular-resolution is required to recognize the role of support by micro-turbulence at small-scales. In Sect. \ref{sect:virial} it was pointed out that turbulence within the MDCs at global scales ($\sim$0.1 pc) tends to be smaller than required by the turbulent core model. Here we can further investigate this trend at smaller scales, where
the level of micro-turbulence is traced by the line-dispersion of the individual spectral components recognized in the previous section. Figure \ref{fig:linewidth} shows these results, where the average velocity dispersions of the individual components decrease from $\sim$1~km~s$^{-1}$ at $\sim$0.1 pc-scale to $\sim$0.6~km~s$^{-1}$ at $\sim$0.03 pc-scale. 


Velocity dispersions towards the central regions (3mm continuum peaks, indicated by contours in Figures \ref{fig:N3}-\ref{fig:N63}) were extracted as well. Interestingly, the average velocity dispersion is larger $\sim$ 0.86~{\kms} towards the center. 

These velocity dispersions are close to the predictions of the \citet{MT03} model. However the line-widths in the central regions might not be representative of the bulk of the central dense material for two reasons. For {\hco}, the high level of depletion makes that only the outskirts are traced, which could have a larger velocity dispersion and {\hcn} may trace only higher temperature gas influenced by stellar feedbacks (outflow shocks). Alternatively, local turbulent support could be compatible with the \citet{MT03} model at these small scales in the central regions. 

On the other hand, we point out that the turbulent line-dispersions of the individual velocity components, which are generally found off-center, represent more the pre-collapse gas and are smaller than the model values. They are found to fall on the Larson-scaling. 


\subsection{Are massive-star forming cores in equilibrium?}

The fundamental question is if MDCs are in a quasi-static equilibrium or are governed by dynamic processes. In the scenario proposed by \citet{MT03} massive cores should evolve in a quasi-static equilibrium, where a high effective turbulent pressure balances gravity. Once gravity becomes stronger than the effective pressure within the core, it collapses monolithically to form a single or a binary object. During the quasi-static evolutionary stages a homogenous medium should develop a central concentration of matter while it is stabilized by its internal micro-turbulence. Comparing our single-dish observations of optically thin emission (\hco), we confirm a centrally peaked distribution of emission in a homogenous medium. 



In the sections above we systematically searched for such a support, that can be micro-turbulence, magnetic fields and rotation. Neglecting magnetic fields as it is not known, we found that: \textit{a)} line-dispersions on a 0.1 pc-scale tend to show that micro-turbulence alone is not sufficient to stabilize MDCs using the turbulent-core model; \textit{b)} line-dispersions at small-scales indicate that the initial conditions may be even less turbulent; \textit{c)} rotation like ordered bulk-motions are not sufficient either to stabilize MDCs (Section~\ref{sect:rotation}). Altogether this gives a coherent view of MDCs being out of equilibrium state, as indicated also by global infall motions and supersonic flows.

\subsection{Dynamical processes in the MDCs}
\label{sect:dynamics}

\subsubsection{Evidences for converging flows}
\label{sect:converging-flows}

Several indications are found for dynamical processes acting at small-scales. The individual velocity components seen in the high resolution \hco ~lines are best understood as individual flows (Sect.~\ref{sect:flows}). Since the MDCs are in probable global collapse, these flows are expected to stream inward. They should then ultimately interact in the central regions to build up high density cores at the stagnation points, creating the seeds for the protostellar objects. 
We estimated the crossing times of these convergent flows by dividing the characteristic size of the core (on average $R_{core}$=0.1 pc) with the characteristic velocity shift measured by the largest separation of the individual spectral components ($\delta {\rm v}$) discussed in Section \ref{sect:flows}: $\tau_{cross} = 9.8\ 10^5\  \frac{R_{core}}{[pc]}\  \frac{\delta {\rm v}^{-1}}{[km~s^{-1}]}$.
In Table \ref{tab:sd_params2} we list these parameters and indicate the free-fall times ($\tau_{ff}$) from \citetalias{B09}. 
(For CygX-N63 the crossing time estimate is based on the velocity dispersion of the spectra, since the individual velocity components are less clear.) To calculate the crossing times we adopted a correction for a possible inclination effect assuming an average angle to the line-of-sight of $\sim$57.3$^\circ$ (which is the average angle assuming random distribution of orientation angles). We give also the range of $\tau_{cross}$ for variations of the angle by $\sim$20$^\circ$ to give a hint of the effect of the inclination uncertainty.

We find that all crossing times are comparable to the free-fall times, which indicates that these flows are the main driver for the evolution of MDCs, and can clearly play the major role in building up material of which massive protostars form.

The origin of these small scale flows can just be a natural consequence of the large scale turbulent nature of the parent clumps, and clouds. 
At pc-scale \citet{Schneider_prep} showed that massive structures, like the DR21 filament itself, can be formed by dynamical processes and converging flows, pointing out that at sub-parsec scales dynamical processes are important as well, leading to a complex view of a hierarchical dynamics from pc-scale down to the protostar scale.   

\subsubsection{Importance of competitive accretion}
\label{sect:comp}

In a broad view of the competitive accretion scenario \citep{BB06}, the above discussed high level of dynamics 
necessary leads to competition for mass between the protostars.
Even if such small-scale flows converging to the central parts of MDCs were revealed in the previous sections, it is not clear
 whether these flows continue to build up the mass reservoirs of the protostars. The time-scale to have a new prestellar/protostellar seed formed by convergent flows
 is given by the typical velocity shears observed in the most dynamical regions of the MDCs at a scale of the order of the separation between the protostars ($\sim$5000 AU in \citetalias{B09}). The largest velocity shifts observed over this projected distance are of the order of 3~km~s$^{-1}$  which lead to a typical time-scale of $\sim10^4\,$yr , a factor of a few smaller than the crossing time at the scale of the MDCs. This means that during a crossing time of the MDCs, several protostars could form by the convergence of several small scale flows. This time-scale of $\sim10^4\,$yr is also only slightly larger than the typical free-fall time derived for the continuum fragments reported in Paper I (on average of 4000~yr). If the protostellar cores collapse in a few free-fall times like in low-mass star-forming regions, the formation time-scale could be of the same order of magnitude than the collapsing time, and the mass input from flows during collapse can be significant. This therefore points to the possibility of a significant competitive accretion in the MDCs of Cygnus X.

 \begin{table}
\centering                          
\caption{Parameters corresponding to the identified flows. The velocity difference indicates the relative velocity of the flows on the global scale of the MDCs, which was used to calculate the crossing time-scales.}
\begin{tabular}{c c c c c c }        
\hline\hline                 
Source & $\delta \rm{v}$  &    $\tau_{cross}$ &    $\tau_{ff}$ \\    
            & [km~s$^{-1}$]                   &  $\times 10^4$  [yr] & $\times 10^4$   [yr] \\ 
\hline                        
   CygX-N3   &  2.5       &   2.4 - 3.8          & 6.0  \\     
   CygX-N12 &  3 & 2.0 - 3.2    & 5.9  \\
   CygX-N48 &  3.5    &    1.7 - 2.7           & 6.1 \\
   CygX-N53 &  1.1    &    5.4 - 8.7           & 7.6 \\
   CygX-N63 &  0.5   &     11.8 - 19.1         & 3.4 \\
\hline                                  
\end{tabular}
\label{tab:sd_params2}      
\end{table}

\subsubsection{Origin of the diversity between the 5 MDCs}
\label{sect:diversity}

\begin{figure}
\includegraphics[width=9cm]{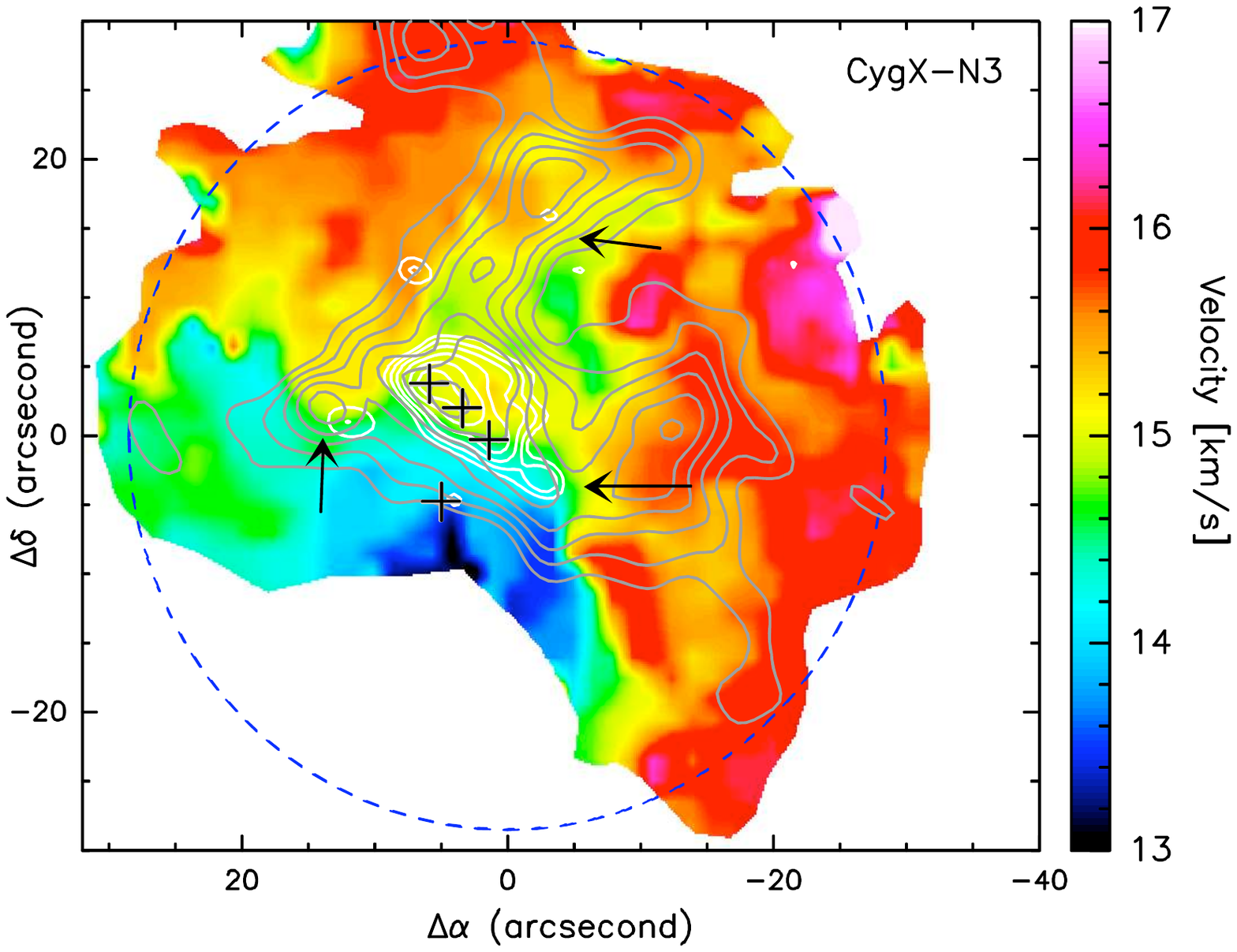}  
\caption{The velocity field map for CygX-N3 is shown with overlays of integrated intensity (contour levels are from 20 to 50$\sigma$ by 5$\sigma$. White contours show the 3mm continuum emission. Arrows indicate the shears associated with weak continuum emission.}
\label{fig:shear}
\end{figure}

From the fragmentation properties \citepalias{B09} and a kinematical point of view (this work) our sample of MDCs is diverse. This diversity can reflect either different evolutionary stages or different initial conditions for the MDCs. 

In CygX-N12, -N53, -N63 the gravitational potential is dominated by a few, but massive fragments in the center, corresponding to more than 30\% of the total mass. 
For these cases, the self-gravity of the MDC is dominated by these few fragments which could then collect most of the remaining flowing mass if the convergent flows are still active.
In CygX-N3 and -N48, on the other hand, the mass in the already formed fragments is a smaller fraction of the total mass, therefore
the local gravitational field is not dominated by the observed fragments, but more by the self-gravity of the mass reservoir of the MDC. These MDCs are either intrinsically different from CygX-N12, -N53, and -N63 or they are younger and will evolve to similar MDCs dominated by a few massive fragments.
The high resolution velocity field in \hco ~should indicate whether or not the fragments are still collecting mass from their parent MDCs, and therefore may help to discriminate the two possibilities. 

In CygX-N3, the 3 most massive fragments are located on a significant shear in velocity which indicate that they can still gain mass from the MDC reservoir. At the same time three shears associated with significant \hco ~emission peaks are also visible and could therefore correspond to newly forming protostars (all three correspond also to weak continuum peaks as indicated by arrows in Figure~\ref{fig:shear}). The relative intensity of the shears and associated \hco ~peaks suggest that a large number of protostellar seeds collect mass from the MDC mass-reservoir leading to the formation of a whole cluster of protostars.

In CygX-N12, several velocity components imply a complex kinematics. 
Since 70~\% of the total mass of the MDC is still available, there is a potential for CygX-N12 to further evolve towards a state similar to CygX-N53 with two main massive fragments in the central regions.

In CygX-N48, the observed shears are intense and are located in the central regions close to, but not necessary exactly associated with the main central fragments. This is like if the first fragments are not yet clearly decoupled in the central, very active region. This is therefore a good candidate MDC to be in an early phase of evolution, perhaps leading later on to a MDC similar as CygX-N53.

In CygX-N53, most of the total mass is contained in two main, central fragments. No clear shears are observed in the central regions. Some indications of shears are seen in the southwest outskirts of the MDCs, though they are less convincingly associated with the MDC (this MDC is located in the north part of the massive DR21 filament). This MDC is therefore probably evolved with no significant flowing gas present in the central regions. 

Finally, CygX-N63 is a special case with only one central peak of emission in the continuum such as if it were not sub-fragmented, and with a global collapse towards this single object. The kinematics shows a rotation-like gradient which may correspond to a large scale shear, revealing that the object still collects mass. 
Interestingly, besides this smooth shift or a weak shear seen towards the main continuum emission, a small scale shear is observed in the south-west edge of the field coinciding with weak 3mm continuum emission, that could there lead to the formation of a new protostar.

We have therefore three MDCs (CygX-N12, N53, N48) which could correspond to three cores of similar nature seen at three different evolutionary stages: CygX-N48 corresponding to a younger and CygX-N53 to a more evolved phase. CygX-N3 seems to be forming a cluster of at least 6 protostars with the total mass shared between them without a dominating massive one. Finally CygX-N63 can be seen as a special case: a single collapsing object not displaying strong dynamics in a collapsing, large scale envelope. Alternatively, it can be seen as the extreme case of the massive star forming MDCs discussed here (CygX-N12, N53, and N48) with only one protostar dominating the collapse. We therefore propose that the initial conditions may explain the differences between CygX-N3, CygX-N12/N48/N53, and CygX-N63 while the evolutionary state would be the explanation for the observed differences between CygX-N12, N48, and N53.

\section{Summary \& Conclusions}

We presented a detailed study of high-density gas tracers towards 5 MDCs in Cygnus-X focusing on their kinematic properties on global ($\sim$0.1 pc) and small scales ($\sim$0.03 pc). Our main findings are:


   \begin{enumerate}
     \item  On $\sim$0.1 pc-scales dense gas traced by \hco~and \hcop~shows infall signature in CygX-N12, -N48, -N53 and -N63. The velocity fields indicate coherent, organized bulk motions towards CygX-N3, -N53, -N63 with a rotation-like pattern, while towards CygX-N12 and -N48 no global velocity gradients are dominant.
          
     \item Simultaneous radiative transfer modeling of {\hco} and {\hcop} shows that due to depletion effects the emission is tracing more the bulk of material around the protostars ($\sim$ 10$^5$-10$^6$ cm$^{-3}$). Depletion is found to be most significant in CygX-N63.

     \item At high angular-resolution we find a better correspondence of {\hcn} with continuum peaks than {\hco}. 
     
     \item High angular-resolution maps with the PdBI in {\hco} reveal significant substructure at small-scales and in all MDCs several line components are found, which are disentangled into small-scale coherent flows with intrinsic velocity gradients and shears.
           
     \item Analysis of line-dispersion of the high-angular resolution {\hco} spectra question if the level of micro-turbulence at small-scales is enough to provide sufficient support against gravitational collapse. 
     
     \item At high angular-resolution the larger scale global coherent motions are resolved into individual coherent flows with filamentary structures, showing velocity shifts and more importantly shears/discontinuity in velocity towards the continuum peaks. The relative difference in velocity position of the flows give dynamical time-scales of the order of the free-fall time-scale for CygX-N3, -N12,  -N48, -N53. Only for CygX-N63 we find an order of magnitude larger dynamical time-scales.
     
     \item Small-scale velocity shears may indicate $\sim$10$^4$~yr dynamical time-scales for new protostellar seeds built up by small-scale converging flows.

     \item Comparing the fragmentation and the kinematic properties of our sample, they point to different evolutionary stages for CygX-N48, -N12 and -N53, while different initial conditions may explain the differences towards CygX-N3 and -N63.

   \end{enumerate}

\begin{acknowledgements}
The authors would like to thank the anonymous referee for its critical review of the article and for the useful comments and suggestions which have lead to a signiÞcant improvement of the text. The authors also wish to thank to V. Ossenkopf for providing the latest version of {\sl Simline}. T. Csengeri acknowledges support from the FP6 Marie-Curie Research Training Network ÕConstellation: the origin of stellar massesÕ (MRTN-CT-2006-035890). Part of this work was supported by the ANR (Agence Nationale pour la Recherche) project ÒPROBeSÓ, number ANR-08-BLAN-0241
(LA). 

\end{acknowledgements}

\begin{appendix} 
\section{Details on radiative transfer modeling}
\label{app:simline}

\begin{figure}[!htpb]
   \centering
   \includegraphics[width=8cm]{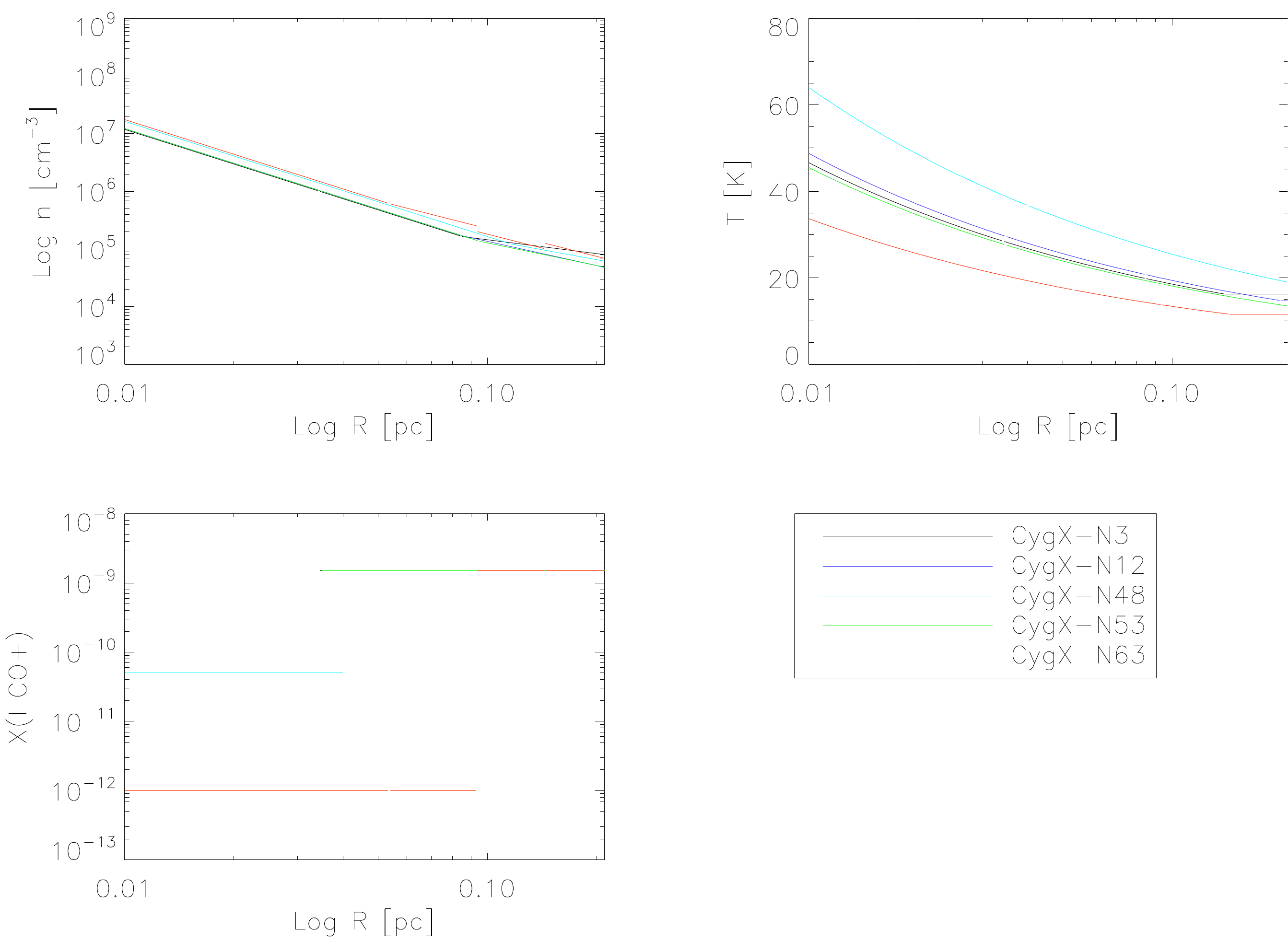}
  \caption{Profiles of density, temperature and {\hcop} abundance. Colors indicate the different MDCs.}
         \label{fig:profiles}
   \end{figure}

To model the molecular line emission of MDCs, we consider a core embedded in its parent molecular clump with smooth transition of the physical parameters. 
For each core we used a simple model with the same assumptions: the mass and the radius is taken from continuum data of \citet{M07} and the corresponding 90\% mass and radius were used to constrain the density profile with a power-law exponent of $\alpha_1$$\sim$-2 (similarly as done in \citealp{Schneider_prep}). NH$_3$ measurements by \citet{Wienen} give constrains on the temperature and the size of the NH$_3$ cores which correspond to the measured temperature. Using these parameters as mass-averaged temperature within a size slightly larger than the size of the MDCs ($\sim$~0.2 pc) the temperature profile is then constrained. The cores are embedded in a larger structure, a molecular clump ($\sim$0.3-0.6 pc), whose parameters were taken from \citet{M07}. We use the size of the clump for the outer radius and constrain the density profile with a shallower exponent ($\alpha_2 \sim$ -1.0 to -1.6). First estimates for turbulent widths are taken from the gaussian line-fitting of the {\hco} lines. Figure~\ref{fig:profiles} shows the density, temperature and abundance profiles used for the models.

We use an abundance of $X($\hcop$)$$\sim$10$^{-9}$ and introduce depletion at a given density. We both vary the density at which depletion occurs (10$^5$-10$^6$ cm$^{-3}$) and the level of depletion (0-1000). Summarizing, we keep the depletion density and the depletion factor as a free parameter and constrain the best fit using all the spectra extracted along the radius (see the results in Figures \ref{fig:sim-n3} -- \ref{fig:sim-n63}).  Thus by fitting a radial profile, we restrict the degeneracy of the model. The explored parameter-space is the same for each MDC and the best fit is determined by comparing the reduced $\chi^2$ values of the fit. From this modeling we obtain the opacity ($\tau$) and the density, where depletion occurs.

\begin{figure}[!htpb]
   \centering
   \includegraphics[width=8cm]{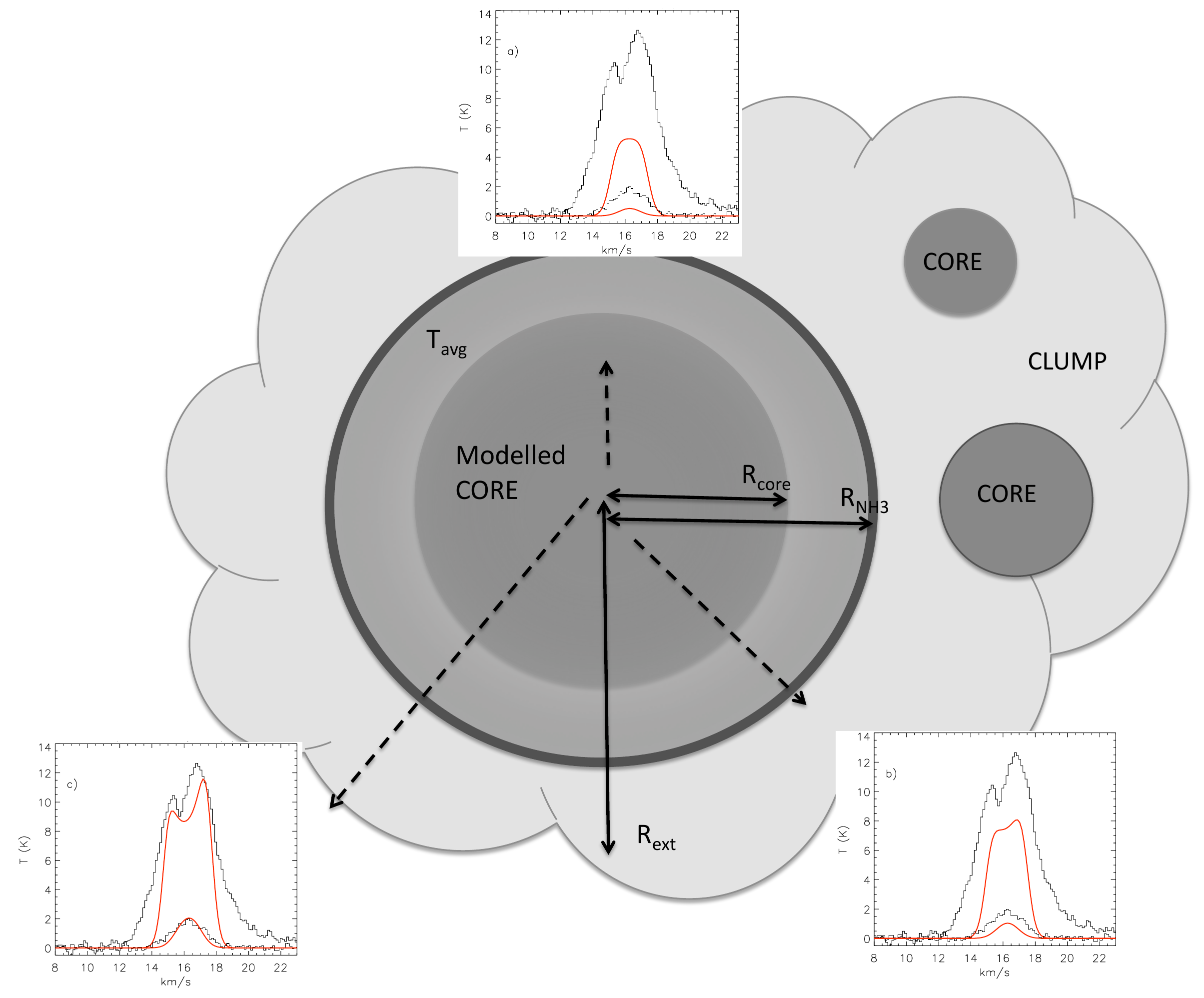}
  \caption{The scheme of the physical model. Model spectra are produced for CygX-N3 in 3 variations, {\sl a)} modeling only the core, {\sl b)} the core corresponding to the NH$_3$ measurements, {\sl c)} modeling the core and its embedding clump.}
         \label{fig:sim-model}
   \end{figure}

\begin{figure*}[ht]
   \centering
   \includegraphics[width=18cm]{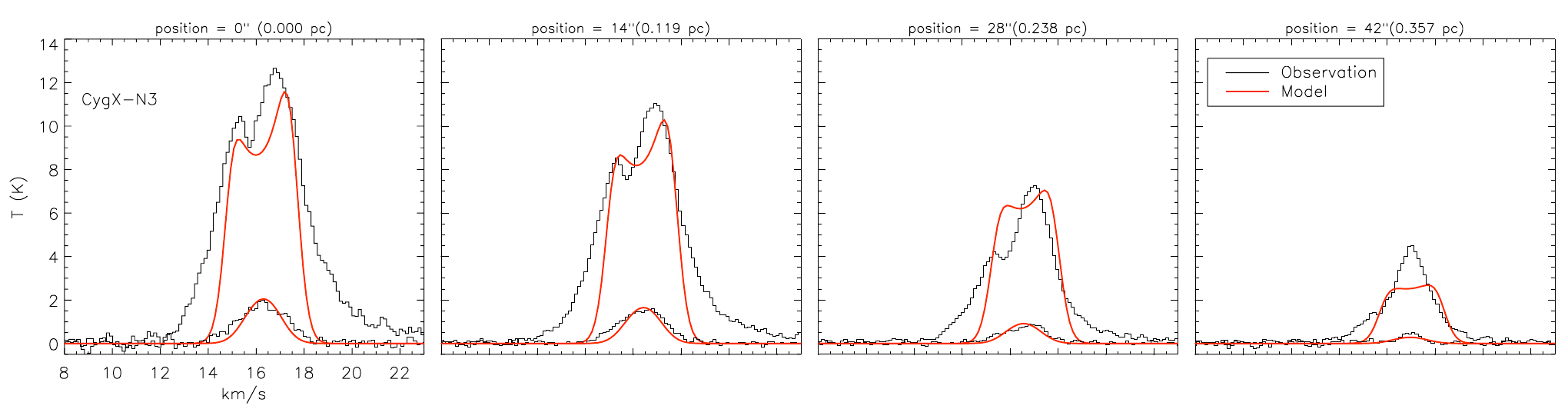}
  \caption{{\hcop} and {\hco} mean spectra computed over concentric annuli in steps of 14{\mbox{$^{\prime\prime}$}} with respect to the map center towards CygX-N3. The black line shows the observations, the red line corresponds to the model.}
         \label{fig:sim-n3}
   \end{figure*}

\begin{figure*}[ht]
   \centering
   \includegraphics[width=18cm]{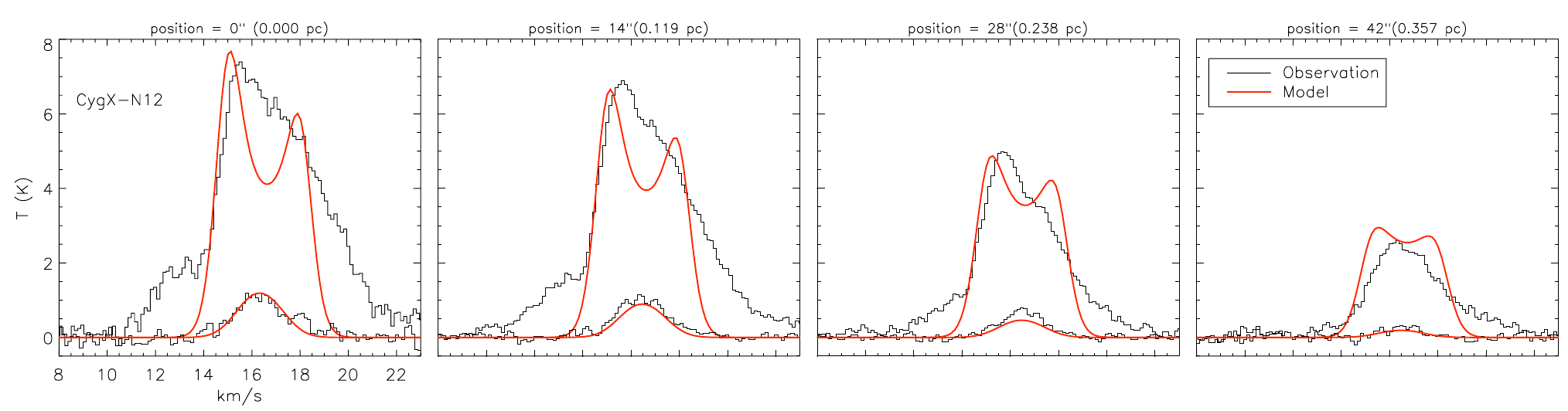}
  \caption{Same as Figure \ref{fig:sim-n3} for CygX-N12.}
         \label{fig:sim-n12}
   \end{figure*}

\begin{figure*}[ht]
   \centering
   \includegraphics[width=18cm]{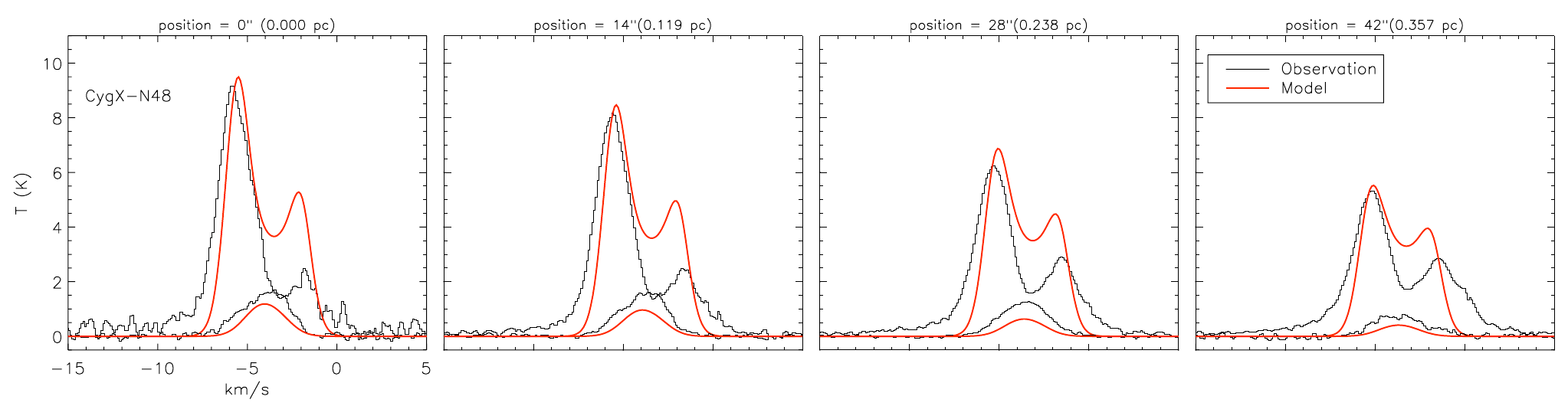}
  \caption{Same as Figure \ref{fig:sim-n3} for CygX-N48.}
         \label{fig:sim-n48}
   \end{figure*}

\begin{figure*}[ht]
   \centering
   \includegraphics[width=18cm]{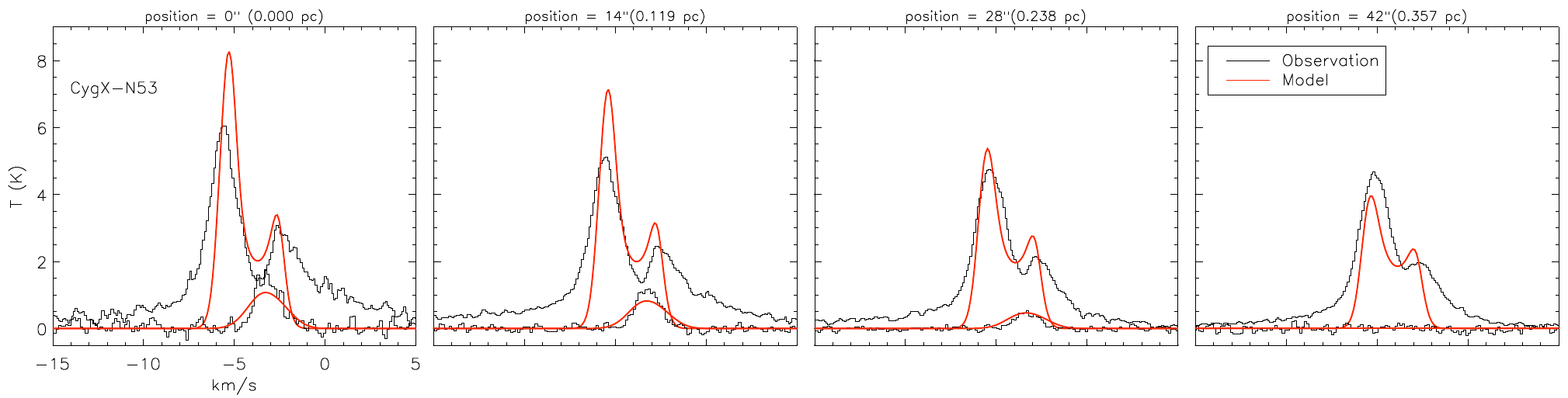}
  \caption{Same as Figure \ref{fig:sim-n3} for CygX-N53.}
         \label{fig:sim-n53}
   \end{figure*}

\begin{figure*}[]
   \centering
   \includegraphics[width=18cm]{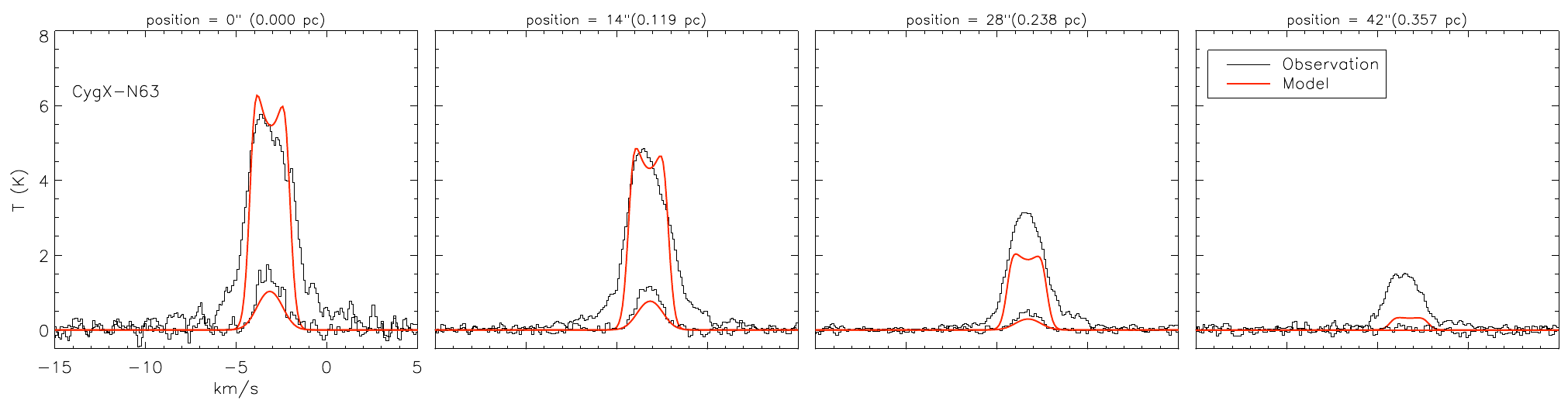}
  \caption{Same as Figure \ref{fig:sim-n3} for CygX-N63.}
        \label{fig:sim-n63}
  \end{figure*}

Figure \ref{fig:sim-model} shows a sketch of the model, the core embedded in a clump and the models of the spectra we get at certain radius. This shows the emission coming only from the core (\textit{a}), then within the radius, where NH$_3$ observations were taken (\textit{b}) and finally the spectra taken through the whole clump (\textit{c}) illustrating that a significant part of the emission is originating from 10$^4$-10$^6$ cm$^{-3}$ densities. Figure \ref{fig:profiles} shows the obtained density, temperature and abundance profiles, respectively.

\paragraph{CygX-N3}

The {\hcop} spectra show red-shifted asymmetry in the line profiles, which may be an indication of expanding gas. Broad line-wings are present due to outflowing gas, which we do not aim to reproduce in our model. We confirm with \textsl{Simline} that \hcop~is optically thick and we obtain a reasonably good fit with an expansion velocity of $\sim$0.3~km~s~$^{-1}$ to the central spectra. The radial fitting of the profile is less good indicating that we either overestimate the optical thickness of the line, or the expansion velocity is not constant. The spectral decomposition shows a second line component at the same velocity of the red-shifted peak, which may indicate, that expansion is not necessarily the reason for the line profile, but an optically thin second component at the same velocity may produce similar line profile as well. Nevertheless it is beyond the scope of this work to perform such a detailed modeling. 


\paragraph{CygX-N12}

Indications for contribution of more than one line-components can be seen in the spectra of CygX-N12, in addition to broad line-wings, which limits our approach to perform a good modeling for this core. Nevertheless our results reproduce the line ratio of {\hcop} and \hco. Note, that the spectra show a shallower shoulder on the red part, which can be indication of infalling material. Therefore we include 0.2~km~s~$^{-1}$ as infall velocity in the model.  

\paragraph{CygX-N48 and -N53}

The clump associated with CygX-N48 and CygX-N53 was modeled in detail in \citet{Schneider_prep}. Here we use the same physical model for all the MDCs, which is adapted to model more precisely the cores embedded within the clump, that has not been considered in \citet{Schneider_prep}. 

\paragraph
{CygX-N63}
CygX-N63 has the weakest \hcop~and \hco~emission compared to the other MDCs and the morphology of the high-resolution emission confirms a hole centered on the continuum peak. Our best model requires a high depletion factor and a lower depletion density ($\sim$2~10$^5$ cm$^{-3}$). Note that the line-profile is similar to CygX-N12, a shallow red shoulder of the spectra may indicate slight infall-motions that we model with 0.1~km~s~$^{-1}$ infall velocity.

\section{Decomposition of spectral line components in \hco}
\label{sec:spec_decomp}

\begin{figure*}[!ht]
   \includegraphics[width=9cm]{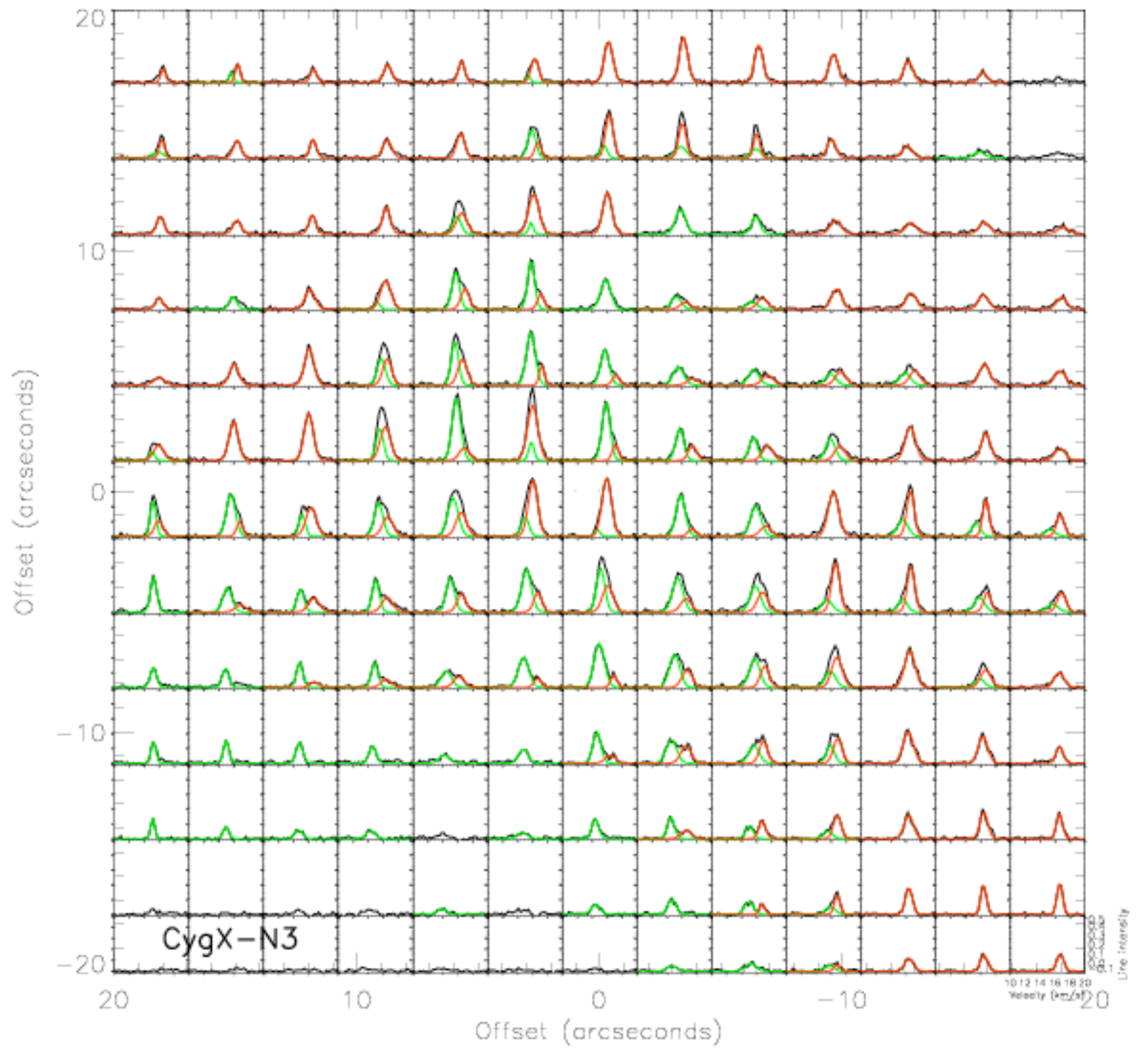}
   \includegraphics[width=9cm]{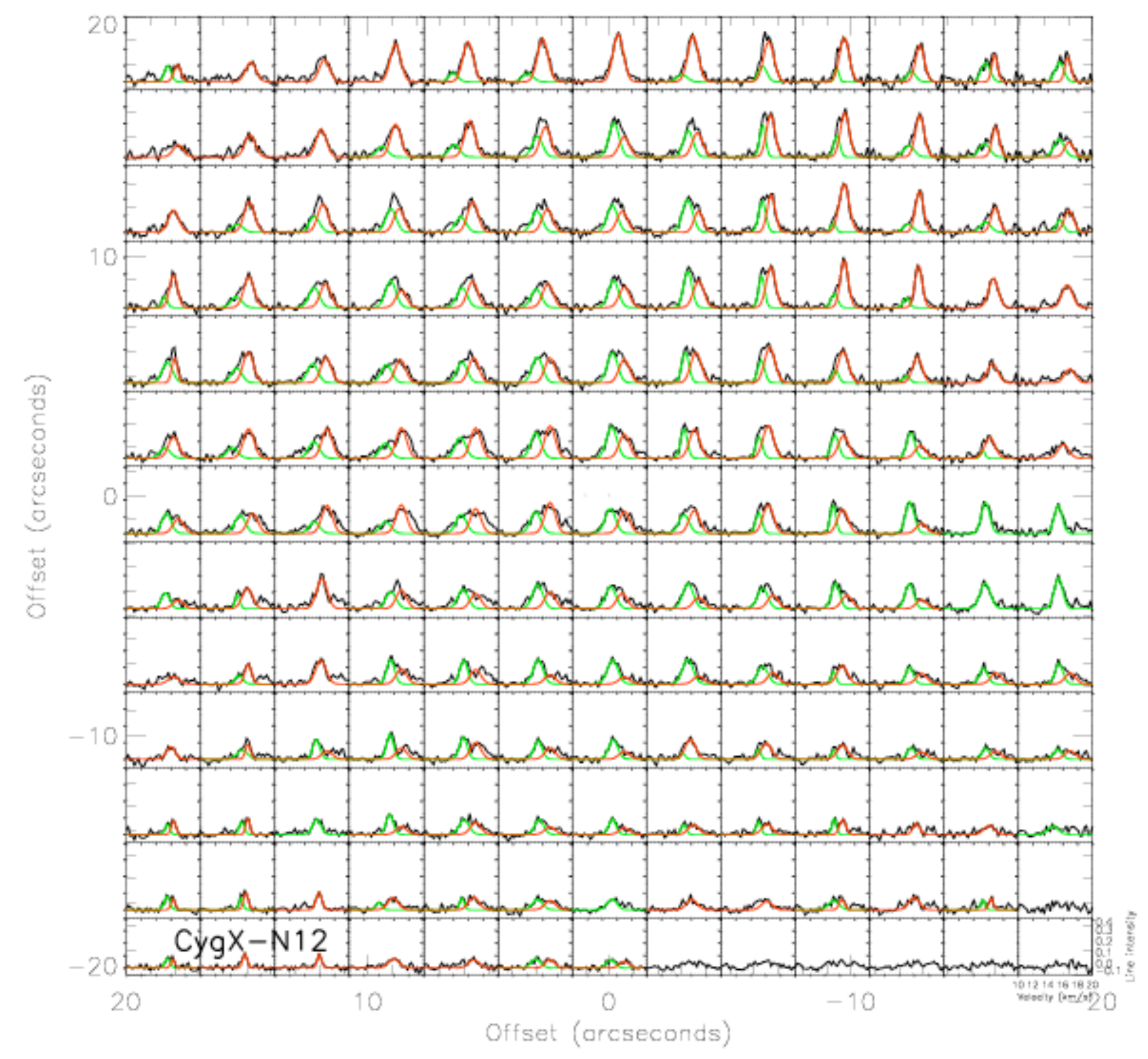}
   \includegraphics[width=9cm]{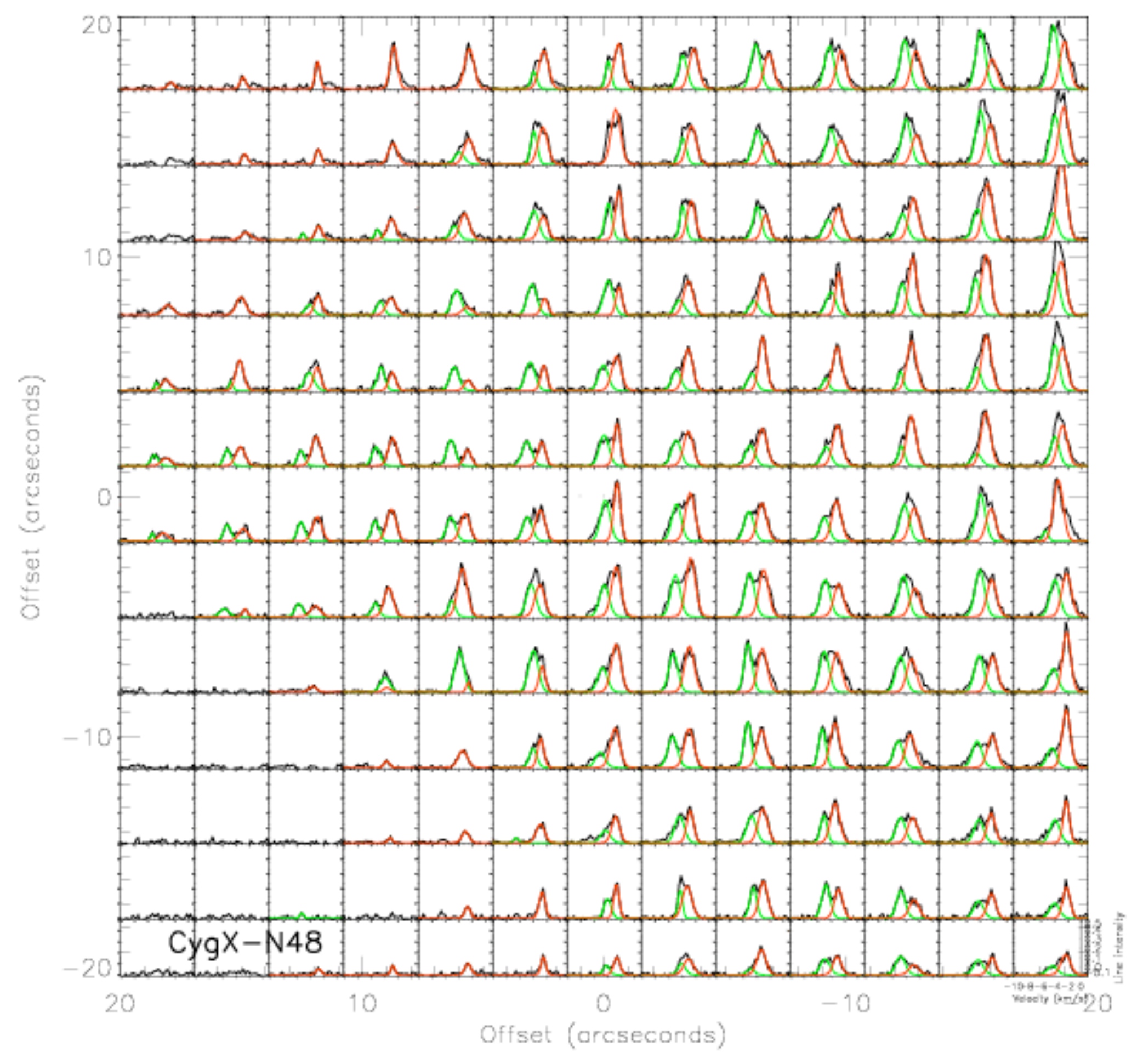}
   \includegraphics[width=9cm]{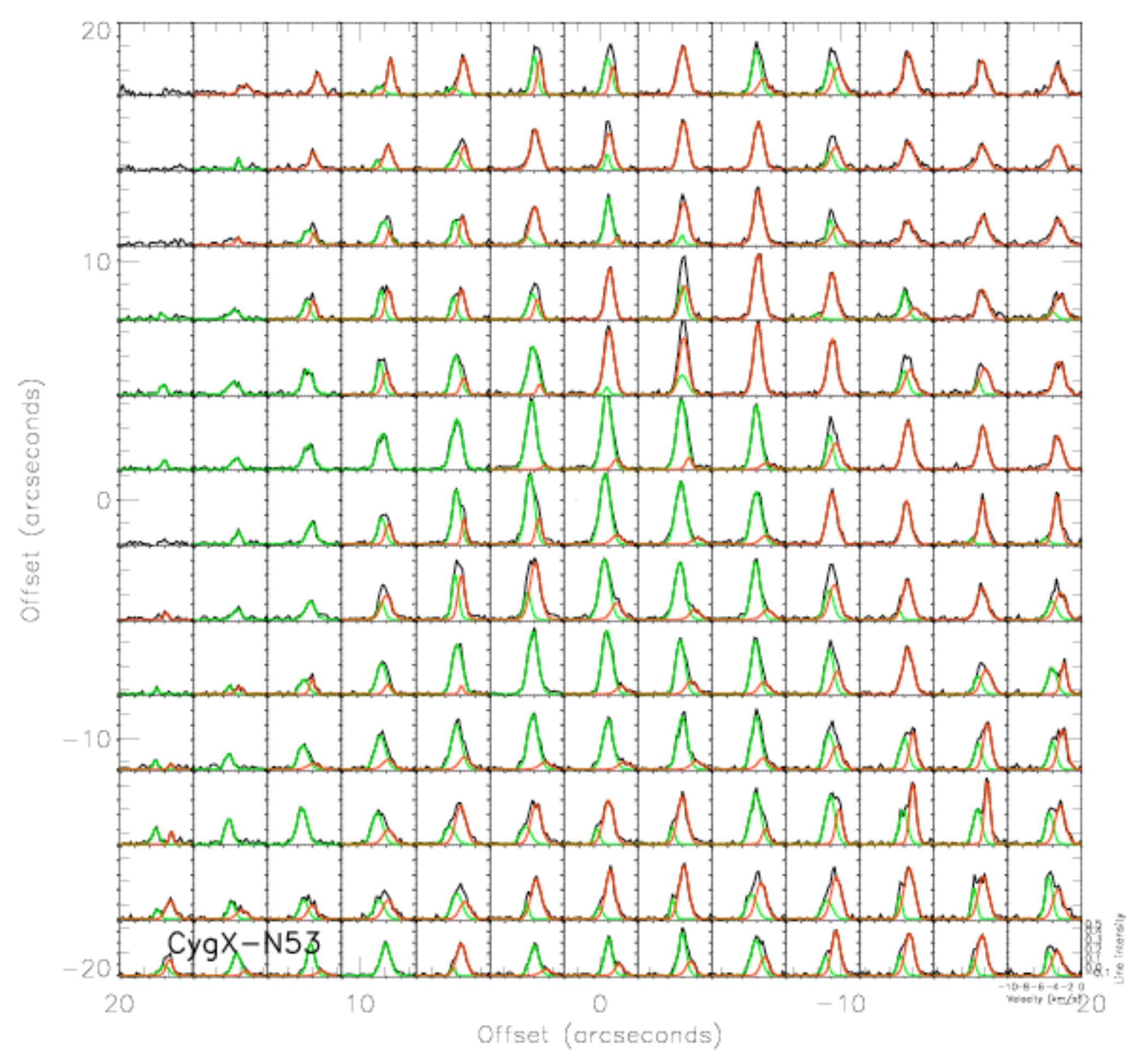}

   \includegraphics[width=9cm]{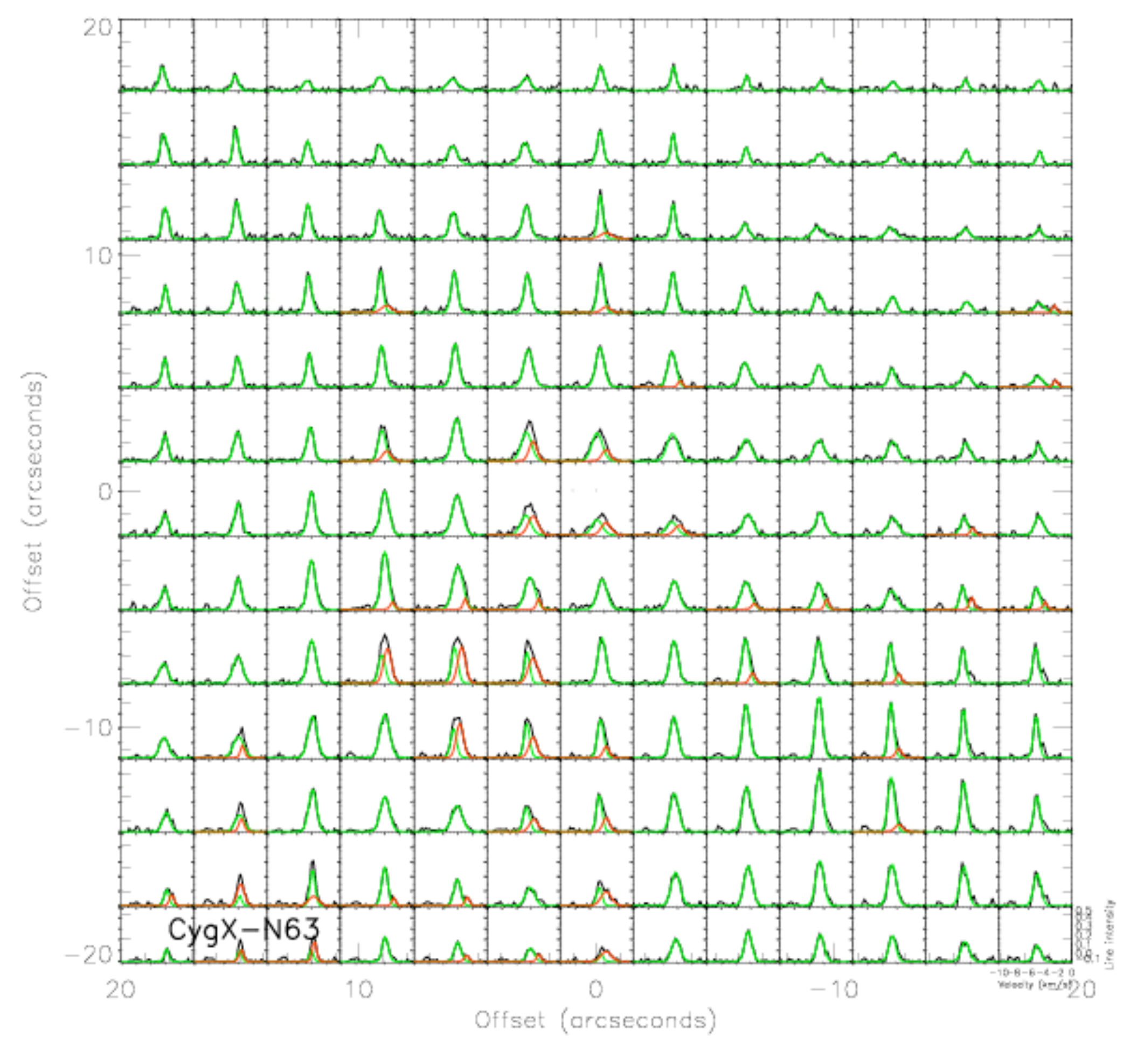}
  \caption{Spectral decomposition of the line-components seen in 30m+PdBI \hco spectra maps. Black line shows the observed spectra, while red and green lines show the resulting fits of two components. We only show the components with an S/N $\textgreater$5.}
         \label{fig:sp_dec}
   \end{figure*}

To  disentangle the individual components seen in the spectral line maps (Figure \ref{fig:sda} {\sl b}) a double gaussian fit was performed to each spectra with an IDL routine using {\sl MPFIT}. Figure \ref{fig:sp_dec} shows the resulting two component fits in green and red. We recall that the noise towards the edge of the map is larger. Individual components are fitted in a fixed velocity range with respect to the v$_{lsr}$. A line is fitted when the intensity is above S/N ratio of $\textgreater$5, otherwise no second component is fitted. It is a tendency that for CygX-N3, -N53 and -N63 the components are best separated in the outer parts of the core, while in the central part only a single gaussian is fitted. Especially, in core CygX-N63, most of the spectra are dominated by one strong component. This tendency is also consistent with the assumption that the double-peaked line-profiles are not due to an opacity effect (thus to self-absorption), because one would expect this to be significant in the positions of the strongest emission associated with the central parts, which is clearly not the case. For CygX-N12 and -N48 the routine picks up well the two components across the whole field. Note that the line-positions of CygX-N48 are well corresponding to the position of the two components seen in {\hcn} (within $\pm$ 0.3~km~s$^{-1}$).
 
We summarized the fit parameters in Table \ref{table:sp_fit}. The identified components in the spectra correspond to an average (where the line intensity is larger, than 5 S/N), since they are not fixed in position but show systematic shifts in velocity.  
\begin{table}
\caption{Result of spectral decomposition of IRAM 30m+PdBI {\hco} lines.}
\begin{tabular}{c c c c c}        
\hline\hline                 
Source & \multicolumn{2}{c}{peak velocity} (v$_{lsr}$) & \multicolumn{2}{c}{line-dispersion ($\sigma$)}  \\    
            & 1st comp. & 2nd comp.  & 1st comp. & 2nd comp.   \\    
            & [km~s$^{-1}$] & [km~s$^{-1}$]  &[km~s$^{-1}$] & [km~s$^{-1}$]   \\    
\hline                        
   CygX-N3   &   15.33 &  16.75 & 0.56 & 0.58 \\     
   CygX-N12 &   15.36 &  16.87 & 0.6   & 0.68   \\
   CygX-N48 &  -4.46 & -2.75 & 0.65 & 0.65   \\
   CygX-N53 & -4.43 & -3.17 & 0.55 &  0.54 \\
   CygX-N63 & -4.39 & -2.89 & 0.49 &  0.36 \\ 
\hline                                  
\end{tabular}
\label{table:sp_fit}      
\end{table}

\section{Coherent flows seen at high spatial resolution - continued}
\label{app:filaments}

CygX-N12 shows several velocity components in the spectra and as mentioned in Section~\ref{sect:flows} due to their complexity it is not possible to determine their spatial structure in our systematic approach. Position-velocity cuts indicate a complex mixture of several line components and the extracted spectral profile confirm this complexity. Individual velocity components appear in a range of $\sim$2.5-3.5~{\kms}.

CygX-N48 shows a similarly complex mixture of individual velocity components with the individual components separated in a range of $\sim$3.5~{\kms}. 

CygX-N53 may have a similar configuration as CygX-N3, but the separation of the individual components and thus the location of velocity shears is less clear.


    \begin{figure*}
   \centering
    \includegraphics[width=0.85\linewidth]{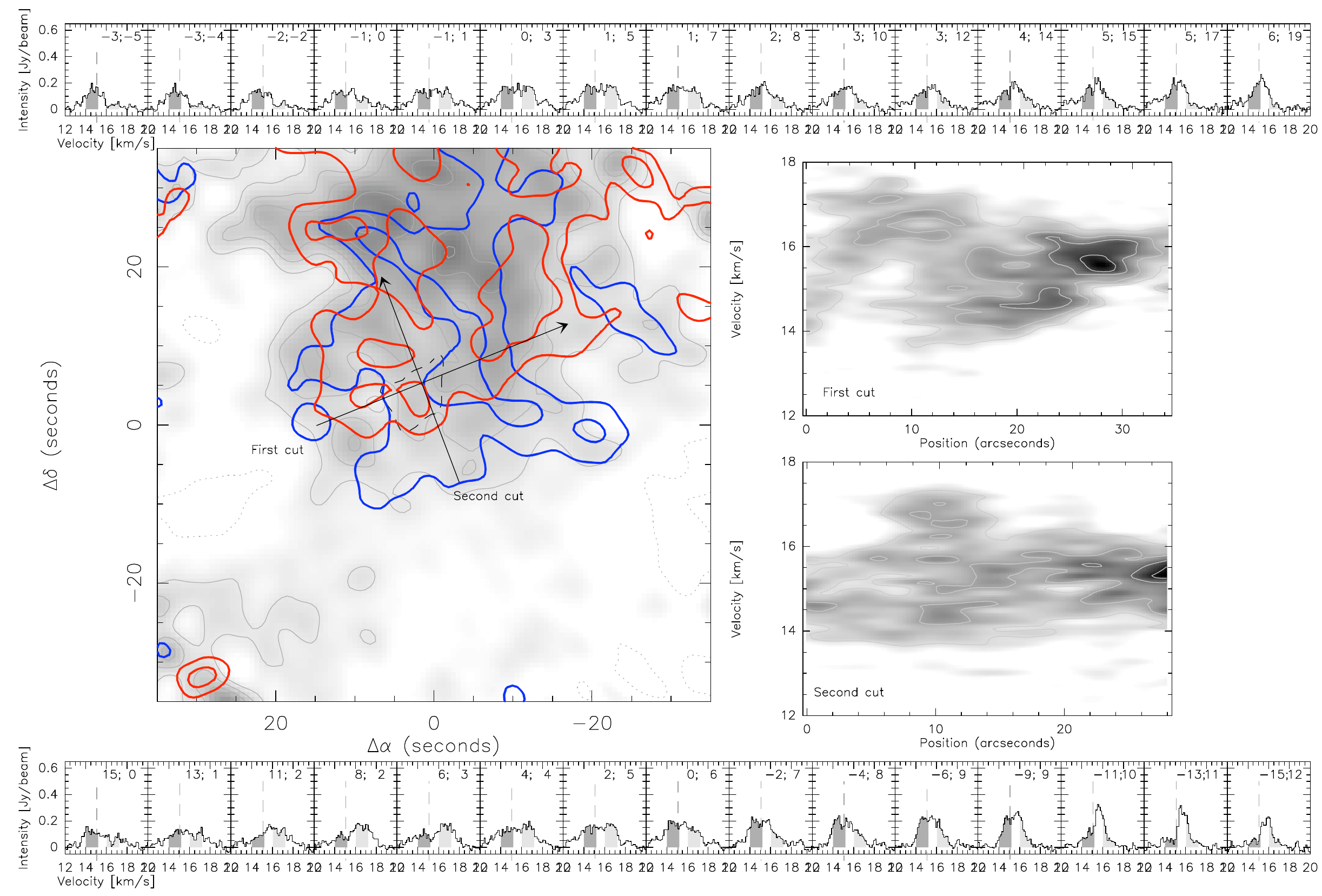}
     \caption{CygX-N12 presented similarly as Figure \ref{fig:N3}. Integration range for the grey scale is 15.1 to 16.1~{\kms}, red contours show an integration range between 16.1 to 17.1~{\kms}, blue contours are integrated between 14.1 to 15.1~{\kms}. Contour levels go from 5$\times$rms noise in steps of 3$\times$rms noise.
}
         \label{fig:N12}
   \end{figure*}

  \begin{figure*}
   \centering
   \includegraphics[width=0.85\linewidth]{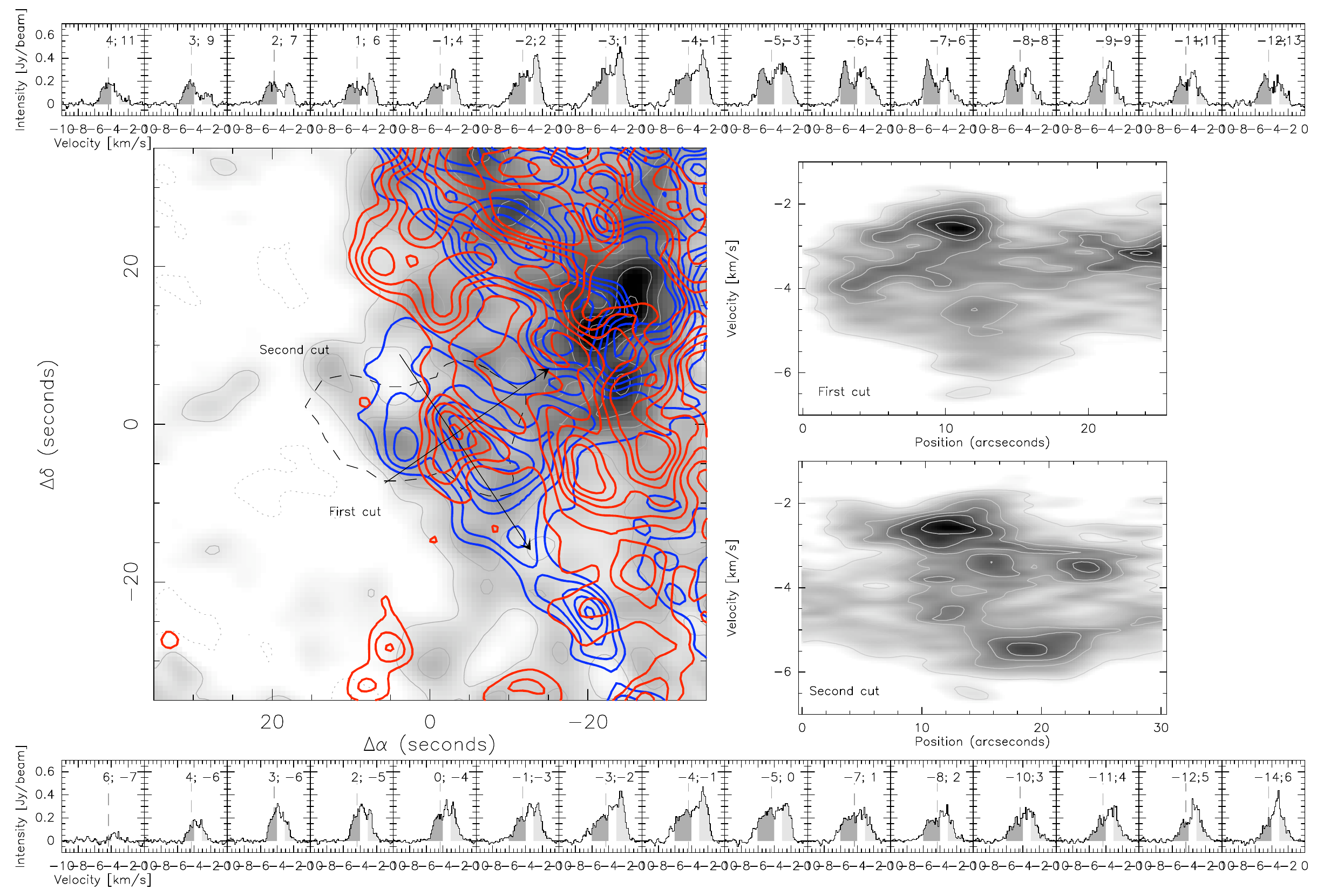}
     \caption{CygX-N48 presented similarly as Figure \ref{fig:N3}. Integration range for the grey scale is $-4$ to $-3$~{\kms}, red contours show an integration range between $-3$ to $-1$~{\kms}, blue contours are integrated between $-6$ to $-4$~{\kms}. Contour levels go from 10$\times$rms noise in steps of 3$\times$rms noise.}
         \label{fig:N48}
   \end{figure*}

   \begin{figure*}
   \centering
    \includegraphics[width=0.85\linewidth]{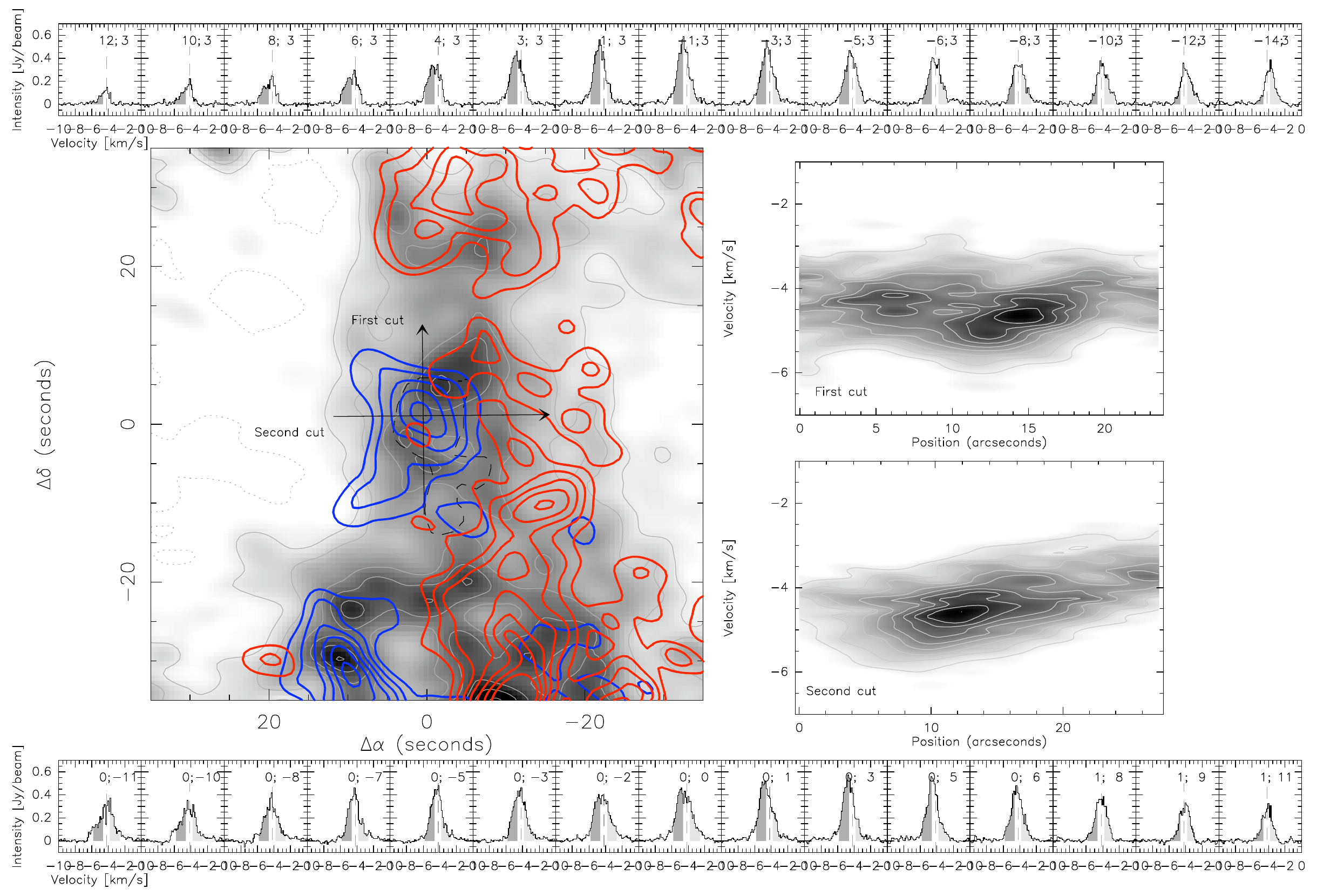}
     \caption{CygX-N53 presented similarly as Figure \ref{fig:N3}. Integration range for the grey scale is $-4.7$ to $-3.7$~{\kms}, red contours show an integration range between $-3.7$ to $-2.7$~{\kms}, blue contours are integrated between $-5.7$ to $-4.7$~{\kms}. Contour levels go from 10$\times$rms noise in steps of 3$\times$rms noise.}
         \label{fig:N53}
   \end{figure*}


%

 
\end{appendix}

\bibliography{h13co_2ndversion} 	

\end{document}